\documentclass[apj]{emulateapj}

\usepackage{apjfonts}
\usepackage{verbatim}
\usepackage{float}
\usepackage[export]{adjustbox}
\usepackage{graphicx}
\usepackage{subfigure}
\usepackage{appendix}

\newcommand{\water}{H$_2$O} 
\newcommand{\methanol}{CH$_3$OH}
\newcommand{\ammonia}{NH$_3$}
\newcommand{\lsun}{$L_\odot$}
\newcommand{\msun}{$M_\odot$}
\newcommand{\rsun}{$R_\odot$}
\newcommand{\HII}{H\,{\sc ii}}
 
\newcommand{\mjb}{mJy~beam$^{-1}$}

\newcommand{\jyb}{Jy~beam$^{-1}$}

\newcommand{\Mjsr}{MJy~sr$^{-1}$}
\newcommand{\kms}{km~s$^{-1}$}  
\newcommand{\mum}{$\mu$m}

\newcommand{\Tk}{$T_{kin}$}
\newcommand{\cc}{cm$^{-3}$}

\newcommand{\chisq}{$\chi^{2}$}
\newcommand{\rstar}{$R_{\star}$}
\newcommand{\tstar}{$T_{\star}$}
\newcommand{\lstar}{$L_{\star}$}
\newcommand{\mstar}{$M_{\star}$}
\newcommand{\age}{$t_{\star}$}

\begin{document}

\title{SOFIA {\it FORCAST} Photometry of 12 Extended Green Objects in the Milky Way}

\author{A. P. M. Towner\altaffilmark{1*,2}, C. L. Brogan\altaffilmark{1}, T. R.  Hunter\altaffilmark{1},
C. J. Cyganowski\altaffilmark{3}, R. K. Friesen\altaffilmark{1}}

\altaffiltext{1}{National Radio Astronomy Observatory, 520 Edgemont Rd, Charlottesville, VA 22903, USA}
\altaffiltext{2}{Department of Astronomy, University of Virginia, P.O. Box 3818, Charlottesville, VA 22903, USA}
\altaffiltext{3}{Scottish Universities Physics Alliance (SUPA), School of Physics and Astronomy, University of St. Andrews, North Haugh, St Andrews, Fife KY16 9SS, UK}
\altaffiltext{*}{A.P.M.T. is a Grote Reber Doctoral Fellow at the National Radio Astronomy Observatory.}

\begin{abstract}
Massive young stellar objects are known to undergo an evolutionary phase in which high mass accretion rates drive strong outflows. A class of objects believed to trace this phase accurately is the GLIMPSE Extended Green Object (EGO) sample, so named for the presence of extended 4.5 $\mu$m emission on sizescales of $\sim$0.1 pc in \textit{Spitzer} images. 
We have been conducting a multi-wavelength examination of a sample of 12 EGOs with distances of 1 to 5 kpc. In this paper, we present mid-infrared images and photometry of these EGOs obtained with the SOFIA telescope, and subsequently construct SEDs for these sources from the near-IR to sub-millimeter regimes using additional archival data.
We compare the results from greybody models and several publicly-available software packages which produce model SEDs in the context of a single massive protostar. 
The models yield typical \rstar\/$\sim$10 \rsun, \tstar\/$\sim$10$^3$ to 10$^4$ K, and \lstar\/$\sim$1~$-$~40~$\times$~10$^3$ \lsun; the median $L/M$ for our sample is 24.7 \lsun/\msun.
Model results rarely converge for \rstar\/ and \tstar, but do for \lstar, which we take to be an indication of the multiplicity and inherently clustered nature of these sources even though, typically, only a single source dominates in the mid-infrared. 
The median $L/M$ value for the sample suggests that these objects may be in a transitional stage between the commonly described ``IR-quiet'' and ``IR-bright'' stages of MYSO evolution. The median $T_{dust}$ for the sample is less conclusive, but suggests that these objects are either in this transitional stage or occupy the cooler (and presumably younger) part of the IR-bright stage.
\end{abstract}

\section{Introduction}
Massive young stellar objects (MYSOs) are challenging to observe due to their comparative rarity and short-lived natal phase, large distances from Earth, and highly-obscured formation environments.
Early observations of suspected MYSOs were performed mostly with large beams, and probed size scales ranging from cores to clumps and clouds \citep[$\sim$0.1~pc, $\sim$1~pc, and $\sim$10~pc, respectively; see][]{Kennicutt2012}. Detailed descriptions of early surveys for MYSOs and their results can be found in, e.g., \citet{Molinari1996}, \citet{Sridharan2002}, and \citet{Fontani2005}. Follow-up observations with improved sensitivity and spatial resolution, such as interferometric radio and millimeter observations, revealed that many of the objects originally identified as ``MYSOs'' were actually sites in which multiple protostars were forming simultaneously \citep[e.g.][to name just a few]{Hunter2006,Cyganowski2007,Vig2007,Zhang2007}.
This predilection for forming in clustered environments means that the study of high-mass protostars is necessarily the study of proto\textit{clusters}: clusters of protostars with a range of masses and in a variety of evolutionary stages.
Current theories of high mass star formation differ in their predictions of the aggregate properties of these protoclusters, such as mass segregation (if any), sub-clustering of the protostars, and stellar birth order \citep[e.g.][]{VazquezSemadeni2017,Banerjee2017,Bonnell2006,McKee2003}.
It is therefore necessary to consider each high-mass protostar in combination with its environment.

Extended Green Objects (EGOs) were first identified by \citet{egocat} using data from the Galactic Legacy Infrared Midplane Survey Extraordinaire \citep[GLIMPSE,][]{GLIMPSE2003,GLIMPSE2009} project. EGOs are named for their extended emission in the 4.5~\mum\/ \textit{Spitzer} IRAC band (commonly coded as ``green'' in three-color RGB images), which is due to shocked H$_2$ from powerful protostellar outflows \citep[e.g.,][]{Marston2004}. 
Follow-up observations of $\sim$20 EGOs with the Karl G. Jansky Very Large Array (VLA) by \citet{Cyganowski2009} established both the presence of massive protostars (traced by 6.7~GHz Class II \methanol\/ masers) and shocked molecular gas indicative of outflows (traced by 44~GHz Class I \methanol\/ masers).  
The causal link between accretion and ejection \citep{Frank2014} thus implies that these objects contain protostars undergoing active accretion, and the maser data indicate that these protostars are massive.  The youth of the massive protostars within these EGOs was confirmed by deep (at that time) VLA continuum observations \citep{Cyganowski2011b}, which yielded only a few 3.6~cm detections, and by later VLA 1.3~cm continuum observations \citep{Towner2017}, which revealed primarily weak ($<$1 \mjb), compact emission. 
The low detection rates and integrated flux densities of the centimeter continuum emission in these sources demonstrate that any free-free emission is weak, consistent with a stage prior to the development of ultracompact HII regions.

Given that high-mass stars form in clusters, it is likely that EGOs are signposts for protoclusters rather than isolated high-mass protostars, though the level of multiplicity of massive sources ($>$8\msun) and overall cluster demographics remain open questions. Millimeter dust continuum observations of EGOs with $\sim$3$\arcsec$ resolution, suggest that the number of massive protostars per EGO is typically one to a few \citep[e.g.][]{Cyganowski2012,Cyganowski2011a, Brogan2011}. 
However, the precise physical properties of protoclusters traced by EGO emission - such as total mass, luminosity, and massive protostellar multiplicity - remain largely unexplored in EGOs as a class.

The infrared emission from EGOs, and indeed MYSOs in general, is often challenging to characterize due to the presence of high extinction from their surrounding natal clumps (as they are still deeply embedded), and confusion from 
more evolved sources nearby. The latter issue has been particularly affected by the relatively poor angular resolution ($> 1\arcmin$) that has heretofore been available at mid- and far-infrared wavelengths, where the high extinction can be overcome. Yet these wavelengths contain crucial information
as hot dust, shocked gas, and polycyclic aromatic hydrocarbons (PAHs) all emit in this regime. Scattered light originating from the protostar itself may also sometimes escape through outflow cavities and would likewise be visible in the infrared. Thus mid-infrared wavelengths are a crucial component of the Spectral Energy Distribution (SED) which is a useful tool for constraining important source properties such as mass, bolometric luminosity, and temperature.

These properties are of particular interest for MYSOs, as
recent analysis of the Herschel InfraRed Galactic Plane Survey
\citep{Elia2017} and a full census of the properties of ATLASGAL Compact Source Catalog (CSC) objects \citet{Urquhart2018} shows how the luminosity to mass ratio $L/M$ of protostellar clumps can be used to both qualitatively and quantitatively discriminate between the different evolutionary stages of pre- and protostellar objects. 
In theoretical terms, $L/M$ is tied to evolutionary state primarily due to abrupt changes in luminosity during different stages of MYSO/clump evolution \citep[see, e.g., the stages described in][]{Hosokawa2009,Molinari2008}.

In this paper, we present new data that directly address the questions of the multiplicity and physical properties (temperature, mass, and luminosity) of the massive protoclusters traced by EGOs.
We have utilized the unique capabilities of the Stratospheric Observatory for Infrared Astronomy \citep[SOFIA,][]{Temi2014} to image a well-studied sample of 12 EGOs at two mid-IR wavelengths: 19.7 and 37.1 \mum\/
with the necessary sensitivity ($\sim$0.05 to $\sim$0.25 \jyb) and angular resolution ($\sim$ 3$\arcsec$) to detect and resolve the mid-infrared emission from the massive protocluster members.
By combining these results with ancillary multi-wavelength archival data, we create well-constrained SEDs from the near-infrared through submillimeter regimes. We then use three SED modelling packages published by \citet{R06}, \citet{R17}, and \citet{ZT18}, to constrain physical parameters \citep[see, e.g.,][]{CarinaNebula,soma}.
In \S~\ref{obs_all}, we describe our targeted SOFIA observations and the observational details of the archival data at each wavelength. In \S~\ref{results_all}, we describe our aperture-photometry procedures for each data set and discuss our detection rates and trends. We also present sets of multi-scale, multiwavelength images for each object in order to better demonstrate their small- and large-scale properties and overall environments.
In \S~\ref{analysis}, 
we compare the physical parameters obtained from the various SED modeling methods, including $L/M$, which help to place EGOs into a broader evolutionary context.
In \S~\ref{future} we discuss the implications of our results, and outline future investigations.

\section{The Sample \& Observations}
\label{obs_all}
In this paper, we conduct a multiwavelength aperture-photometry study of 12 EGOs using the SOFIA Faint Object infraRed CAmera for the SOFIA Telescope \citep[FORCAST][]{Herter2012}. We use new SOFIA FORCAST 19~\mum\/ and 37~\mum\/ observations in conjunction with publicly-available archival datasets from {\it Spitzer}, {\it Herschel}, and the Atacama Pathfinder EXperiment\footnote{This publication is based on data acquired with the Atacama Pathfinder Experiment (APEX). APEX is a collaboration between the Max-Planck-Institut fur Radioastronomie, the European Southern Observatory, and the Onsala Space Observatory.} (APEX) telescope, to model the SED of the dominant protostar in each of our target EGOs. 
Details of source properties for our sample are listed in Table~\ref{sourceproperties}.

\begin{deluxetable*}{lcccccccc}
\tablecaption{EGO Source Properties} 
\tablecolumns{9} 
\tabletypesize{\footnotesize} 
\tablehead{ \colhead{Source} & \colhead{$V_{\rm LSR}^{\rm a}$} & \colhead{Distance$^{\rm b}$} & \colhead{EGO$^{\rm c}$} & \colhead{IRDC$^{\rm d}$} & \colhead{H$_2$O$^{\rm e}$} & \multicolumn{3}{c}{CH$_3$OH Masers (GHz)$^{\rm f}$}\\
              & \colhead{(\kms\/)}            & \colhead{(kpc)}             & \colhead{Cat} &      & \colhead{Maser} & \colhead{6.7$^{\rm g}$} & \colhead{44$^{\rm h}$} & \colhead{95$^{\rm i}$}
}
\startdata 
G10.29$-$0.13 & 14 & 1.9                                    & 2 & Y  &  Y  & Y  & Y  & Y\\ 
G10.34$-$0.14 & 12 & 1.6                                    & 2 & Y  &  Y  & Y  & Y  & Y\\ 
G11.92$-$0.61 & 36 & 3.38$^{+0.33}_{-0.27}$ (3.5)           & 1 & Y  &  Y  & Y  & Y  & Y\\ 
G12.91$-$0.03 & 57 & 4.5                                    & 1 & Y  &  Y  & Y  & ?  & Y\\ 
G14.33$-$0.64 & 23 & 1.13$^{+0.14}_{-0.11}$ (2.3)           & 1 & Y  &  Y  & ?  & Y  & Y\\ 
G14.63$-$0.58 & 19 & 1.83$^{+0.08}_{-0.07}$ (1.9)           & 1 & Y  &  Y  & Y  & ?  & Y\\ 
G16.59$-$0.05 & 60 & 3.58$^{+0.32}_{-0.27}$ (4.2)           & 2 & N  &  Y  & Y  & ?  & Y\\ 
G18.89$-$0.47 & 66 & 4.2                                    & 1 & Y  &  Y  & Y  & Y  & Y\\ 
G19.36$-$0.03 & 27 & 2.2                                    & 2 & Y  &  N  & Y  & Y  & Y\\ 
G22.04$+$0.22 & 51 & 3.4                                    & 1 & Y  &  Y  & Y  & Y  & Y\\ 
G28.83$-$0.25 & 87 & 4.8                                    & 1 & Y  &  Y  & Y  & Y  & ?\\ 
G35.03$+$0.35 & 53 & 2.32$^{+0.24}_{-0.20}$ (3.2)           & 1 & Y  &  Y  & Y  & Y  & Y 
\enddata 
\tablenotetext{a}{LSRK velocities are the single dish NH$_3$ (1,1) values from \citet{Nobeyama}.}
\tablenotetext{b}{Distances without errors are estimated from the LSRK velocity and the Galactic rotation curve parameters from \citet[][]{Reid2014}. Parallax distances (with their uncertainties) are given where available from \citet[][]{Reid2014} and references therein, with the kinematic distance in parentheses for comparison. All kinematic distances are the near distance. The uncertainty on each kinematic distance is assumed to be 15\%, based on the median percent difference between the parallax-derived and kinematic distances from the five sources which have both.}    
\tablenotetext{c}{This is the Table number of the EGO in \citet{egocat}. In that paper, Tables 1 \& 2 list ``likely'' EGOs for which 5-band (3.6 to 24 \mum\/) or only 4.5 \mum\/ {\em Spitzer} photometry can be measured, respectively.}
\tablenotetext{d}{Coincidence of EGO with IRDC as indicated by \citet{egocat}.}
\tablenotetext{e}{Water maser data from the \citet{Nobeyama} Nobeyama 45-m survey of EGOs.}
\tablenotetext{f}{Sources for which we could find no information in the literature are indicated by ``?".}
\tablenotetext{g}{The 6.7~GHz maser detection information comes from \citet{Cyganowski2009} using the VLA, except for G12.91$-$0.03, G14.63$-$0.58, and G16.59$-$0.05, which come from \citet[][and references therein]{Green2010} observations using the Australia Telescope Compact Array (ATCA).}
\tablenotetext{h}{Information for 44~GHz masers come from the VLA and were taken from \citet[][]{Cyganowski2009}, except for G14.33$-$0.64, which comes from \citet{Slysh1999}.}
\tablenotetext{i}{Most information for 95~GHz masers was taken from \citet{Chen2011} using the Mopra 22~m telescope. The exceptions are G14.33$-$0.64 from \citet{Valtts2000} using Mopra, G16.59$-$0.05 from \citet{Chen2012} using the Purple Mountain Observatory 13.7~m telescope, and G35.03+0.35 from \citet{Kang2015} using the Korean VLBA Network.}
\label{sourceproperties}
\end{deluxetable*}

\subsection{SOFIA FORCAST Observations: 19.7 \& 37.1~\mum}
\label{sofiaobs}
We used SOFIA FORCAST to observe our 12 targets simultaneously at 19.7~\mum\/ and 37.1~\mum. 
Observations were performed in the asymmetric chop-and-nod imaging observing mode C2NC2. 
The measured\footnote{These FWHM are the average values in dual-channel mode for each wavelength as measured by the SOFIA team since Cycle 3. More information can be found in the Cycle 5 Observer's Handbook on the SOFIA website at https://www.sofia.usra.edu/science/proposing-and-observing/sofia-observers-handbook-cycle-5} FWHM are 2$\farcs$5 at 19.7~\mum\/ and 3$\farcs$4 at 37.1~\mum. At the nearest (1.13 kpc) and farthest (4.8 kpc) source distances, these FWHM correspond to physical size scales of 2,830 to 12,000 au at 19.7~\mum\/ and 3,840 to 16,300 au at 37.1~\mum. The instantaneous field of view (FOV) of FORCAST is 3$\farcm$4 $\times$ 3$\farcm$2, with pixel size $\theta$ = 0$\farcs$768 after distortion correction. This FOV corresponds to 1.1 $\times$ 1.1 pc at a distance of 1.13 kpc, and 4.8 $\times$ 4.5 pc at a distance of 4.8 kpc.
Table~\ref{sofia_obstable} summarizes observation information for each EGO. 
The project's Plan ID is 04\_0159.

\begin{deluxetable*}{lcccccc}[!hbt]
\tablecaption{SOFIA FORCAST Observing Parameters}
\tablefontsize{\scriptsize}
\tablehead{
\colhead{Source} & \multicolumn{2}{c}{Pointing Center (J2000)} & \colhead{Obs. Date$^{a}$} & \colhead{TOS$^{b}$} & \multicolumn{2}{c}{$\sigma$ (MAD)$^{c}$}\\
 & \colhead{RA} & \colhead{Dec} & & \colhead{(s)} & \colhead{37~\mum} & \colhead{19~\mum}
}
\startdata
G10.29$-$0.13 & 18:08:49.2 & -20:05:59.3 & 2016 July 13 & 502  & 0.26   & 0.08\\
G10.34$-$0.14 & 18:08:59.9 & -20:03:37.3 & 2016 Sept 27 & 626  & 0.30   & 0.07\\
G11.92$-$0.61 & 18:13:58.0 & -18:54:19.3 & 2016 July 12 & 604  & 0.22   & 0.07\\
G12.91$-$0.03 & 18:13:48.1 & -17:45:41.3 & 2016 July 19 & 1000 & 0.18   & 0.04\\
G14.33$-$0.64 & 18:18:54.3 & -16:47:48.3 & 2016 July 12 & 593  & 0.24   & 0.07\\
G14.63$-$0.58 & 18:19:15.3 & -16:29:57.3 & 2016 July 13 & 641  & 0.22   & 0.07\\
G16.59$-$0.05 & 18:21:09.0 & -14:31:50.3 & 2016 July 20 & 810  & 0.19   & 0.04\\
G18.89$-$0.47 & 18:27:07.8 & -12:41:38.3 & 2016 Sept 27 & 626  & 0.25   & 0.06\\
G19.36$-$0.03 & 18:26:25.7 & -12:03:56.3 & 2016 Sept 20 & 285  & 0.46   & 0.09\\
G22.04$+$0.22 & 18:30:34.6 & -09:34:49.3 & 2016 Sept 20 & 642  & 0.21   & 0.05\\
G28.83$-$0.25 & 18:44:51.2 & -03:45:50.3 & 2016 Sept 27 & 470  & 0.26   & 0.07\\
G35.03$+$0.35 & 18:54:00.4 & +02:01:15.7 & 2016 Sept 22 & 500  & 0.29   & 0.08
\enddata
\tablenotetext{a}{All July observations were performed on flights from Christchurch, New Zealand; all September observations were performed on flights from Palmdale, CA, USA.}
\tablenotetext{b}{This column lists the total time on source (TOS) for each target. The original proposal called for 600~s of integration on each source. For four sources, 600~s could not be achieved due to either high clouds (G19.36) or telescope issues (G10.29, G28.83, G35.03). G12.91 was a shared observation with another group whose observations required additional integration time.}
\tablenotetext{c}{The background noise of the SOFIA images is non-Gaussian in the majority of sources. This column gives the scaled ${\it MAD} = 1.482\times{\it MAD}$ values for all sources, where {\it MAD} is the median absolute deviation from the median background pixel value; {\it MAD} must be multiplied by 1.482 to become rms-like. Aperture photometry was performed using cutoffs based on {\it MAD} for all sources. {\it MAD} values listed here are in \jyb.}
\label{sofia_obstable}
\end{deluxetable*}

Data calibration and reduction are performed by the SOFIA team using the SOFIA data-reduction pipeline\footnote{The FORCAST Data Handbook can be found on the SOFIA website at https://www.sofia.usra.edu/science/proposing-and-observing/data-products}.
After receipt of the Level 3 data products (artifact-corrected, flux-calibrated images), we converted our images from Jy~pixel$^{-1}$ to \jyb\/ in order to more easily perform photometric measurements in CASA \citep{McMullin2007}.
Conversion was accomplished by using the CASA task {\tt immath} to multiply each image by the beam-to-pixel conversion factor $X_\lambda = (beam \ area)/(pixel \ area)$. This factor depends on beam size and pixel size, and therefore is different for each wavelength.
The beam-to-pixel conversion factors are X$_{19.7 \mu m}$ = 12.0067 pixels/beam and X$_{37.1 \mu m}$ = 22.2076 pixels/beam.

\subsection{Archival Data}
\subsubsection{Spitzer IRAC (GLIMPSE) Observations: 3.6, 5.8, \& 8.0~\mum}
All of our EGO targets were originally selected due to their extended emission at 4.5~\mum\/ as seen in  {\it Spitzer} GLIMPSE images.
In order to constrain the SEDs of the driving sources themselves, we used the archival {\it Spitzer} observations at 3.6~\mum, 5.8~\mum, and 8.0~\mum\/ (bands I1, I3, and I4, respectively) from the GLIMPSE project \citep{GLIMPSE2003,GLIMPSE2009}.
The point response function (PRF) of the IRAC instrument varies by band and position on the detector.
The mean FWHM in bands I1, I3, and I4 are 1$\farcs$66, 1$\farcs72$, and 1$\farcs$88, respectively, as detailed in \citet{Fazio2004}.
All archival GLIMPSE data were downloaded from the NASA/IPAC Infrared Science Archive (IRSA) Gator Catalog List.
The images returned by the archive are all in units of \Mjsr.

\subsubsection{Spitzer MIPS (MIPSGAL) Observations: 24~\mum}
\label{mipsobs}
We utilized archival 24~\mum\/ data from the MIPSGAL survey to provide additional mid-IR constraints on our SEDs for 9 of our 12 targets.
For the remaining 3 targets (G14.33, G16.59, G35.03), MIPSGAL 24~\mum\/ data could not be used for the second task due to saturated pixels in the regions of interest.
MIPSGAL images have a native brightness unit of \Mjsr, and were converted to \jyb\/ by first multiplying each image by 1$\times$10$^6$ (to convert from MJy to Jy) and then multiplying by the solid angle subtended by the 6$\farcs$0 $\times$ 6$\farcs$0 MIPS beam at 24~\mum.
Technical details of the MIPS instrument can be found in \citet{RiekeMIPS}.
For details of the MIPSGAL observing program, see \citet{mipsgal2009} and \citet{mipsgal2015}.
All MIPSGAL data were downloaded from the IRSA Gator Catalog List.

\subsubsection{Herschel PACS (Hi-GAL) Observations: 70 \& 160~\mum}
We used archival 70~\mum\/ and 160~\mum\/ data from the Herschel Infrared Galactic Plane Survey \citep[Hi-GAL,][]{Molinari2016}, observed with the {\it Herschel} Photoconductor Array Camera and Spectrometer \citep[PACS;][]{Poglitsch2010} instrument, to probe the far-IR portion of the spectrum. 
These data were originally observed as part of the Herschel Hi-GAL project \citep{Molinari2010,Molinari2016} between 2010 October 25 and 2011 November 05.
The observations were performed in parallel mode with a scan speed of 60$''$/s.
Beam sizes, which are dependent on observing mode, were $\theta_{70\mu m}$ = 5.8$'' \times$ 12.1$''$ and $\theta_{160\mu m}$ = 11.4$'' \times$ 13.4$''$ as reported in \citet{Molinari2016}.
The native brightness unit of the Hi-GAL data is \Mjsr.
Therefore, these images were converted to \jyb\/ using the same method as in \S\ref{mipsobs}.

We chose to use the Hi-GAL data over the archival PACS data available on the European Space Agency (ESA) Heritage Archive due to the additional astrometric and absolute flux calibration performed by the Hi-GAL team, as detailed in \citet{Molinari2016}.
All Hi-GAL data were obtained from the Hi-GAL Catalog and Image Server on the Via Lactea web portal\footnote{http://vialactea.iaps.inaf.it/vialactea/eng/index.php}.

\subsubsection{APEX LABOCA (ATLASGAL) Observations: 870~\mum}
We used archival 870~\mum\/ observations from the APEX Telescope Large Area Survey of the Galaxy \citep[ATLASGAL,][]{Schuller2009} to populate the submillimeter portion of the SED. The data were retrieved from the ATLASGAL Database Server\footnote{\mbox{http://atlasgal.mpifr-bonn.mpg.de/cgi-bin/ATLASGAL\_DATABASE.cgi}}.
The ATLASGAL beam size is 19$\farcs$2 $\times$ 19$\farcs$2; additional observational details can be found in \citet{Schuller2009}.
These images were already in units of \jyb, and thus required no unit conversion.

\section{Results}
\label{results_all}
\begin{figure*}[hbt]
    \centering
    \subfigure{
    \includegraphics[width=0.9\textwidth]{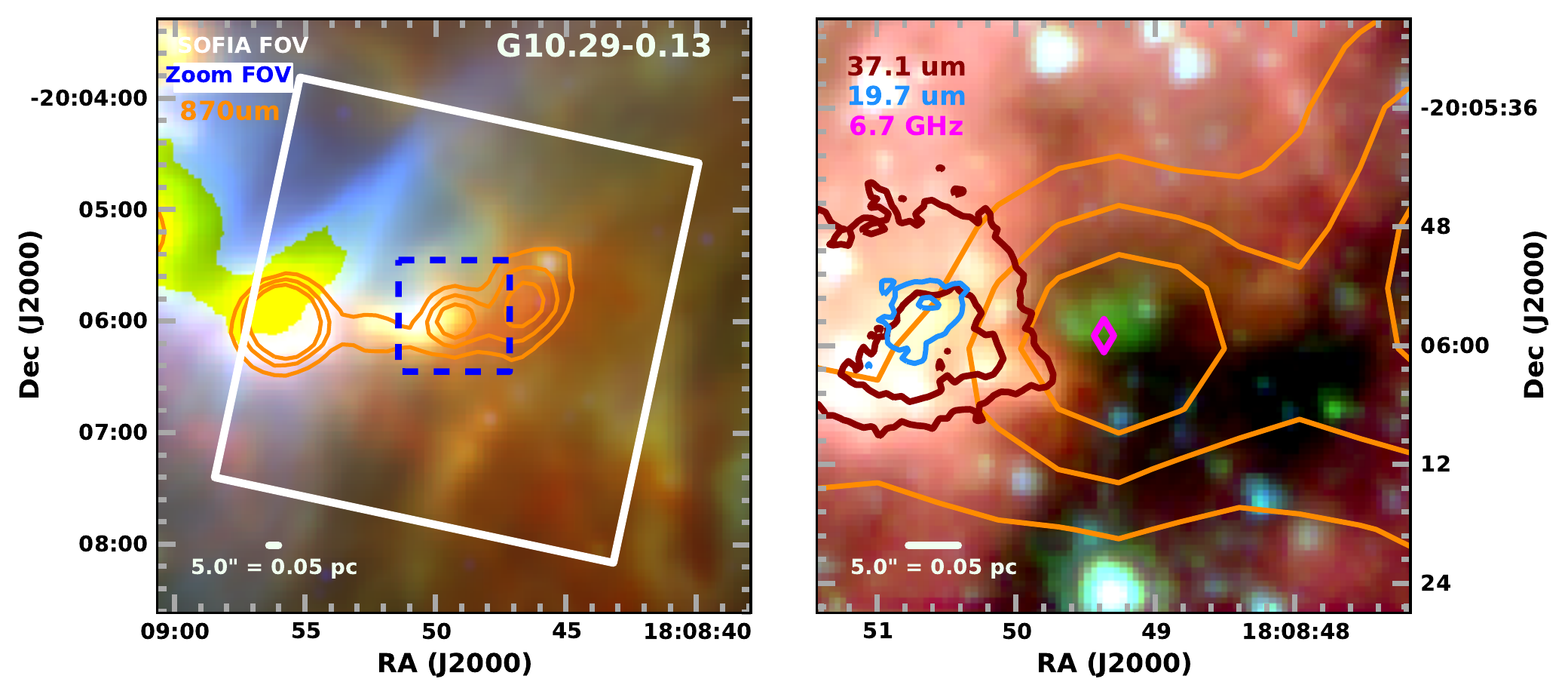}}
    \\
    \centering
    \subfigure{
    \includegraphics[width=0.9\textwidth]{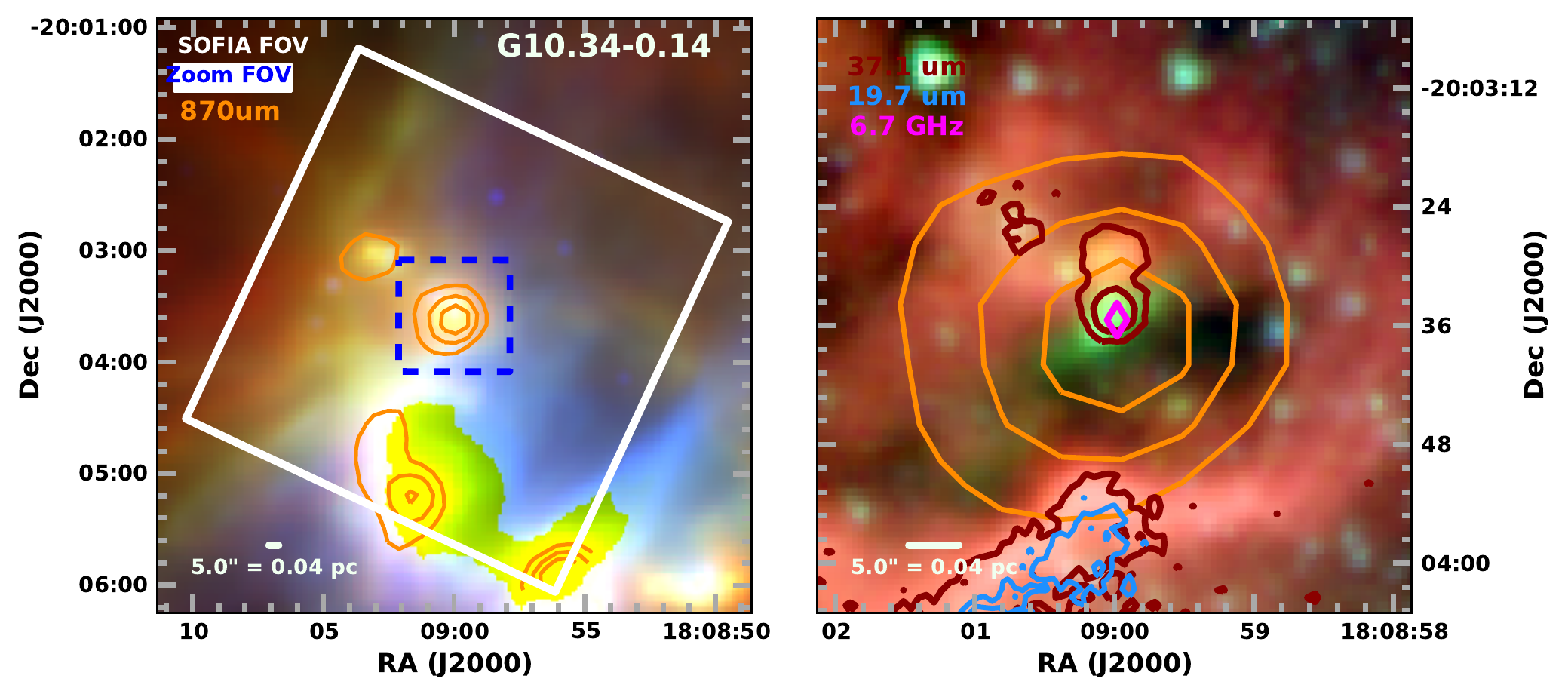}}
    \\
    \centering
    \subfigure{
    \includegraphics[width=0.9\textwidth]{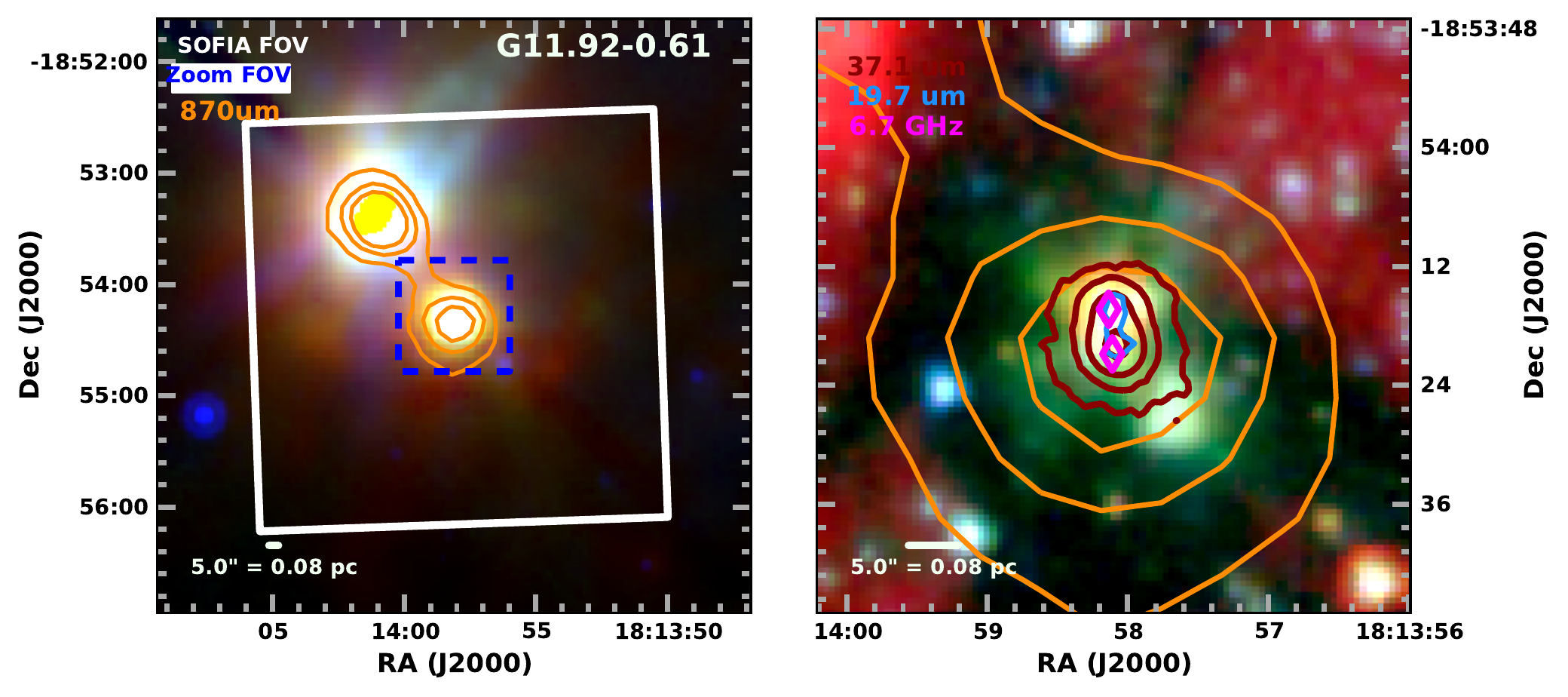}}
    \caption{RGB images for EGO sources. The left panel for each source shows the 160~\mum, 70~\mum, and 24 \mum\/ wavelengths mapped to R, G, and B, respectively, with 870~\mum\/ contours overlaid in magenta. The ATLASGAL contour levels are [0.25, 0.5, 0.75]$\times$I$_{max}$, where $I_{max}$ is the peak intensity value of the ATLASGAL data in each FOV. The solid white boxes show the size and approximate orientation of the SOFIA FOV, and the dashed blue boxes show the size and position of the FOV of the zoomed images, shown in the right-hand panels. The right-hand panels show the 8.0~\mum, 4.5~\mum, and 3.6~\mum\/ wavelengths mapped to R, G, and B, respectively; the extended green emission shows the extent of each EGO. SOFIA FORCAST 19.7~\mum\/ and 37.1~\mum\/ contours are overlaid in blue and red, respectively, and the positions of known 6.7~GHz Class II \methanol\/ masers are denoted by magenta diamonds. SOFIA contour levels are [5,15,45,125,250]$\times \sigma$ for 37.1~\mum, and [4,8,16,28]$\times \sigma$ for 19.7~\mum, where $\sigma$ is the scaled MAD. The $I_{max}$ values for G10.29$-$0.13, G10.34$-$0.14, and G11.92$-$0.61 are 3.31 \jyb, 4.39 \jyb, and 4.01 \jyb, respectively.}
    \label{3color_1}
\end{figure*}

\begin{figure*}
    \centering
    \subfigure{
    \includegraphics[width=0.9\textwidth]{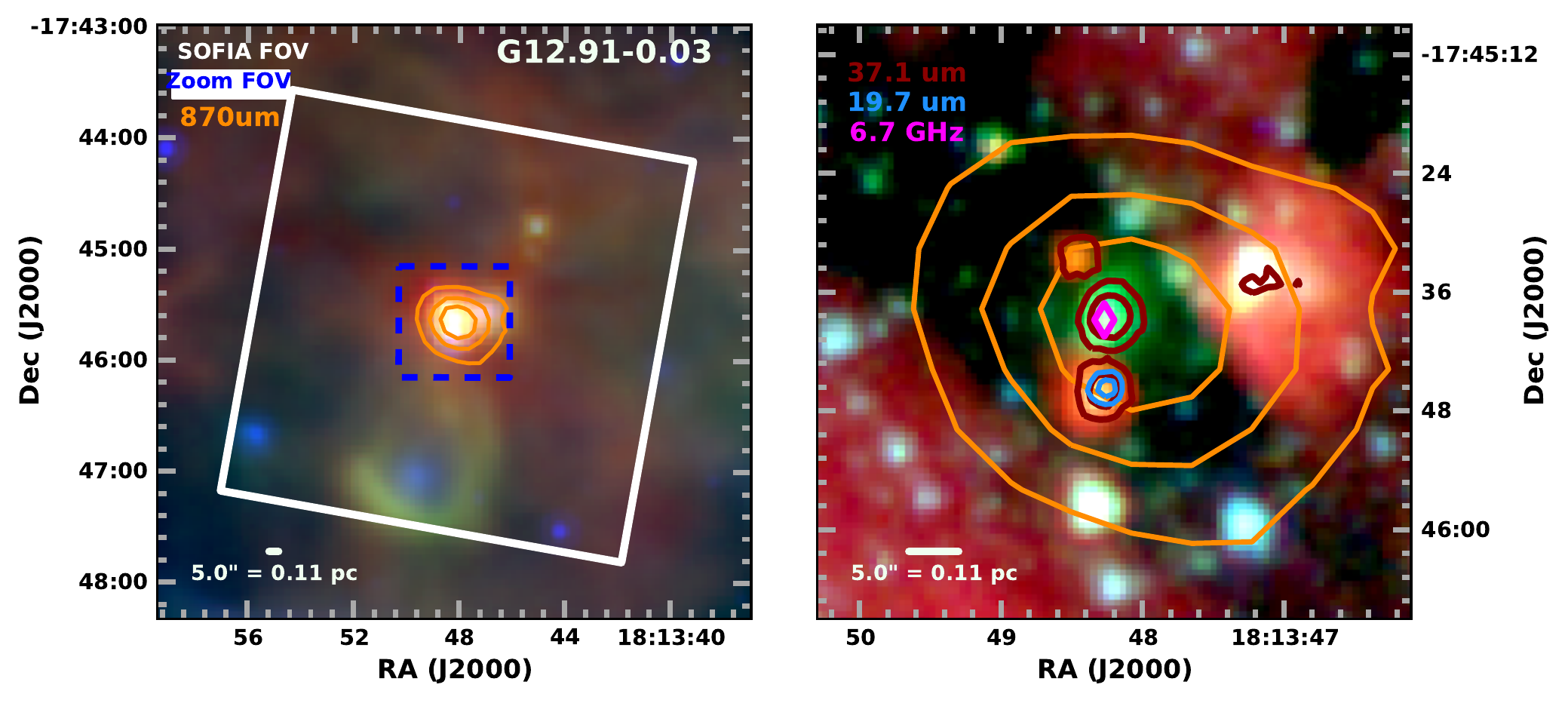}}
    \\
    \centering
    \subfigure{
    \includegraphics[width=0.9\textwidth]{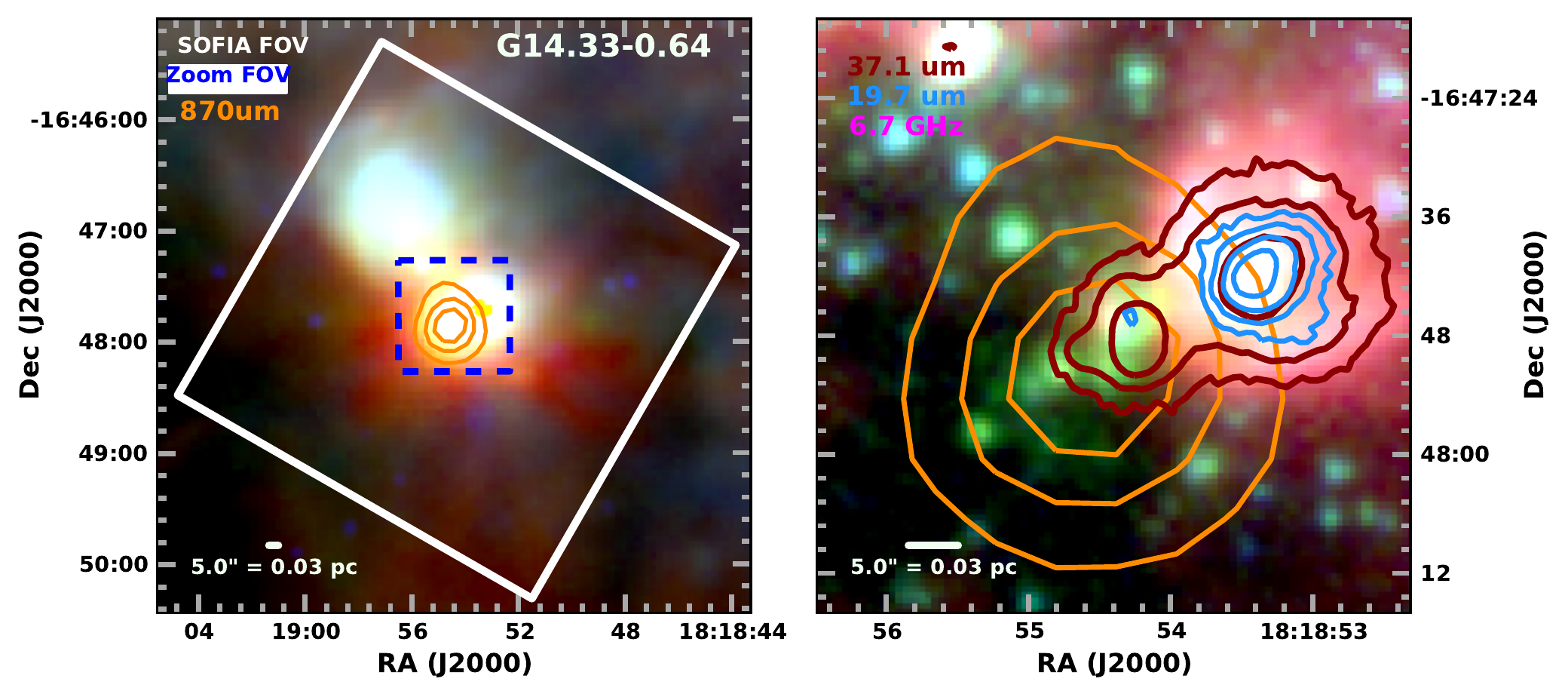}}
    \\
    \centering
    \subfigure{
    \includegraphics[width=0.9\textwidth]{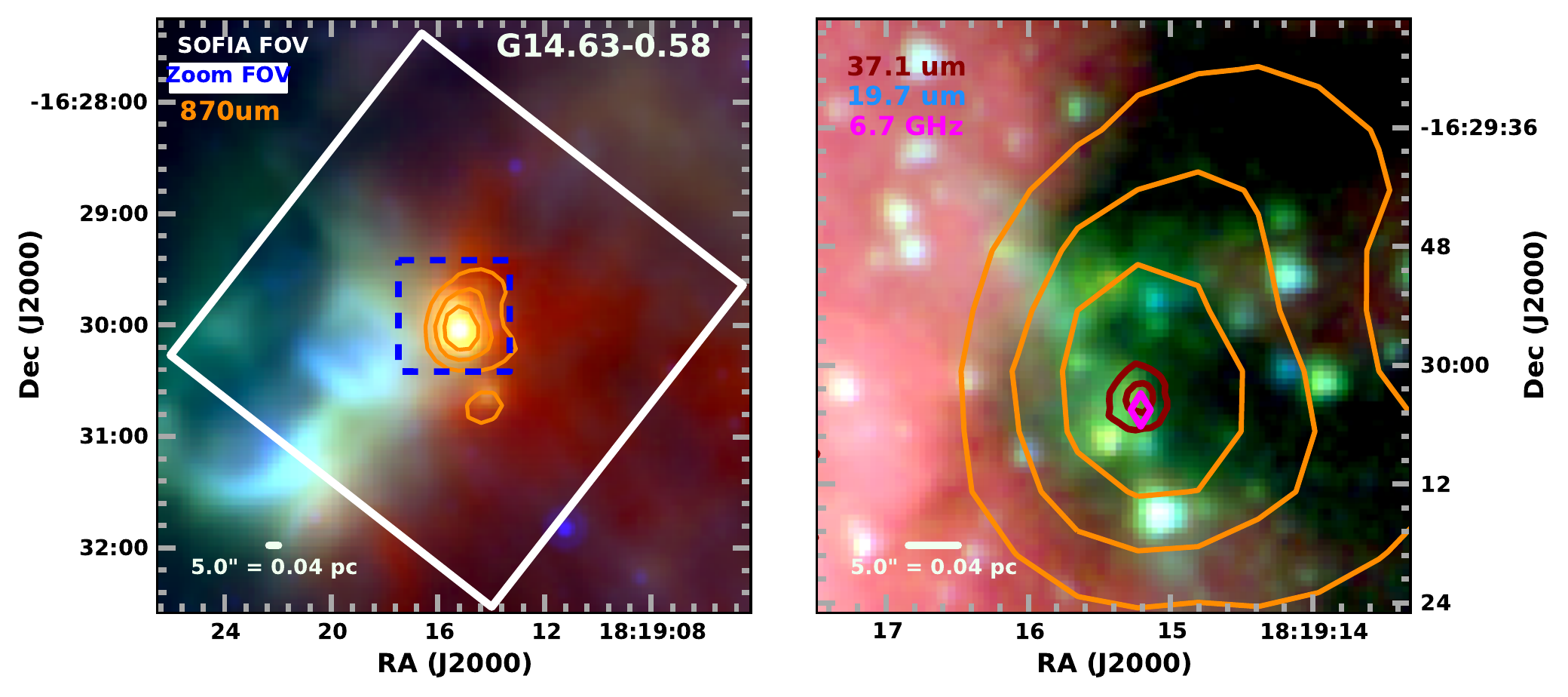}}
    \caption{RGB images for EGO sources. See Figure~\ref{3color_1} for key. The $I_{max}$ values for G12.91$-$0.03, G14.33$-$0.64, and G14.63$-$0.58 are 2.78 \jyb, 12.98 \jyb, and 4.35 \jyb, respectively.}
    \label{3color_2}
\end{figure*}

\begin{figure*}[hbt]
    \centering
    \subfigure{
    \includegraphics[width=0.9\textwidth]{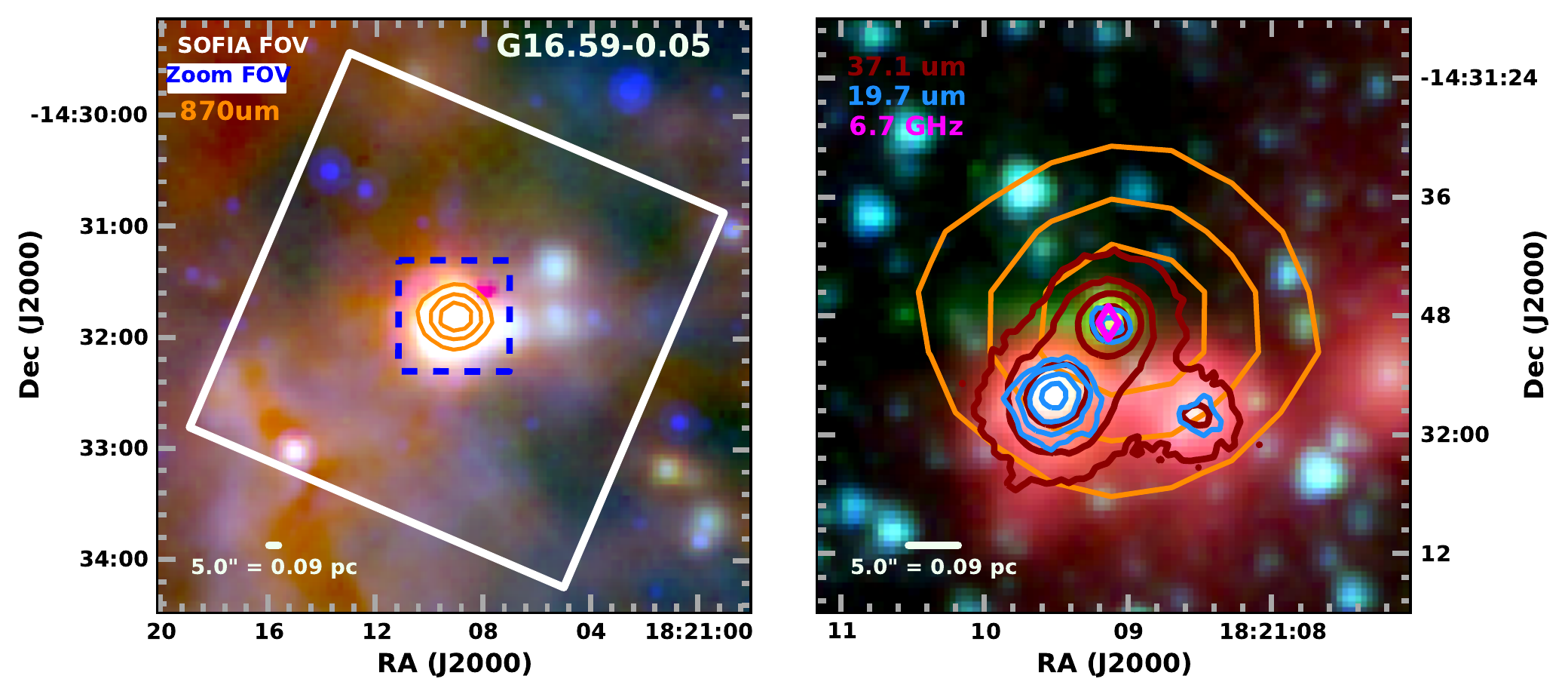}}
    \\
    \centering
    \subfigure{
    \includegraphics[width=0.9\textwidth]{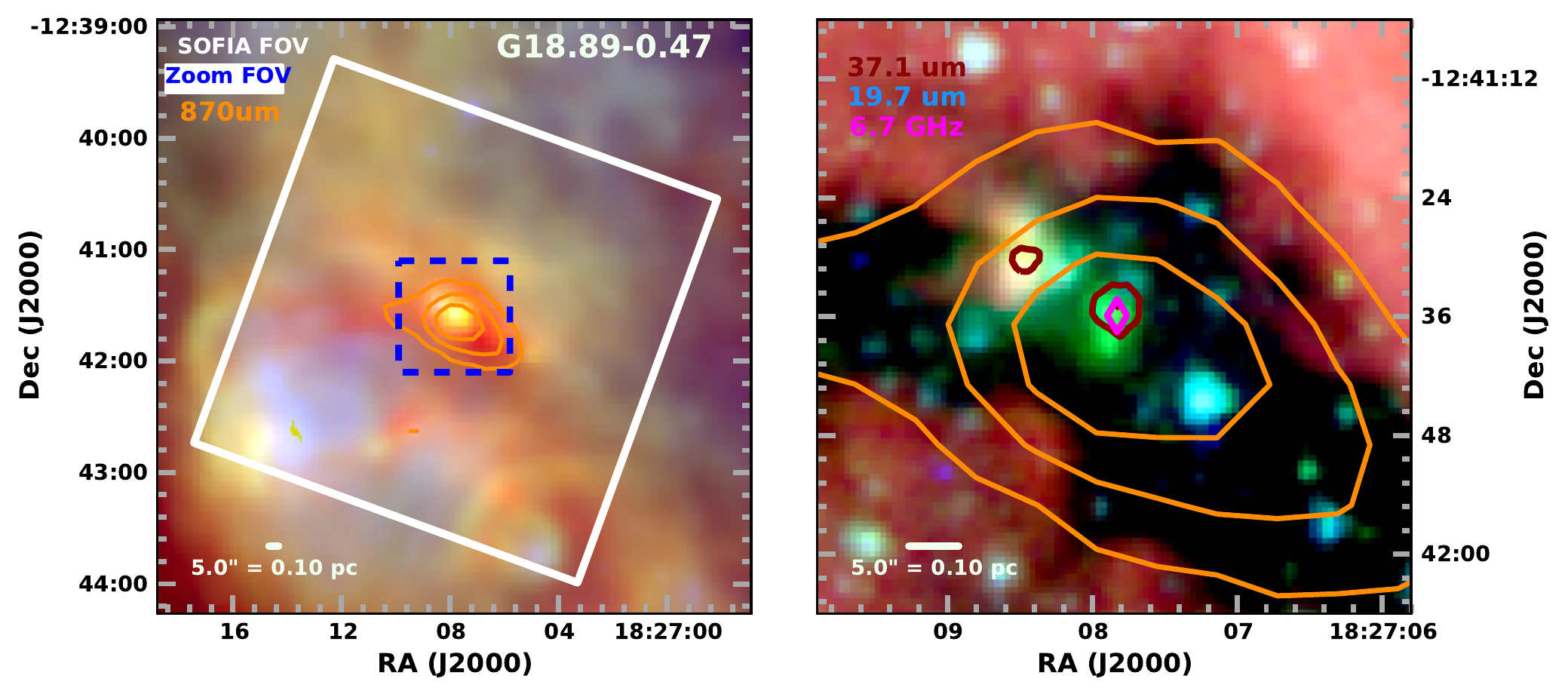}}
    \\
    \centering
    \subfigure{
    \includegraphics[width=0.9\textwidth]{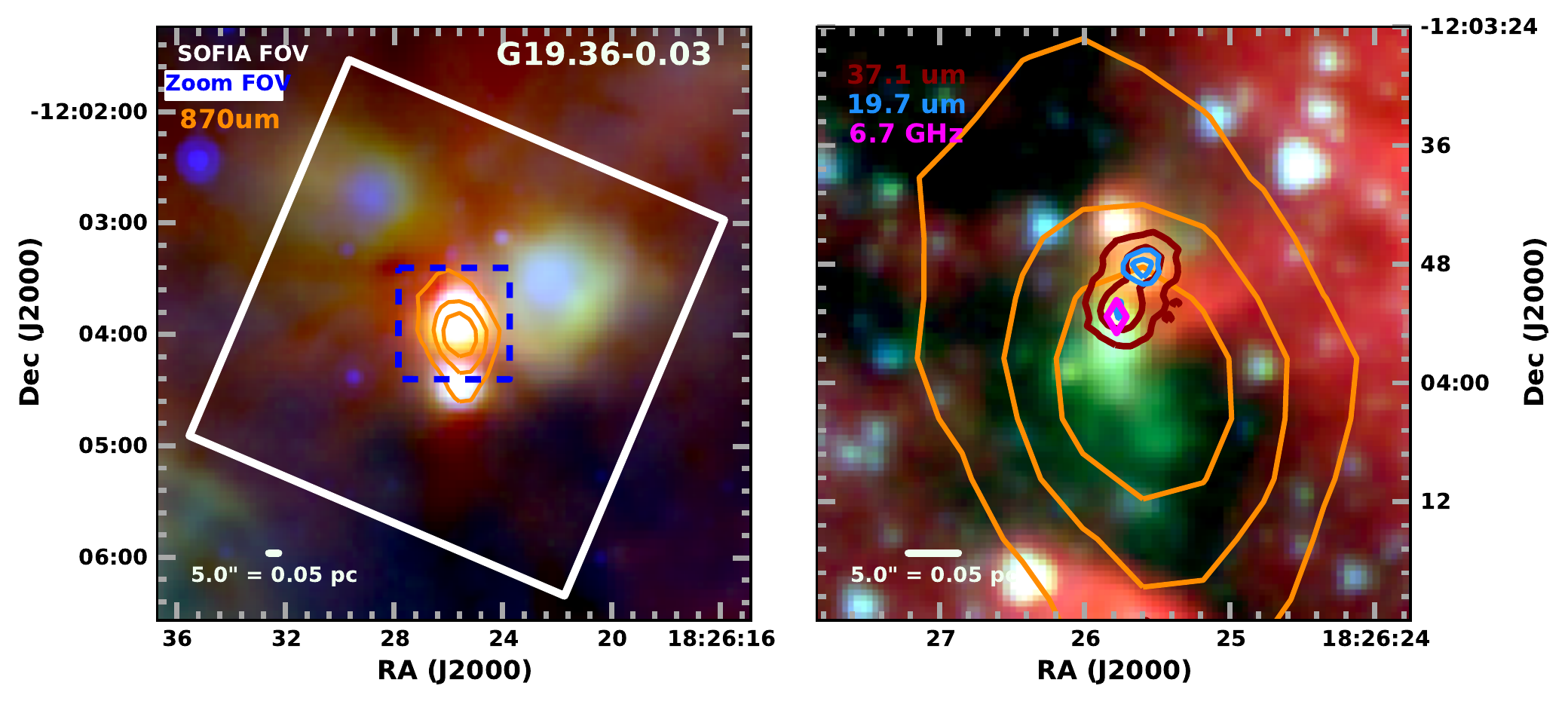}}
    \caption{RGB images for EGO sources. See Figure~\ref{3color_1} for key. The $I_{max}$ values for G16.59$-$0.05, G18.89$-$0.47, and G19.36$-$0.03 are 5.13 \jyb, 3.30 \jyb, and 2.90 \jyb, respectively.}
    \label{3color_3}
\end{figure*}

\begin{figure*}
    \centering
    \subfigure{
    \includegraphics[width=0.9\textwidth]{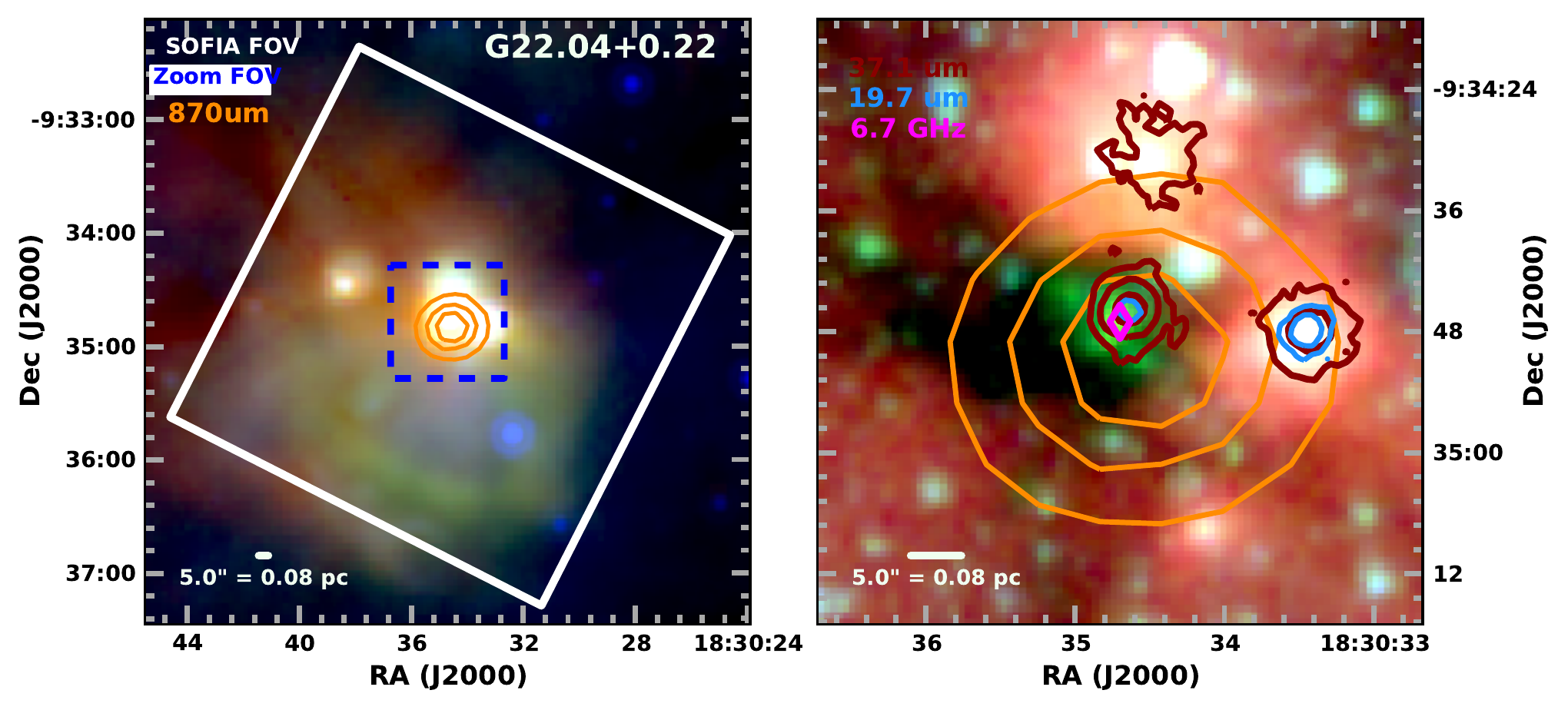}}
    \\
    \subfigure{
    \includegraphics[width=0.9\textwidth]{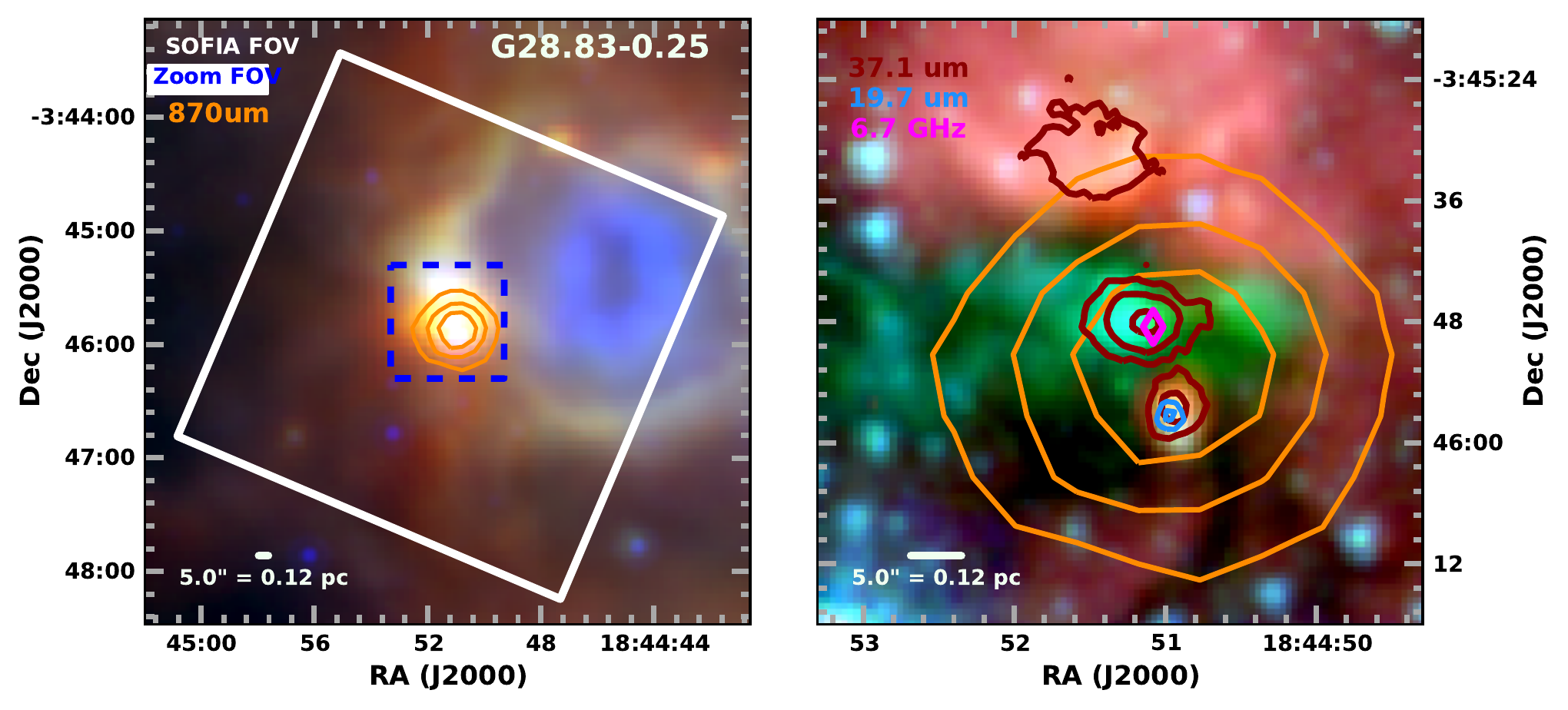}}
    \\
    \centering
    \subfigure{
    \includegraphics[width=0.9\textwidth]{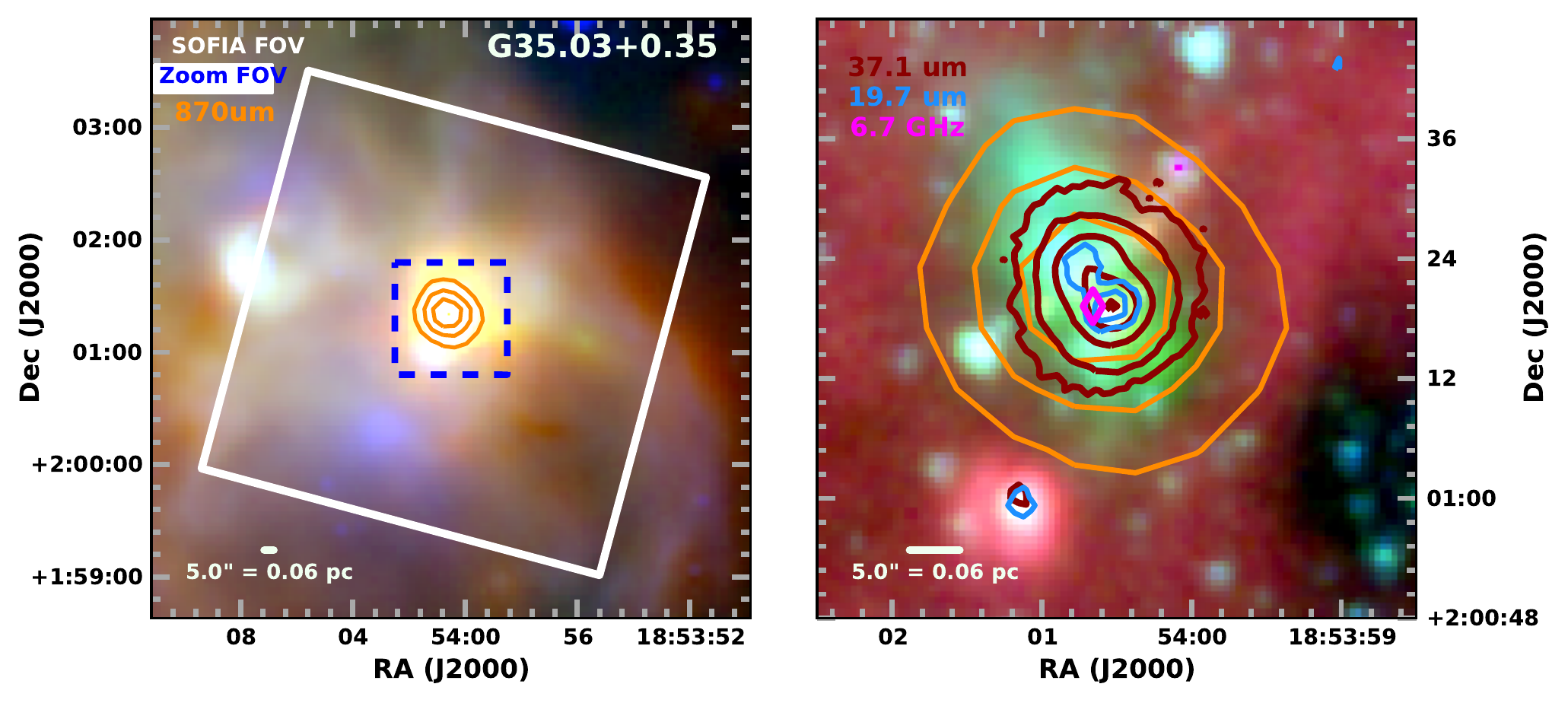}}
    \caption{RGB images for EGO sources. See Figure~\ref{3color_1} for key. The $I_{max}$ values for G22.04$+$0.22, G28.83$-$0.25, and G35.03$+$0.35 are 3.33 \jyb, 4.08 \jyb, and 4.91 \jyb, respectively.}
    \label{3color_4}
\end{figure*}

Figures~\ref{3color_1} through \ref{3color_4} show pairs of three-color (RGB) images for each source. The left-hand panels show a 5$\farcm$3 field of view with 160, 70, and 24 \mum\/ data mapped to R, G, and B, respectively, and with 870~\mum\/ contours overlaid. These panels show the large-scale structure of the cloud and overall environment in which each EGO is located. The right-hand panels all have a 1$\farcm$0 FOV with the {\it Spitzer} IRAC 8.0, 4.5, and 3.6~\mum\/ data mapped to R, G, and B, respectively; the extended green emission in these images shows the extent of each EGO. SOFIA FORCAST 19.7 and 37.1~\mum\/ contours and ATLASGAL 870~\mum\/ contours are overlaid, and 6.7~GHz \methanol\/ masers \citep{Cyganowski2009} are marked with diamonds. These panels show the small-scale structure and detailed NIR and MIR emission of each EGO, how this emission relates to the larger-scale 870~\mum\/ emission, and the locations of any associated markers of MYSOs, such as 6.7~GHz \methanol\/ masers.


Below we discuss in detail the photometric methodology used for each band for the SED analysis.
Because the angular resolution and sensitivity - and hence level of confusion - vary significantly among the different observations, we have elected to use a photometry method best suited for each particular wavelength in order to minimize (as much as feasible) contamination from unrelated sources. In the following sections we describe in some detail how the photometry was done for each wavelength.

\subsection{SOFIA FORCAST Photometry}
\label{sofiaphot}
Table \ref{sofia_fluxes} shows the photometry for the 19.7 and 37.1~\mum\/ SOFIA images. In this section, we describe the SOFIA astrometry, source selection, and photometry in more detail.

\paragraph{Astrometry}
The SOFIA images required additional astrometric corrections. While the relative astrometry between the 19.7 and 37.1~\mum\/ data was accurate to less than one pixel, the absolute astrometry of the SOFIA data varied considerably. Relative to the {\it Spitzer} MIPS 24~\mum\/ data, the positions of the SOFIA images varied by up to $\sim3\arcsec$.
In order to properly register the SOFIA images, we selected field point sources that were present in both the 24~\mum\/ images and either the 37.1 and 19.7~\mum\/ images, fit a 2-dimensional gaussian to that point source in both the 24~\mum\/ and SOFIA frame and applied the calculated position difference to both SOFIA images. In most cases, we were able to find a position match with the 24~\mum\/ data in only one of the two SOFIA frames, and relied on the sub-pixel relative astrometry between the two SOFIA images in order to correct the non-matched frame. Post-astrometric correction, we consider the absolute astrometric accuracy of the SOFIA images to be dominated by the absolute position uncertainty of the MIPS 24~\mum\/ images: $\sim$1$\farcs$4.

\paragraph{Mid-IR Source Selection and Nomenclature}
\label{nomenclature}
We limit our analysis to those mid-IR sources we consider to be plausibly associated with the protocluster in which the EGO resides (with some exceptions described below). For short, we call these sources ``EGO-associated.''
In this context, ``EGO-associated'' means one of two things: a) the mid-IR source is coincident with the extended 4.5~\mum\/ emission of the EGO and is therefore likely tracing some aspect of the EGO driving source in the mid-IR, or b) the mid-IR source lies outside the 5$\sigma$ level of the 4.5~\mum\/ emission but is still near to the EGO, and it is unclear whether the source is related or is a field source.
In order to create a self-consistent system for selecting sources in the latter category, we establish two criteria: 
i) the source must lie above the 25\% peak intensity level of the ATLASGAL emission and ii) it must be detected at both 19.7 and 37.1~\mum.
If a mid-IR detection is not within the bounds of the 4.5~\mum\/ emission and does not meet both criteria i) and ii), then it is considered to be a field source.

One source, G14.33$-$0.64\_b meets neither criteria and is likely a field source. It is coincident with the known \HII\/ region IRAS 18159$-$1648. However, it was necessary to explicitly fit this source in order to get accurate flux density results for the EGO-associated sources.

For a given EGO field, source ``a'' is always the brightest EGO-associated source at 37~\mum\/, source ``b'' is the second-brightest at 37~\mum\/ (of all analyzed sources for that FOV), and so on in order of decreasing brightness.
The 37~\mum\/ source name designations are used for all the wavelengths analyzed in this paper.

\paragraph{Photometry}
After source selection, we fit each source with 2-dimensional gaussian functions using the CASA task {\tt imfit} in order to determine the total flux density, peak intensity, and major and minor axes. We then applied a multiplicative correction factor (an ``aperture correction'') to each fitted flux in order to account for the deviation of the SOFIA PSF from a true gaussian.
Our detailed procedure was as described below.

We first selected emission-free regions in each image in order to determine the background noise levels.
These emission-free regions are identical for all three mid-IR data sets (SOFIA 37.1~\mum\/ and 19.7~\mum, and MIPS 24~\mum) for a given source. However, the SOFIA images in particular have background levels that typically do not show noise variations about zero.
Therefore, we chose to use the scaled MAD as an estimate of the noise (1.482$\times$MAD, where MAD is the median absolute deviation from the median), rather than the rms or standard deviation. With the exception of the ATLASGAL data, all data sets analyzed in this work have noise variations that are not centered about zero. Therefore, we have used the scaled MAD for all data sets for the sake of consistency. From this point forward, the ``$\sigma$'' symbol refers to the scaled MAD whenever we are estimating or discussing background noise levels of the images.

We then performed the fitting for each source using {\tt imfit}. We iteratively refined each fit (e.g. by holding certain parameters, such as source position, fixed during the fit) until we determined the fit to be satisfactory.
We declared a fit to be satisfactory once the absolute value of the residual intensities of all pixels in the central Airy disk were below 4$\times$MAD of the residual image, with the majority below 2$\times$MAD.
In cases where source parameters are held fixed, {\tt imfit} does not return an uncertainty for those specific parameters, so the uncertainty is due entirely due to user choice of source position, size, etc. For these fits, our position uncertainties are 0.01 pixels, and uncertainties in the major and minor axes or position angles are 0.1$^{\circ}$.
All the uncertainties for parameters held fixed during the fit are listed in italics in Table~\ref{sofia_fluxes}.

Finally, we determined a wavelength-dependent multiplicative correction factor to the {\tt imfit} flux results. The SOFIA PSF is an obscured Airy diffraction pattern - its central bright disk has a slightly narrower width than a standard Airy diffraction pattern due to the effect of a central obscuration in the light path (the secondary mirror). However, {\tt imfit} only fits 2-dimensional Gaussians. In effect, it fits a Gaussian to the central Airy disk and ignores the surrounding Airy rings. These correction factors are effectively serving as ``aperture corrections'' for our data; the only difference is that they are corrections to the fitted flux values returned by {\tt imfit}, rather than corrections to direct measurements. As Airy diffraction patterns are wavelength-dependent, we calculated separate aperture corrections for our 19.7~\mum\/ and 37.1~\mum\/ data.
While 
the best practice in aperture photometry would be to measure the PSF of an unrelated, isolated point source in each field and then apply that PSF correction to the data, we found that almost none of our fields contained an unrelated point source, much less one bright enough to measure the PSF with any confidence. Instead, we employed the procedure described below.

We first created four 100$\times$100-pixel Airy diffraction patterns using the optical properties of the SOFIA telescope (primary and secondary mirror size and separation, etc.) at each of our two wavelengths. The PSFs are sampled with 0$\farcs$768 pixels, the same as the FORCAST instrument. At this pixel size, the total grid is 76$\farcs$8 in diameter; this is $\sim$23 times the FWHM at 37.1~\mum\/ (3$\farcs$4) as quoted in the Handbook, and $\sim$31 times the FWHM at 19.7~\mum\/ (2$\farcs$5). Although Airy-disk diffraction patterns mathematically extend to infinity, on a practical level, our synthetic PSFs had to be truncated to a particular size; we considered $>$20 times the quoted FWHM to be sufficient. The four PSFs for a given wavelength are mathematically identical, but each center position is given either zero- or half-pixel offsets in both the x and y directions. This effectively gives us four different sampling scenarios for the PSF. This was done to account for the fact that the peak of a given point source might not always fall neatly onto a single pixel, but instead might be sampled relatively equally between two or even four pixels. We then used {\tt imfit} to fit the central disk of each of these four PSFs, and compared the flux returned by {\tt imfit} for the central disk alone to the flux measured within an aperture of radius 50 pixels (38$\farcs$4 at 0$\farcs$768 per pixel; 50 pixels was the largest aperture radius available to us for a 100$\times$100-pixel grid). We calculated the ratio of measured to fitted fluxes for each of the four PSF grids for one wavelength, and took the mean of these ratios as our aperture correction factor for that wavelength.
The aperture correction at 37~\mum\/ is 1.17 $\pm$ 0.02, and the aperture correction at 19~\mum\/ is 1.11 $\pm$ 0.04, where the uncertainties are the standard deviation of the four measured-to-fitted flux ratios at each wavelength.

Table \ref{sofia_fluxes} shows the aperture-corrected {\tt imfit} results for our 37~\mum\/ and 19~\mum\/ data.
Non-detections are noted as upper limits.
Our detection rate at 37~\mum\/ is 92\%; the only target for which we did not detect any 37~\mum\/ emission is G10.29-0.13. Overall, we detect 24 separate 37~\mum\/ sources in our 12 targets.
Our detection rate at 19~\mum\/ is slightly lower - we detect 19~\mum\/ emission in only 9 of our 12 targets, for a detection rate of 75\%. Overall, we detect 18 separate 19~\mum\/ sources in our 12 targets. 

\begin{deluxetable*}{lccccccc}[!hbt]
\tablecaption{SOFIA FORCAST 19.7~\mum\/ and 37.1~\mum\/ Fitted Flux Densities}
\tablefontsize{\scriptsize}
\tablehead{
\colhead{EGO} & \colhead{Source$^{a}$} & \multicolumn{2}{c}{Coordinates (J2000)$^{b}$} & \multicolumn{2}{c}{Fitted Size$^{b}$} & \colhead{19~\mum\/ Flux$^{c}$} & \colhead{37~\mum\/ Flux$^{c}$}\\
 & & \colhead{RA (h m s)} & \colhead{Dec ($^{\circ}$ $\arcmin$ $\arcsec$)} & \colhead{Major $\times$ Minor ($\arcsec \times \arcsec$)} & \colhead{PA ($^{\circ}$)} & \colhead{Density (Jy)} & \colhead{Density (Jy)} 
}
\startdata
G10.29$-$0.13	&	a	&	\nodata                     &	\nodata                     &	\nodata                                     &	\nodata     &	<0.40	        &	<1.3	\\
G10.34$-$0.14	&	a	&	18:09:00.001 (0.003)	    &	-20:03:34.53 (0.04)	        &	1.41 $\times$ 1.01 (0.31 $\times$ 0.45)	    &	97 (31)	    &	<0.36	        &	14 (3)\\
	            &	b	&	18:09:00.02  (0.02) 	    &	-20:03:28.8 (0.2)	        &	4.55 $\times$ 2.97 (0.71 $\times$ 0.61)	    &	97 (15)	    &	0.7 (0.2)	    &	9 (2)	\\
G11.92$-$0.61	&	a	&	18:13:58.078 (0.002)	    &	-18:54:20.16 (0.04)	        &	2.63 $\times$ 2.39 ({\it 0.1} $\times$ 0.05)	    &	32 (9)	    &	1.1 (0.4)	    &	57 (12)	\\
	            &	b	&	18:13:58.113 (0.008)	    &	-18:54:16.25 (0.09)	        &	3.41 $\times$ 2.59 (0.14 $\times$ 0.13)	    &	100 (6)	    &	0.7 (0.3)	    &	25 (5)	\\
G12.91$-$0.03	&	a	&	18:13:48.227 (0.004)	    &	-17:45:38.46 (0.06)	        &	2.17 $\times$ 1.58 (0.26 $\times$ 0.30)	    &	148 (17)	&	0.2 (0.1)	    &	9 (2)	\\
	            &	b	&	18:13:48.276 (0.005)	    &	-17:45:45.91 (0.09)	        &	2.01 $\times$ 1.49 (0.40 $\times$ 0.48)	    &	160 (29)	&	0.7 (0.2)	    &	6 (1)	\\
	            &	c	&	18:13:48.44 (0.01)  	    &	-17:45:32.5 (0.2)	        &	1.43 $\times$ 1.11 (0.84 $\times$ 0.91)	    &	35 (140)	&	<0.20	        &	2.2 (0.6)	\\
G14.33$-$0.64	&	a	&	18:18:54.232 (0.003)	    &	-16:47:48.40 (0.06)	        &	5.40 $\times$ 3.41 (0.08 $\times$ 0.06)	    &	4.2 (1.3)	&	0.7 (0.3)	    &	85 (17)	\\
	            &	b	&	18:18:53.36 (0.01)  	    &	-16:47:42.3 (0.1)	        &	11.42 $\times$ 10.23 (0.15 $\times$ 0.13)	&	107 (5)	    &	22 (5)	        &	200 (41)	\\
	            &	c	&	18:18:54.64 (0.01)	        &	-16:47:49.6 (0.1)	        &	point source	                            &	pt. src.	&	<0.33	        &	7 (1)	\\
G14.63$-$0.58	&	a	&	18:19:15.225 (0.009)	    &	-16:30:03.3 (0.1)	        &	3.60 $\times$ 2.91 (0.41 $\times$ 0.39)	    &	59 (22)	    &	<0.34	        &	10 (2)	\\
G16.59$-$0.05	&	a	&	18:21:09.124 (0.002)	    &	-14:31:48.79 (0.03)	        &	2.47 $\times$ 2.41 (0.04 $\times$ 0.04)	    &	89 (29)	    &	0.7 (0.2)	    &	67 (14)	\\
	            &	b	&	18:21:09.509 {\it (0.001)}  &	-14:31:55.88 {\it (0.01)}   &	6.14 $\times$ 5.18 ({\it 0.1} $\times$ 0.09)  &	141 (4)	    &	5 (1)	        &	68 (14)	\\
	            &	c	&	18:21:08.511 {\it (0.001)}  &	-14:31:57.92 {\it (0.01)}   &	1.91 $\times$ 1.91 ({\it 0.1} $\times$ {\it 0.1}) &	167 (360)	&	0.8 (0.3)	&	6 (1)\\
G18.89$-$0.47	&	a	&	18:27:07.835 (0.008)	    &	-12:41:35.1 (0.1)	        &	1.93 $\times$ 1.85 (0.70 $\times$ 0.72)	    &	167 (307)	&	<0.31	        &	5 (1)	\\
	            &	b	&	18:27:08.46 (0.03)	        &	-12:41:29.8 (0.5)	        &	5.0  $\times$ 3.3 (1.7 $\times$ 1.5)	    &	23 (31)	    &	<0.31	        &	4 (1)	\\
G19.36$-$0.03	&	a	&	18:26:25.750 (0.005)	    &	-12:03:52.57 (0.07)	        &	2.69 $\times$ 2.48 (0.28 $\times$ 0.28)	    &	70 (52)	    &	0.3 (0.1)	    &	24 (5)	\\
	            &	b	&	18:26:25.591 (0.009)	    &	-12:03:47.9 (0.1)	        &	3.78 $\times$ 2.56 (0.41 $\times$ 0.39)	    &	84 (12)	    &	1.7 (0.5)	    &	20 (4)	\\
G22.04$+$0.22	&	a	&	18:30:34.635 (0.003)	    &	-09:34:45.74 (0.04)	        &	2.13 $\times$ 1.57 (0.13 $\times$ 0.15)	    &	135 (9)	    &	0.4 (0.2)	    &	21 (4)\\
	            &	b	&	18:30:33.43 (0.01)	        &	-09:34:47.9 (0.2)	        &	4.81 $\times$ 4.34 (0.40 $\times$ 0.38)	    &	95 (32)	    &	1.8 (0.4)	    &	14 (3)	\\
G28.83$-$0.25	&	a	&	18:44:51.138 {\it (0.001)}	&	-03:45:48.05 {\it (0.01)}   &	4.41 $\times$ 2.53 (0.16 $\times$ 0.14)	    &	87.5 {\it (0.1)} &	0.3 (0.2)	&	32 (7)\\
	            &	b	&	18:44:50.938 (0.008)	    &	-03:45:56.5 (0.2)	        &	2.61 $\times$ 1.34 (0.42 $\times$ 0.63)	    &	159 (12)	&	1.1 (0.3)	    &	8 (2)	\\
G35.03$+$0.35	&	a	&	18:54:00.524 (0.003)	    &	+02:01:19.16 (0.04)	        &	3.64 $\times$ 3.10 (0.03 $\times$ 0.03)	    &	115 (2)	    &	1.5 (0.4)	    &	160 (31)	\\
	            &	b	&	18:54:00.700 (0.007)	    &	+02:01:23.2 (0.1)	        &	5.47 $\times$ 3.91 (0.09 $\times$ 0.07)	    &	131 (2)	    &	0.8 (0.3)       &	100 (20)	
\enddata
\tablenotetext{a}{The listed source positions and fitted sizes are from the 37.1~\mum\/ fit results only.}
\tablenotetext{b}{G10.29$-$0.13 has ``\nodata'' in place of coordinates and fitted size because it was a non-detection at both wavelengths.}
\tablenotetext{c}{Upper limits are given for sources that have no emission above 5$\sigma$. In these cases, the listed upper limit is the 5$\sigma$ value for the FOV.}
\label{sofia_fluxes}
\end{deluxetable*}

The uncertainties on the integrated flux density values are the quadrature sum of three values: the fitted-flux uncertainties returned by the imfit task, the uncertainty of our measured aperture corrections, and the absolute flux calibration uncertainty for the SOFIA FORCAST data.
\citet{Herter2012} quote an absolute flux calibration accuracy to within 20\% of the total integrated flux for a given object, and that is the value we adopt here.
The uncertainties on the integrated flux densities returned by {\tt imfit} are set by the background noise level, which we set to the scaled MAD for each source and wavelength during the fitting procedure. 
The uncertainties of our aperture correction factors are discussed above.

\subsection{Photometry of Archival Data}
\subsubsection{Spitzer IRAC Photometry}
In order to constrain the near-infrared portion of the SEDs, we chose to perform aperture photometry for our targets at 3.6~\mum, 5.8~\mum\/, and 8.0~\mum\/ (the IRAC I1, I3, and I4 bands, respectively) using CASAViewer.
However, because the flux in these bands likely includes emission from some sources or processes unrelated to our sources of interest, and because the SED models we employ in \S~\ref{seds} do not include emission from PAHs, we chose to include these data as upper limits. 

We obtained the necessary IRAC images from the NASA/IPAC Gator Catalog List, and aperture corrections were applied to each measurement according to the table on page 27 of the IRAC Instrument Handbook\footnote{https://irsa.ipac.caltech.edu/data/SPITZER/docs/irac/iracinstrumenthandbook/}. 
We did not include measurements in the 4.5~\mum\/ (I2) band because the emission in this band is extended in all cases (this was the original classification criterion for this object type).

The background noise level for the IRAC bands, as for all other wavelengths, is the scaled MAD within an emission-free region in each image. The emission-free regions were identical for all three IRAC bands used.
For each source with significant ($>$5$\sigma$) emission at 37.1~\mum, we measured the integrated IRAC band flux within a circular aperture centered on the 37.1~\mum\/ coordinates. We also measured the flux within an annulus of corresponding size. Aperture and annulus sizes were chosen based on the aperture corrections listed in the IRAC Instrument Handbook and the FWHM of each source in each band. For a given source, we used the same aperture for all three IRAC bands (i.e. we did not modify the size of the aperture with wavelength); we chose the smallest aperture that would successfully fit a source in all three bands.  
Each integrated flux measurement was corrected for background emission by subtracting the product of the median intensity value within the annulus and the size of the aperture from the direct aperture-flux measurement.
After this subtraction, we applied the appropriate aperture corrections as listed in the IRAC Instrument Handbook. All aperture and annulus radii, aperture corrections, and corrected fluxes for our sources are listed in Table~\ref{irac_fluxes}.

Due to the very crowded nature of these fields in the IRAC bands and the generally clustered nature of our sources, it was sometimes necessary to use annuli for local background subraction that were not centered on our sources.
When this was necessary, we chose isolated stars within the same field of view and centered our annuli on those sources.
We were careful to choose annulus stars of similar or lower brightness than the source in question.
Choosing a star of equal or lower brightness for background subtraction would only have the effect of {\it increasing} the measured flux density.
While it does sacrifice some precision, allowing the measured flux density to perhaps be artificially increased maintains the self-consistency of the photometry, as the data from these bands will only be used as upper limits.

\begin{deluxetable*}{lccccccccc}[!hbt]
\tablecaption{IRAC Integrated Flux Densities}
\tablefontsize{\scriptsize}
\tablehead{
\colhead{EGO} &\colhead{Source} & \colhead{Aperture$^{a}$} & \colhead{Annulus$^{a}$} & \multicolumn{3}{c}{Aperture Correction$^{b}$} & \multicolumn{3}{c}{Aperture-Corrected Flux Density (mJy)}\\
 & & \colhead{Radius ($\arcsec$)} & \colhead{Radii ($\arcsec$)} & \colhead{3.6~\mum} & \colhead{5.8~\mum} & \colhead{8.0~\mum} &  \colhead{3.6~\mum} & \colhead{5.8~\mum} & \colhead{8.0~\mum}
}
\startdata
G10.29$-$0.13   & a     & 3.6   & 3.6 to 8.4       & 1.125 & 1.135 & 1.221      & 0.4 (0.3)   & 15 (3)      & 22 (8)\\
G10.34$-$0.14   & a	    & 3.6   & 3.6 to 8.4       & 1.125 & 1.135 & 1.221      & 16.0 (0.8)  & 94 (6)      & 99 (14)\\
                & b     & 3.6   & 3.6 to 8.4       & 1.125 & 1.135 & 1.221      & 11.2 (0.6)  & 137 (7)     & 350 (21)\\
G11.92$-$0.61   & a     & 2.4   & 2.4 to 7.2       & 1.215 & 1.366 & 1.568      & 10.0 (0.4)  & 66 (3)      & 40 (2)\\	
                & b     & 2.4   & 2.4 to 7.2       & 1.215 & 1.366 & 1.568      & 22.9 (0.9)  & 193 (7)     & 152 (6)\\
G12.91$-$0.03   & a     & 3.6   & 3.6 to 8.4       & 1.125 & 1.135 & 1.221      & 8.7 (0.4)   & 52 (3)      & 20 (5)\\
                & b	    & 4.8   & 14.4 to 24.0     & 1.070 & 1.076 & 1.087      & 1.3 (0.2)   & 55 (4)      & 130 (13)\\
                & c     & 3.6   & 3.6 to 8.4       & 1.125 & 1.135 & 1.221      & 0.5 (0.2)   & 15 (2)      & 34 (5)\\
G14.33$-$0.64   & a     & 4.8   & 14.4 to 24.0     & 1.070 & 1.076 & 1.087      & 6.6 (0.4)   & 86 (4)      & 136 (9)\\
                & b     & 12.0  & 14.4 to 24.0     & 1.000 & 1.000 & 1.000      & 141 (6)     & 1600 (70)   & 4300 (180)\\
                & c     & 3.6   & 3.6 to 8.4       & 1.125 & 1.135 & 1.221      & 2.1 (0.2)   & 26 (1)      & 31 (3)\\
G14.63$-$0.58   & a     & 2.4   & 2.4 to 7.2       & 1.215 & 1.366 & 1.568      & 0.9 (0.1)   & 10.5 (0.1)  & 13 (2)\\
G16.59$-$0.05   & a     & 3.6	& 3.6 to 8.4       & 1.125 & 1.135 & 1.221      & 3.3 (0.3)   & 108 (5)     & 107 (7)\\
                & b     & 6.0   & 6.0 to 12.0      & 1.060 & 1.063 & 1.084      & 60 (3)      & 740 (30)    & 1840 (80)\\
                & c     & 6.0   & 6.0 to 12.0      & 1.060 & 1.063 & 1.084      & 43 (2)      & 330 (15)    & 820 (40)\\
G18.89$-$0.47   & a     & 4.8   & 14.4 to 24.0     & 1.070 & 1.076 & 1.087      & 4.6 (0.6)   & 46.2 (0.4)  & 20 (11)\\
                & b     & 4.8   & 14.4 to 24.0     & 1.070 & 1.076 & 1.087      & 22 (1)      & 104 (6)     & 200 (15)\\
G19.36$-$0.03   & a     & 4.8   & 14.4 to 24.0     & 1.070 & 1.076 & 1.087      & 24 (1)      & 151 (7)     & 190 (12)\\
                & b     & 3.6   & 3.6 to 8.4       & 1.125 & 1.135 & 1.221      & 11.0 (0.5)  & 101 (4)     & 220 (10)\\
G22.04$+$0.22   & a     & 4.8   & 14.4 to 24.0     & 1.070 & 1.076 & 1.087      & 1.4 (0.3)   & 30 (2)      & 26 (6)\\
                & b     & 9.6	& 14.4 to 24.0     & 1.011 & 1.011 & 1.017      & 79 (4)      & 791 (32)    & 1980 (83)\\
G28.83$-$0.25   & a     & 4.8   & 14.4 to 24.0     & 1.070 & 1.076 & 1.087      & 23 (1)      & 77 (5)      & 17 (8)\\
                & b     & 4.8   & 14.4 to 24.0     & 1.070 & 1.076 & 1.087      & 8.0 (0.6)   & 91 (5)      & 220 (13)\\
G35.03$+$0.35   & a     & 3.6   & 3.6 to 8.4       & 1.125 & 1.135 & 1.221      & 17.0 (0.8)  & 140 (6)     & 58 (6)\\
                & b     & 3.6   & 3.6 to 8.4       & 1.125 & 1.135 & 1.221      & 30 (1)      & 121 (6)     & 34 (6)
\enddata
\tablenotetext{a}{These columns list the radii of the aperture and annuli used for aperture photometry for each source. Radii are listed in arcseconds. Pixel scale is 1$\farcs$2/pixel for all IRAC bands.}
\tablenotetext{b}{Aperture correction factors are from the IRAC Instrument Handbook.}
\label{irac_fluxes}
\end{deluxetable*}

The uncertainties on the integrated flux densities are the quadrature sum of three values: the background noise levels, the absolute flux calibration uncertainty for the IRAC bands, and the uncertainty in the aperture-correction values.
The background noise levels are discussed above.
The IRAC Instrument Handbook quotes an absolute flux calibration accuracy to within 3\% of the total integrated flux for a given object, and that is the value we adopt here.
Additionally, the Handbook quotes an absolute aperture-correction accuracy to within 2\% of the total aperture-correction factor.

\subsubsection{Spitzer MIPS Photometry}
We used CASA's {\tt imfit} task to determine the integrated flux densities of our targets at 24~\mum\/ using the same fitting procedure described in \S~\ref{sofiaphot}. 
Due to the MIPS 24~\mum\/ images' significant Airy rings for point sources (up to 22\% of the total integrated flux according to the MIPS Instrument Handbook\footnote{http://irsa.ipac.caltech.edu/data/SPITZER/docs/mips/mipsinstrumenthandbook/}), the MIPS fit results require an aperture correction similar to that discussed in \S~\ref{sofiaphot}.
Fortunately, the MIPS images, unlike our SOFIA images, contain a plethora of isolated point sources with which to measure the PSF directly.

In order to determine the value of the necessary aperture correction, we performed the {\tt imfit} fitting procedure described in \S~\ref{sofiaphot} on five isolated, relatively bright point sources with fluxes listed in the MIPSGAL Point Source Catalog \citep{mipsgal2015}.
We selected the sources to span a range of colors and 24~\mum\/ flux densities.
As with the SOFIA sources, we considered fits to be ``satisfactory'' when the absolute value of the residuals within the Airy disk were all under 4$\times$MAD of the residual image, with the majority under 2$\times$MAD.
We compared the integrated fluxes returned by the {\tt imfit} task to those listed in the MIPSGAL Point Source Catalog.
We found a consistent aperture correction value of 1.59 $\pm$ 0.00893.
Table \ref{mips_standardstars} shows the positions, catalog fluxes, fitted flux results, and calculated flux ratios for these five standard stars.

\begin{deluxetable*}{lccccc}[!hbt]
\tablecaption{MIPS 24~\mum\/ Standard Star Fitted and Catalog Fluxes, \& Flux Ratios}
\tablefontsize{\scriptsize}
\tablehead{
\colhead{Star$^{a}$} & \multicolumn{2}{c}{Coordinates (J2000)$^{b}$} & \colhead{Catalog Flux} & \colhead{Fitted Flux} & \colhead{Flux Ratio}\\
 & \colhead{RA (h m s)} & \colhead{Dec ($^\circ$ $\arcmin$ $\arcsec$)} & \colhead{(mJy)} & \colhead{(mJy)} &
}
\startdata
1 & 18:30:32.40 & -09:35:47.25 & 1950 (39) & 1220 (40) & 1.60\\
2 & 18:30:12.78 & -09:36:47.99 & 1980 (36) & 1250 (38) & 1.58\\
3 & 18:30:53.95 & -09:39:51.27 & 2960 (55) & 1870 (71) & 1.58\\
4 & 18:30:46.32 & -09:32:28.89 & 1230 (23) & 770 (23)  & 1.60\\
5 & 18:30:48.45 & -09:36:00.11 & 780 (14)  & 490 (17)  & 1.59
\enddata
\tablenotetext{a}{Coordinates and both fitted (this work) and catalog \citep{mipsgal2015} flux densities for five bright, isolated point sources in the MIPSGAL Point Source Catalog.}
\tablenotetext{b}{Listed coordinates are from the MIPS Point Source Catalog \citep{mipsgal2015}.}
\label{mips_standardstars}
\end{deluxetable*}

We then applied the fitting procedure and measured aperture correction to our science targets.
As for the SOFIA {\it FORCAST} data, sometimes certain parameters (source position, size, etc.) were held fixed during the fitting procedure; these cases are noted in Table~\ref{mips_flux}.
Our results are listed in Table~\ref{mips_flux}, which presents
the final, aperture-corrected fitted flux results (to be used in the SED fitting) as well as the initial, un-corrected {\tt imfit} flux results.

As with the SOFIA data, the uncertainties on the integrated flux density values are the quadrature sum of the uncertainties returned by the {\tt imfit} task, the uncertainty of our calculated aperture-correction value, and the absolute flux calibration uncertainty for the MIPS data.
The MIPS Instrument Handbook quotes an absolute flux calibration accuracy to within 5\% of the total integrated flux of a given object.
The uncertainties on both the peak intensity and the integrated flux density returned by {\tt imfit} are set by the background noise level, which we set to the scaled MAD during the fitting procedure.
The uncertainty of the calculated aperture correction value we take to be the standard deviation of the five measured values: 8.93$\times$10$^{-3}$.

\begin{deluxetable*}{lccccc}
\tablecaption{MIPS 24~\mum\/ Aperture-Corrected Fitted Flux Densities}
\tablefontsize{\scriptsize}
\tablehead{
\colhead{EGO} & \colhead{Source} & \multicolumn{2}{c}{Coordinates (J2000)$^{a}$} & \colhead{Fitted Flux$^{b}$} & \colhead{Aperture-Corrected$^{c}$}\\
 & & \colhead{RA (h m s)} & \colhead{Dec ($^{\circ}$ $\arcmin$ $\arcsec$)} & \colhead{Density (Jy)} & \colhead{Fitted Flux (Jy)}
}
\startdata
G10.29$-$0.13   & a & \nodata                       & \nodata                   &  \nodata          & <0.28       \\
G10.34$-$0.14   & a & 18:08:59.989 (0.004)          & -20:03:34.97 (0.06)       &   0.77 (0.03)     & 1.23 (0.08) \\
                & b & 18:09:00.017 (0.003)          & -20:03:28.75 (0.05)       &   1.51 (0.04)     & 2.4 (0.1) \\
G11.92$-$0.61   & a & 18:13:58.065 {\it (0.001)}    & -18:54:21.26 {\it (0.01)} &   2.484 (0.003)   & 4.0 (0.2) \\
                & b & 18:13:58.122 {\it (0.001)}    & -18:54:14.97 {\it (0.01)} &   2.262 (0.003)   & 3.6 (0.2) \\
G12.91$-$0.03   & a & 18:13:48.233 (0.002)          & -17:45:38.19 (0.03)       &   1.11 (0.02)     & 1.77 (0.09) \\
                & b & 18:13:48.283 (0.001)          & -17:45:46.21 (0.01)       &   1.50 (0.01)     & 2.4 (0.1) \\
                & c & 18:13:48.469 (0.004)          & -17:45:31.77 (0.04)       &   0.28 (0.01)     & 0.45 (0.03) \\
G14.33$-$0.64   & a & \nodata                       & \nodata                   &  confused         & \nodata     \\
                & b & \nodata                       & \nodata                   &  saturated        & \nodata     \\
                & c & \nodata                       & \nodata                   &  confused         & \nodata     \\
G14.63$-$0.58   & a & 18:19:15.221 (0.001)          & -16:30:03.26 (0.02)       &  0.683 (0.009)    & 1.09 (0.06) \\
G16.59$-$0.05   & a & \nodata                       & \nodata                   &  saturated        & \nodata     \\
                & b & \nodata                       & \nodata                   &  saturated        & \nodata     \\
                & c & \nodata                       & \nodata                   &  confused         & \nodata     \\
G18.89$-$0.47   & a & 18:27:07.82 (0.01)            & -12:41:35.1 (0.2)         &  0.34 (0.04)      & 0.54 (0.07) \\
                & b & 18:27:08.45 (0.01)            & -12:41:29.5 (0.2)         &  0.64 (0.07)      & 1.0 (0.1) \\
G19.36$-$0.03   & a & 18:26:25.782 {\it (0.001)}    & -12:03:53.73 {\it (0.01)} &  2.356 (0.005)    & 3.8 (0.2) \\
                & b & 18:26:25.569 {\it (0.001)}    & -12:03:48.13 {\it (0.01)} &  3.371 (0.006)    & 5.4 (0.3) \\
G22.04$+$0.22   & a & 18:30:34.627 (0.001)          & -09:34:46.24 (0.02)       &  2.23 (0.02)      & 3.6 (0.2) \\
                & b & 18:30:33.432 {\it (0.001)}    & -09:34:48.39 {\it (0.01)} &  2.79 (0.02)      & 4.4 (0.2) \\
G28.83$-$0.25   & a & 18:44:51.136 (0.001)          & -03:45:47.845 (0.009)     &  2.230 (0.008)    & 3.6 (0.2) \\
                & b & 18:44:50.931 (0.001)          & -03:45:56.506 (0.009)     &  2.362 (0.008)    & 3.8 (0.2) \\
G35.03$+$0.35   & a & \nodata                       & \nodata                   &  saturated        & \nodata     \\
                & b & \nodata                       & \nodata                   &  saturated        & \nodata          
\enddata
\tablenotetext{a}{Source coordinates are the fitted coordinates returned by {\tt imfit}. Sources that have \nodata\/ values in place of coordinate values are either undetected at 24~\mum\/ (G10.29$-$0.13) or suffer from saturation and/or confusion (G14.33$-$0.64, G16.59$-$0.05, G35.03$+$0.35). Sources with position uncertainties in italics (G11.92$-$0.61, G19.36$-$0.03, G22.04$+$0.22\_b) had their coordinates held fixed during the fitting procedure, so the position uncertainties come not from the {\tt imfit} results but from the uncertainty in the choice of source position (usually of order 0.01 pixels, or, 0$\farcs$0125).}
\tablenotetext{b}{These are the fitted fluxes directly returned by {\tt imfit}; they have not been corrected for aperture effects. Sources with ``\nodata'' are nondetections at 24~\mum. Sources listed as ``saturated'' are saturated at 24~\mum. Sources listed as ``confused'' are not saturated at 24~\mum\/ but suffer from an angular confusion problem, usually with a nearby saturated source.}
\tablenotetext{c}{These are the aperture-corrected fitted flux densities, where the applied aperture correction is 1.59 $\pm$ 0.00893, as calculated in Table~\ref{mips_standardstars}. We use the data in this column for constructing our SEDs. Sources listed as ``\nodata'' could not be fit at 24~\mum\/ due to saturation and/or confusion issues and thus have no fitted flux value to which to apply an aperture correction.}
\label{mips_flux}
\end{deluxetable*}

\subsubsection{Hi-GAL and ATLASGAL Photometry}
\label{fir_phot}
Unlike with the near- and mid-infrared data sets, our far-IR data could rarely be considered point-like.
Therefore, instead of fitting gaussians to the emission using {\tt imfit}, we measured the integrated flux of each source within a given intensity level using CASAViewer. The intensity levels were chosen uniquely for each source depending on local background emission and the overall image noise level ($\sigma$). 
Generally, apertures for the ATLASGAL data followed the 5$\sigma$ level. 
Apertures for the Hi-GAL data varied between 60$\sigma$ and 200$\sigma$ at 70~\mum\/ and between 40$\sigma$ and 150$\sigma$ at 160~\mum.
These apertures follow comparatively high contours due to the combination of low scaled {\it MAD} values (typically of order 10$^{-1}$) and, in most cases, relatively bright large-scale ambient emission.
Each integrated flux measurement was corrected for this background emission by subtracting the product of the median intensity value within a local annulus and the size of the aperture from the direct aperture-flux measurement.
The mean and median aperture radii at 70~\mum\/ are 19$\farcs$6 and 18$\farcs$8, respectively, as compared to the HiGAL 70~\mum\/ beam size of 5$\farcs$8 $\times$ 12$\farcs$1. The mean and median aperture radii at 160~\mum\/ are 26$\farcs$3 and 26$\farcs$7, respectively, as compared to the HiGAL 160~\mum\/ beam size of 11$\farcs$4 $\times$ 13$\farcs$4.
Positions and integrated flux values for the Hi-GAL and ATLASGAL data are listed in Table~\ref{fir_fluxes}.

Our far-IR flux uncertainties are the quadrature sum of two values.
First, there is the statistical uncertainty of the measurement itself, which we take to be the product of the background noise level $\sigma$ in \jyb\/ and the square root of the aperture size in beams.
Second, there is the inherent uncertainty of the image due to flux calibration accuracy. \citet{Molinari2016} quote an absolute flux uncertainty of 5\% for the Hi-GAL data, and \citet{Schuller2009} quote an absolute flux uncertainty of 15\% for the ATLASGAL survey.
We adopt these values for our uncertainty calculations for the Hi-GAL and ATLASGAL data, respectively.

\paragraph{Far-IR source selection}
Beginning at 70~\mum, the fluxes of sources that are not dominant at 37.1~\mum\/ (sources ``b'' and ``c'' for each FOV) begin to decrease, in some cases rapidly. This decrease in flux is usually such that, by either 160~\mum\/ or 870~\mum, there is only one dominant source at that wavelength. In all cases, that dominant source is spatially coincident with the location of the brightest source at 37.1~\mum. However, the angular resolution of the FIR data worsens as wavelength increases, so even if there are multiple sources present in the FIR images, the angular resolution may be insufficient to separate them.
Because of the comparatively low resolution of these images, it is not uncommon to see FIR flux that is spatially coincident with one of the ``b'' or ``c'' sources for a given EGO, but neither is it clear that the spatial coincidence is not merely a result of resolution limitations. In cases where the morphology of the 70~\mum\/ or 160~\mum\/ emission was consistent with a single source, we assigned all the emission in that band to source ``a.'' In cases where it was clear that there were multiple sources present in the Hi-GAL data, we took one of two approaches. First, we attempted to fit the emission using multiple gaussian components using {\tt imfit}. If we achieved satisfactory fits with this approach, the fitted fluxes of both sources are listed in Table~\ref{fir_fluxes}. Second, if we could not achieve satisfactory fits with multiple gaussian components, we attempted to estimate the maximum possible amount of flux that could be ascribed to the weaker source. We then performed the photometric procedure described above on the emission as a whole (dominant and weaker source combined) and assigned all of the resulting flux to the dominant 37.1~\mum\/ source, and added the estimated flux from the weaker source to our uncertainty value for the dominant source. For these cases, the measured fluxes are marked in bold in Table~\ref{fir_fluxes}. While imperfect, this method does allow us to at least account for the effects of multiple blended sources even when we cannot satisfactorily deblend the emission itself. 

Source confusion was not an issue in any of the ATLASGAL images, since the ATLASGAL data a) have an angular resolution that is significantly poorer than any of the other data sets, thus potentially blending any individual sources past the point where 
one could recognize separate sources, and b) necessarily probe cooler gas. This effectively means that the emission in the ATLASGAL images originates primarily in the outer regions of the parent clump, which is an identifiably larger physical size scale than those probed by the Hi-GAL and our mid- or near-IR data sets. Due to source morphology in the ATLASGAL data and the aforementioned drop in flux in the FIR for sources that are not the dominant source at 37.1~\mum, we attribute all 870~\mum\/ flux to the single, dominant 37.1~\mum\/ source in all cases.

The effect of these source-selection criteria is that full SEDs are constructed for the brightest 37.1~\mum\/ sources (the ``a'' sources) only, and these SEDS are based on the explicit assumption that these sources are by far the most dominant in the far-IR.

\begin{deluxetable*}{lcp{3.25cm}ccc}
\tablecaption{Hi-GAL 70~\mum\/ \& 160~\mum\/ and ATLASGAL 870~\mum\/ Flux Densities}
\tablefontsize{\scriptsize}
\tablehead{
\colhead{EGO} &\colhead{Source} & \colhead{Hi-GAL$^{a}$} & \colhead{70~\mum\/ Flux$^{b}$} & \colhead{160~\mum\/ Flux$^{b}$} & \colhead{870~\mum\/ Flux}\\
  & & \colhead{70~\mum\/ Notes} & \colhead{(Jy)} & \colhead{(Jy)} & \colhead{(Jy)}
}
\startdata
G10.29$-$0.13   & a & \nodata                               & 75 (6)            & 298 (30)          & 5.0 (0.8) \\
\\
G10.34$-$0.14   & a & \nodata                               & 280 (20)          & 590 (49)          & 6 (1) \\
                & b & assuming all emission from a          & \nodata           & \nodata           & \nodata \\
\\
G11.92$-$0.61   & a & \nodata                               & 640 (33)          & 980 (58)          & 11 (2) \\
                & b & assuming all emission from a          & \nodata           & \nodata           & \nodata \\
\\
G12.91$-$0.03   & a & \nodata                               & 96 (6)            & {\bf 270 (52)}    & 7 (1)\\
                & b & assuming all emission from a          & \nodata           & \nodata           & \nodata \\
                & c & assuming all emission from a          & \nodata           & \nodata           & \nodata \\
\\
G14.33$-$0.64   & a & \nodata                               & {\bf 1130 (190)}  & 1940 (120)        & 32 (4)\\
                & b & assuming all emission from a          & \nodata           & \nodata           & \nodata \\
                & c & assuming all emission from a          & \nodata           & \nodata           & \nodata \\
\\
G14.63$-$0.58   & a & \nodata                               & 130 (8)           & 390 (30)          & 15 (2)\\
\\
G16.59$-$0.05   & a & \nodata                               & 490 (26)          & 740 (48)          & 9 (1)\\
                & b & assuming all emission from a          & \nodata           & \nodata           & \nodata \\
                & c & assuming all emission from a          & \nodata           & \nodata           & \nodata \\
\\
G18.89$-$0.47   & a & sources a and b fit with {\tt imfit}  & 48 (2)            & {\bf 180 (27)}    & 10 (2)\\
                & b & sources a and b fit with {\tt imfit}  & 19.6 (0.2)        & \nodata           & \nodata \\
\\
G19.36$-$0.03   & a & \nodata                               & 250 (14)          & 470 (41)          & 8 (1)\\
                & b & assuming all emission from a          & \nodata           & \nodata           & \nodata \\
\\
G22.04$+$0.22   & a & sources a and b fit with {\tt imfit}  & 204 (10)          & {\bf 400 (94)}    & 5.9 (0.8)\\
                & b & sources a and b fit with {\tt imfit}  & 59 (3)            & \nodata           & \nodata \\
\\
G28.83$-$0.25   & a & \nodata                               & {\bf 510 (65)}    & 870 (60)          & 10 (1)\\
                & b & assuming all emission from a          & \nodata           & \nodata           & \nodata \\
\\
G35.03$+$0.35   & a & \nodata                               & 1350  (71)        & 1230 (84)         & 8 (1)\\
                & b & assuming all emission from a          & \nodata           & \nodata           & \nodata
\enddata
\tablenotetext{a}{This column addresses the confusion of our sources at the 70~\mum\/ wavelength and angular resolution. Sources with the note ``assuming all emission from a'' do have 70~\mum\/ emission coincident with the position of source b and/or c, but we assume the emission to be entirely from or significantly dominated by source a. Sources with the note ``sources a and b fit with {\tt imfit}'' have emission coincident with both source a and source b, and there were two emission regions sufficiently distinguishable at 70~\mum\/ to be fit with the {\tt imfit} tool. These notes only apply to the 70~\mum\/ data.}
\tablenotetext{b}{The sources in {\bf bold} in these two columns suffer from confusion at either 70 or 160~\mum, and were not sufficiently well-separated to be successfully fit with two components {\tt imfit}. In these cases, the uncertainties of the flux densities are increased to reflect this effect. The precise method by which the uncertainties account for the confusion issue is discussed in detail in-text in \S~\ref{fir_phot}.}
\label{fir_fluxes}
\end{deluxetable*}

\subsection{Images and Trends}
\label{3colorfigs}
Figures~\ref{3color_1} through \ref{3color_4} show the detected SOFIA 19.7 and 37.1~\mum\/ emission in the vicinity of each EGO.
We detected 37.1~\mum\/ emission in all twelve fields; in eleven cases, this emission was associated with the EGO. This is in itself a high detection rate. However, we detect an average of only two sources per target, of which only one, on average, is actually associated with the EGO. This suggests that, rather than detecting multiple protostars within each protocluster, we are typically detecting only the dominant source in each EGO. Likewise, we detect 19.7~\mum\/ emission in nine of our twelve fields, but it is only associated with the target EGO in eight cases. We detect more 19.7~\mum\/ emission toward sources that are not associated with the target EGOs than emission toward sources that are (10/18 not associated versus 8/18 that are). At 19.7~\mum, we still detect an average of two sources per target. 
Taken together, these trends suggest that our target protoclusters are still quite young and/or deeply-embedded; this would explain the trend of overall dominance by a single source, as well as the poorer detection rate of even these dominant sources at 19.7~\mum.

Of our 37.1~\mum\/ sources, all but one are located entirely within the 25\% ATLASGAL contours of the clump associated with the target EGO, for a total of 23 37.1~\mum\/ sources within eleven ATLASGAL clumps (G10.29$-$0.13 has no 37.1~\mum\/ emission toward the EGO itself, so its ATLASGAL clump is not counted). This is an average of slightly more than two mid-infrared sources per clump. The one 37.1~\mum\/ source not located within an ATLASGAL clump is G14.33$-$0.64\_b, which has some extended emission within the 870~\mum\/ contours but is centered outside of it; our source G14.33$-$0.64\_b is the known \HII\/ region IRAS 18159$-$1648 \citep{Jaffe1982}.

Eleven of our sources are located in IRDCs; the only exception is G16.59$-$0.05. Eleven sources are known to be coincident with 6.7~GHz \methanol\/ masers (references for maser detections are in the tablenotes of Table~\ref{sourceproperties}); the remaining source, G14.33$-$0.64, has no published 6.7~GHz data at the time of writing.
Three sources - G10.29$-$0.13 and G10.34$-$0.14 \citep[near the W31 \HII\/ region G10.32$-$00.15, see][]{Westerhout1958}, and G28.83$-$0.25 \citep[near N49, see][]{Wink1982} - are adjacent to are known \HII\/ or UC\HII\/ regions.

\subsection{Mid-infrared Multiplicity}
There is some evidence of multiplicity at mid-infrared wavelengths for nearly all of our targets, with G10.29$-$0.13 (lacking any mid-IR detection) and  G14.63$-$0.58 being the only exceptions. The evidence for mid-IR multiplicity for the other sources falls generally into two categories: individual EGO-related sources (i.e. within the boundaries of extended 4.5~\mum\/ emission) that have unresolved substructure at the angular resolution of our SOFIA data, and sources that have nearby ($\lesssim$ 10$\arcsec$) 37.1~\mum\/ detections which are not within the extended 4.5um emission of the EGO, and whose association with the EGO is unclear. We discuss each category in greater detail in the following sections.  The naming convention of the new detections is described in \S~\ref{sofiaphot}.

\subsubsection{EGO Sources with Unresolved Substructure at 37.1~$\mu$m}
The dominant EGO-related sources in G11.92$-$0.61, G14.33$-$0.58, G28.83$-$0.25, G35.03$+$0.35 exhibit elongated, unresolved 37.1 \mum\/ emission suggestive of multiplicity at scales $\lesssim 5\arcsec$ (the SOFIA angular resolution is $3\farcs4$). Below
we explore how the mid-IR emission compares to existing high resolution centimeter to millimeter data.
This comparison helps inform the nature of the emission at each wavelength. Mid-IR emission may trace both hot cores and outflow cavities, while centimeter emission can trace both free-free emission (e.g. \HII\/ region, ionized jet) and the long-wavelength end of the Rayleigh-Jeans tail of dust emission. Millimeter observations (in this context) primarily serve to identify individual cores from dust continuum emission. By comparing the emission from these different wavelength regimes, we can attempt to disentangle the possible sources of mid-IR emission in these objects.

\paragraph{G11.92$-$0.61}
G11.92$-$0.61 is elongated roughly N-S at 37.1~\mum\/, and shows two distinct sources at 19.7~\mum\/ which lie along the axis of the 37.1~\mum\/ elongation (Fig.~\ref{3color_1}). 
The southern and northern mid-IR sources (G11.92$-$0.61\_a and G11.92$-$0.61\_b) are coincident with the (sub)millimeter protostellar sources MM1 and MM3, respectively  \citep{Cyganowski2011a,Cyganowski2017}.  Both MM1 and MM3 are associated with 6.7~GHz \methanol\/ masers \citep[a signpost of massive star formation][]{Cyganowski2009,Cyganowski2011a}, and both have also been detected at centimeter wavelengths \citep[the centimeter sources are designated CM1 and CM2][]{Cyganowski2011b,Cyganowski2014,Moscadelli2016,Ilee2016,Towner2017}. 

To further explore how sensitive the SOFIA data are to the presence of multiple protostellar sources, we turn to high-angular resolution, high-sensitivity millimeter data. Atacama Large Millimeter/submillimeter Array (ALMA) observations of G11.92$-$0.61 (1.05~mm, 0$\farcs$49 $\times$ 0$\farcs$34 synthesized beam) by \citet{Cyganowski2017} reveal at least eight 1.05~mm sources within a 5$\arcsec$ radius of the peak of the 37.1~\mum\/ emission, two of which correspond to MM1 and MM3). Of these eight, the authors estimate that six are low-mass objects, one is intermediate- or high-mass (MM3), and one is high-mass (MM1). Indeed, follow-up observations of MM1 at 1.3~mm using ALMA, with a synthesized beam of 0$\farcs$106 $\times$ 0$\farcs$079, find that this source is likely a proto-O star whose circumstellar disk dynamics yield an enclosed mass of M$_{\rm enc}\sim$ 40 $\pm$ 5\msun\/ \citep{Ilee2018}.
These radio data suggest that the mid-IR morphology of G11.92$-$0.61 is dominated by the two intermediate to massive protostellar sources (MM1 and MM3), rather than, e.g., a poorly-resolved outflow cavity. This result also indicates that our SOFIA data are sensitive to massive protostellar multiplicity, though as expected the mid-IR data are not sensitive to lower mass (and luminosity) protocluster members (also see \S~\ref{sensitivity}).

\paragraph{G14.33$-$0.64}
The dominant EGO-related source G14.33$-$0.64\_a (Fig.~\ref{3color_2}) is slightly elongated N-S at 37.1~\mum, and there is a 19.7~\mum\/ detection associated with the northern portion of the elongation.  The brightest component, G14.33$-$0.64\_b, is coincident with the known evolved \HII\/ region IRAS~18159$-$1648. In order to achieve satisfactory fits to the 37.1~\mum\/ emission toward G14.33$-$0.64\_a, it is necessary to fit a third component.  G14.33$-$0.64\_c is located $\sim4\arcsec$ east-southeast of G14.33$-$0.64\_a, is faint at both 37.1~\mum\/ and 24~\mum, and is undetected at 19.7~\mum.

\citet{Towner2017} report significant JVLA 1.3~cm (4$\farcs$6 $\times$ 2$\farcs$5 beam) emission coincident with G14.33$-$0.64\_a and the \HII\/ region G14.33$-$0.64\_b, as well as a marginal detection at the location of G14.33$-$0.64\_c (to within stated position uncertainties), though they were unable to get a satisfactory fit for its (weak) 1.3~cm flux density. Unfortunately, there are no published high-angular resolution millimeter continuum data for this source, though the mid-IR and centimeter data hint that there may be at least two massive protostars coincident with the EGO.

\paragraph{G28.83$-$0.25}
G28.83$-$0.25\_a is elongated E-W, consistent with unresolved substructure; this source is not detected at 19.7~\mum\/ (Fig.~\ref{3color_4}).
Based on the 37.1~\mum\/ emission alone, it is unclear whether this elongation is indicative of multiple unresolved sources or is due to a different cause, such as an unresolved outflow cavity. Interestingly, the elongation follows the same axis as the extended 4.5~\mum\/ emission, which is thought to be due to outflow activity.
\citet{Towner2017} detect two 1.3~cm continuum sources toward this EGO: one is coincident with the peak of the 37.1~\mum\/ emission (called CM2), and one that is coincident with the extended ``spur'' on the western edge of G28.83$-$0.25\_a (called CM1). Both 1.3~cm sources are unresolved at the angular resolution of the 1.3~cm data ($\sim$3$\arcsec$). Both sources are also reported by \citet{Cyganowski2011b} at 3.6~cm with $\sim$1$\arcsec$ resolution. Comparing the two centimeter wavelengths, \citet{Towner2017} suggest that either CM2 has a steeper free-free SED than CM1 or has a higher contribution from dust. If free-free emission is present, then the E-W elongation at both 1.3~cm and 37.1~\mum\/ suggests that this emission could be due to an ionized jet. In the absence of higher-resolution MIR images, and comparable millimeter wavelength data we cannot definitively attribute the elongation in this source to either outflow activity or multiple unresolved protostellar sources.

\paragraph{G35.03$+$0.35}
The 37.1~\mum\/ emission for G35.03$+$0.35 is elongated NE-SW and is indicative of at least two unresolved sources (Fig.~\ref{3color_4}); at 19.7~\mum, the emission is resolved into two distinct sources, which lie along the major axis of the 37.1~\mum\/ elongation.  When observed at 1.3~cm with similar angular resolution ($\sim$3$\arcsec$) to the 37.1~\mum\/ data, the brighter 37.1~\mum\/ source, G35.03$+$0.35\_a, is coincident with compact, unresolved 1.3~cm continuum emission as reported by \citet{Brogan2011,Towner2017}. However, higher angular resolution 3.6~cm VLA observations ($\sim 1$) resolve the continuum emission for G35.03$+$0.35 into at least five distinct, compact centimeter sources \citep{Cyganowski2011b}. Four of these 3.6~cm sources are coincident with the brighter 37.1~\mum\/ source, G35.03$+$0.35\_a, and the unresolved 1.3~cm source. The two strongest of these 3.6~cm sources (CM1 and CM2), trace a known ultra-compact \HII\/ region \citep{Kurtz1994}, and likely a hyper-compact \HII\/ region \citep{Cyganowski2011b}, respectively. Therefore, G35.03$+$0.35\_a, harbors at least two massive protostars. 
With ALMA at 0.87~mm, \citet{Beltran2014} also detect CM1 and CM2, but not CM3, suggesting the latter is not a protostar. More recent high angular resolution JVLA observations of G35.03$+$0.35 (0$\farcs$34 resolution) and analysis of the SEDs by \citet{Sanna2019}, suggest that the hyper-compact \HII\/ region CM2 is driving a powerful outflow and that CM3 corresponds to jet emission launched from CM2. The fifth 3.6~cm source (denoted CM3) lies in the direction of the weaker 19.7 and 37.1~\mum\/ detections, G35.03$+$0.35\_b, but the 19.7~\mum\/ source appears to extend further to the NE than the 3.6~cm emission.  Thus,  G35.03$+$0.35\_b may be tracing an outflow cavity that extends to the NE of G35.03$+$0.35\_a (and the likely powering source CM2).

\subsubsection{37.1~\mum\/ Detections For Which the Association with EGOs is Unclear}
Many of the non-dominant 37.1~\mum\/ sources in our sample (the ``b'' and ``c'' sources) lie close to ($\lesssim$10$\arcsec$) the dominant 37.1~\mum\/ source but outside the bounds of the 4.5~\mum\/ extended emission. They also typically are redder in color than the ``a'' sources. The association of these non-dominant MIR sources with the EGOs is unclear, but understanding this association is an important component of understanding the mid-IR multiplicities in this sample.

One example of such a case is EGO G19.36$-$0.03, which has two sources identified in Table~\ref{sofia_fluxes}. They are fully separable at 19.7~\mum\/ but only marginally separable at 37.1~\mum\/ (Fig.~\ref{3color_3}). The fainter source at 37.1~\mum, G19.36$-$0.03\_b, lies outside the extended 4.5~\mum\/ emission of the EGO. 
Both sources have unresolved 1.3~cm continuum counterparts, reported by \citet{Towner2017}.
G19.36$-$0.03\_b is the stronger source at 1.3~cm, and also has compact emission at 3.6~cm \citep{Cyganowski2011b} with $\sim$1$\arcsec$ resolution.
It is also coincident with MIPS 24~\mum\/ and IRAC emission, and is associated with a line of 44~GHz Class I \methanol\/ masers; 
\citet{Cyganowski2011b} suggest that it is therefore a candidate for an expanding \HII\/ region.
If G19.36$-$0.03\_b is indeed associated with the EGO, then the multiplicity of massive protostars in this EGO is 2. Furthermore, this would make G19.36$-$0.03 an example of a massive protocluster in which multiple stages of high-mass star formation are occurring simultaneously, as noted by \citet{Cyganowski2011b}.
In this case, the classification of the EGO protocluster is significantly impacted by the association (or lack thereof) between the two 37.1~\mum\/ detections.

Other such cases in our sample include G10.34$-$0.14\_b, G18.89$-$0.47\_b, and G12.91$-$0.03\_c.
If every one of our ``b'' and ``c'' sources is truly associated with an EGO (except the \HII\/ region IRAS 18159-1648), then our average multiplicity of massive sources in this sample $-$ at 37.1~\mum\/ with $\sim$3$\arcsec$ resolution $-$ is 1.9. If only half are truly associated, the average multiplicity is 1.4.
These values are roughly in line with the results of \citet{Rosero2018}, who find no strong evidence of high mulitplicity ($>$2 massive sources) in a subset of similar massive protostellar sources from the SOFIA Massive Star Formation Survey (SOMA) sample \citep[see][ and Section 4.3.4 of this work, for a discussion of the SOMA sample and subsamples]{soma}.

In order to properly address 
this multiplicity question, additional observations are needed along the lines of those described above for G11.92$-$0.61 and G35.03$+$0.35.
Such observations must be able to distinguish individual dust cores ($\lesssim$ 0.02 pc spatial resolution) and establish the nature (ionized jet, \HII\/ region, synchrotron, etc.) of the centimeter-wavelength emission. The former allow the identification of individual sources, and the latter allow the differentiation between ionized jets and HC \HII\/ regions.
Indeed, we have observations underway for the majority of the EGOs in this sample with sub-arcsecond resolution in the JVLA C- and K-bands, and with ALMA Band 3 and Band 6. The results of these observations will be published in future work.

\section{Analysis}
\label{analysis}
In order to estimate temperature and mass of the parent clumps, we performed greybody fits to the SEDs in order to derive representative temperatures for each EGO.
Our greybody fits use only the far-IR (Hi-GAL and ATLASGAL) data for each source, and are used to derive dust temperatures that were then used to calculate clump masses based on the ATLASGAL 870~\mum\/ integrated flux densities.
In order to independently assess gas temperature in these clumps, we also examine the gas kinetic temperatures determined by \citet{Nobeyama}, as described below.
Table~\ref{temp_mass} shows the \ammonia\/ and greybody temperature results for each source, along with corresponding estimated masses. The last column also lists the FIR luminosity of each EGO, as returned by the greybody fits.

In order to determine \lstar\/ for each target, we fit the SEDs with several different publicly-available SED models including those published in \citet{R06}, \citet{R17}, and \citet{ZT18}. The different underlying assumptions and components for each model are described below, in order of model publication date. Figure~\ref{g1192_seds_single} and Figures~\ref{g1029_seds} through \ref{g3503_seds} show the SEDs and model fits in the following order for each source: six panels showing \citet{R17} models, one panel showing \citet{R06} models, and one panel showing \citet{ZT18} models. Figures~\ref{g1029_seds} through \ref{g3503_seds} are located in Appendix A.

\begin{deluxetable*}{lcccccc}[!hbt]
\tablecaption{Temperature \& Mass From Greybody Fits and Single-dish \ammonia\/ Observations}
\tablefontsize{\scriptsize}
\tablehead{
\colhead{EGO} & \colhead{Distance$^{a}$} & \multicolumn{2}{c}{Temperatures (K)} & \multicolumn{2}{c}{Masses (M$_{\odot}$)} & \colhead{L$_{FIR}^b$}\\
 & \colhead{(kpc)} & \colhead{T$_{dust}^{c}$} & \colhead{T$_{NH_3}$} & \colhead{Greybody} & \colhead{\ammonia-derived$^{d}$} & \colhead{(10$^3$ \lsun)}
}
\startdata                                                                  
G10.29$-$0.13     & 1.9                             & 24 (1) & 21.19 (0.17) & 76  & 91  & 0.90 \\ 
G10.34$-$0.14     & 1.6                             & 26 (1) & 28.23 (0.38) & 62  & 56  & 1.42\\ 
G11.92$-$0.61     & 3.38$^{+0.33}_{-0.27}$ (3.5)    & 27 (1) & 26.27 (0.19) & 450 & 466 & 12.76\\
G12.91$-$0.03     & 4.5                             & 23 (1) & 23.56 (0.31) & 649 & 627 & 5.49\\ 
G14.33$-$0.64     & 1.13$^{+0.14}_{-0.11}$ (2.3)    & 29 (2) & 25.26 (0.17) & 132 & 159 & 2.84\\ 
G14.63$-$0.58     & 1.83$^{+0.08}_{-0.07}$ (1.9)    & 22 (1) & 20.76 (0.32) & 234 & 254 & 1.31\\ 
G16.59$-$0.05     & 3.58$^{+0.32}_{-0.27}$ (4.2)    & 26 (1) & 20.51 (0.38) & 456 & 636 & 10.57\\
G18.89$-$0.47     & 4.2                             & 22 (1) & 28.24 (0.19) & 879 & 625 & 3.04\\ 
G19.36$-$0.03     & 2.2                             & 26 (1) & 24.90 (0.31) & 147 & 155 & 2.61\\ 
G22.04$+$0.22     & 3.4                             & 26 (1) & 26.71 (0.49) & 257 & 248 & 4.90\\ 
G28.83$-$0.25     & 4.8                             & 26 (1) & 28.27 (0.50) & 851 & 761 & 20.62\\
G35.03$+$0.35     & 2.32$^{+0.24}_{-0.20}$ (3.2)    & 33 (1) & 29.54 (0.92) & 119 & 138 & 9.98   
\enddata
\tablenotetext{a}{Distances shown without uncertainties are estimated from the LSRK velocity and the Galactic rotation curve parameters from \citet{Reid2014}. Parallax distances (with their uncertainties) are given where available from \citet{Reid2014}, and references therein, with the kinematic distance in parentheses for comparison. All kinematic distances are the near distance.}
\tablenotetext{b}{Returned by the greybody fits to Hi-GAL 70~\mum\/ \& 160~\mum\/ and ATLASGAL 870\mum\/ integrated flux densities.}
\tablenotetext{c}{The $T_{dust}$ was derived from greybody fits using a grid of parameters, in which temperature goes in steps of 1 K. Therefore, the uncertainties for all greybody-derived temperatures are 1 K, except for G14.33$-$0.64, which had three adjacent temperatures with the same \chisq\/ value. Here we present the median of those three temperatures, and increase the uncertainty for this source to 2 K.}
\tablenotetext{d}{These are the masses calculated from the ATLASGAL 870~\mum\/ fluxes assuming $T_{dust}$ = $T_{kin}$(NH$_3$).}
\label{temp_mass}
\end{deluxetable*}

\subsection{Temperature and Mass From Dust and \ammonia\/ Emission}
In order to determine the mass of the ATLASGAL clumps (the mass reservoirs) in which our sources are located, we need to know the temperature of the emitting material. This is typically accomplished either by fitting models to molecular line emission (e.g. \ammonia, CH$_3$CN) or by fitting greybody functions to far-IR dust emission. For this work, we chose to employ each method separately and compare results. 

For the gas temperature, we adopt the single-component \ammonia\/ fit results of \citet{Nobeyama}, who performed a \water\/ maser and \ammonia (1,1) through (3,3) inversion-line survey of 94 GLIMPSE-identified EGOs using the Nobeyama Radio Observatory 45-meter telescope. The kinetic temperature (\Tk) results from \citet{Nobeyama} are shown in Table~\ref{temp_mass}.
The \citet{Nobeyama} \ammonia\/ temperatures for our sample\footnote{While \citet{Nobeyama} note that the \ammonia\/ (3,3) masers detected by \citet{Brogan2011} in G35.03$+$0.35 \citep{Brogan2011} are not readily distinguishable as a non-thermal contribution in the Nobeyama data, contamination by (3,3) masers is unlikely to significantly impact the fitted temperatures.} 
have minimum, maximum, and median values of 20.5, 29.5, and 25.8 $\pm$ 2.5 K, respectively, where the uncertainty on the median is the MAD.

In order to estimate dust temperature and derive clump mass, we used the Python package {\tt lmfit} to fit a series of greybody curves to our far-IR (70~\mum, 160~\mum, and 870~\mum) flux densities and thereby derive a temperature T$_{dust}$ for each source. During this procedure, the grain opacity spectral index ($\beta$) was fixed at 1.7 \citep{Brogan2016,Sadavoy2016}. We defined a grid of T$_{dust}$ ranging from 18.0 K to 40.0 K in steps of 1.0 K. For each value of T$_{dust}$, we fit for the opacity at a reference wavelength and computed the corresponding luminosity (L$_{FIR}$).
The best-fit temperature was defined as the temperature for which $\chi^2$ was closest to 1, and the best-fit luminosity was the luminosity corresponding to this best-fit temperature.
Then, we calculated the total gas mass of each source as

\begin{equation}
    M_{gas} = R \bigg(\frac{F_{\nu}D^2}{B_{\nu}(T_{dust})\kappa_{\nu}}\bigg) \bigg(\frac{\tau}{1 - e^{-\tau}}\bigg)
    \label{mass_derivation}
\end{equation}

\noindent where $R = 100$ is the gas-to-dust mass ratio, $F_{\nu}$ is the measured 870~\mum\/ flux density of the source, $D$ is the distance to the source, $B_{\nu}(T_{dust})$ is the blackbody function, $\kappa_{\nu}$ is the dust opacity, and $\tau$ is the optical depth at 870~\mum. The dust opacity was fixed at $\kappa_{870\mu m}$ = 1.85 cm$^2$~g$^{-1}$, which is the value \citet{Schuller2009} interpolate from Table 1 of \citet{Ossenkopf1994} and which is employed by \citet{Cyganowski2017} for their calculation of the mass reservoir of G11.92$-$0.61. 
In all cases, the fitted opacity is sufficiently small at 870~\mum\/ that \( \frac{\tau}{1 - e^{-\tau}} \) $\approx$ 1.
We calculate mean and median $T_{dust}=$25.8~K and 26.0~K, respectively, with a standard deviation of 2.9~K and a MAD of 1.5 K.

While the mean and median values of the two temperature estimates are in statistical agreement, there is a general trend that the temperatures calculated using \ammonia\/ inversion transitions are slightly lower than those calculated from FIR dust emission.
The median difference between the dust- and gas-derived temperatures is only 0.92~K, so the trend is weak and further, more precise investigation is needed in order to make a definitive statement about the implications of such a trend.
However, it should be noted that this {\it is} in broad agreement with the trends noted by \citet{Konig2017} and \citet{Giannetti2017} for the ATLASGAL Top100 sample\footnote{\citet{Konig2017} and \citet{Giannetti2017} are the third and fifth papers, respectively, in a series on the ATLASGAL Top100 sample, which consists of 110 of the brightest submillimeter sources in the ATLASGAL compact source catalog selected to span a full range of evolutionary stages. 
For a description of the sample properties and selection criteria, see \citet{Giannetti2014}.}. Both authors find that dust and \ammonia\/ temperatures are well-correlated for massive star-forming clumps overall, but that \ammonia\/ emission tends to trace gas that is warmer than dust in very cold clumps ($\lesssim$15 K), and gas that is cooler than dust in warmer clumps ($>$15 K). 

The median EGO greybody-derived $T_{dust}$ (26.0~K) is similar to that of the 
the median dust temperature of the Top100 sample \citep[24.7 K, see Table 2 in][]{Konig2017}.
However, \citet{Konig2017} sort the Top100 sources into four subcategories, of which the ``IR-weak'' ($F_{24 \mu m}$ $<$ 2.6 Jy, median $T_{dust}$ = 21.4 K) and ``IR-bright'' ($F_{24 \mu m}$ $>$ 2.6 Jy, median $T_{dust}$ = 28.2 K) samples are the most similar to the EGO sample. 
Indeed, when scaled appropriately for distance\footnote{\citet{Konig2017} and \citet{Giannetti2017} both adopt 2.6~Jy as the 24~\mum\/ IR-weak/bright cutoff as that is the flux density of a B3 star at 4~kpc; for the flux comparison above, we scale the 24~\mum\/ fluxes listed in Table~\ref{mips_flux} for a distance of 4~kpc.}, half of our sources have $F_{24~\mu m}$ $<$ 2.6~Jy, and half have $F_{24~\mu m}$ $>$ 2.6~Jy.
Interestingly, we find that the EGO median dust temperature also falls in between the median $T_{dust}$ of the IR-weak and IR-bright populations, though it is closer to the IR-bright $T_{dust}$.
However, it is notable that the temperature ranges of the two categories are broad: 11.7 to 26.2 K for IR-weak and 21.9 to 35.4 K for IR-bright, with overlap in the 21.9 to 26.2~K range. Indeed, approximately 50\% of the IR-weak and 40\% of the IR-bright sources fall in this overlapping range of $T_{dust}$, so these two subcategories are not distinct with regard to the dust temperature. Interestingly, all of the EGO $T_{dust}$ fall within the Top100 IR-bright range, with the majority (75\%) also falling in the overlap region. 
Two of the EGO-12 sources, G14.63$-$0.58 and G18.89$-$0.47, are included in the Top100 sample, and both are classified as IR-weak. This is consistent with our distance-scaled $F_{24 \mu m}$ discussed above.

The strong overlap of the EGO and IR-bright $T_{dust}$ suggests that the two samples may be drawn from the same parent population, but the clustering of the EGO $T_{dust}$ in the IR-weak/IR-bright overlap region is nontrivial and cannot be discounted.
Based on these competing factors, it seems likely that either a) EGOs preferentially lie somewhere between the IR-weak and IR-bright samples in temperature space, or b) EGOs represent the colder end of the IR-bright sample, but are still only a subset of the IR-bright population and do not constitute a separate population.
Unfortunately, \citet{Konig2017} do not correlate the Top100 sources with sources in the EGO catalogs of \citet{egocat} and \citet{Chen2013}, so we cannot say definitively whether or not EGOs are well-represented in the current Top100 sample.
If they are, this could explain the apparent overlap in population, and if not, a comparison of the properties of EGOs with the Top100 sample would be warranted.

The \citet{Nobeyama} \ammonia\/ temperatures for our sample are slightly warmer than the \ammonia\/ temperatures for both the ``IR-weak'' and ``IR-bright'' subcategories of the Top100 sample \citep[$\sim$18 K and $\sim$22 K, respectively;][]{Giannetti2017}. 
The masses calculated from the two temperatures have mean and median differences of 1.8\% and 4.3\%, respectively, with a maximum difference of 40.6\% and a standard deviation of 17.2\%. 
The mean and median of the ratio of greybody-derived to ammonia-derived mass are 98.2\% and 95.7\%, respectively, again with a standard deviation of 17.2\%.
This difference might indicate that the greybody temperatures systematically produce slightly lower masses than the \ammonia-derived masses, in agreement with the trend noted above that \ammonia\/ generally traces cooler material than dust except in the very coldest environments.
However, we do not find that those sources where greybody fits produce lower masses than the \ammonia\/ fits are systematically the warmest or coldest clumps (using either the greybody or \ammonia\/ temperatures).
Furthermore, the high standard deviations on both of these numbers suggest that the two sets of masses are effectively identical; there is no statistically-significant trend biasing one mass estimate higher than the other.
Given that a sample size of 12 is still well within the regime of small-number statistics, we would strongly caution against over-extrapolating from these particular results - either for or against a particular mass-ratio trend.

Overall, the luminosities we calculate from the greybody fits to our sources are in good agreement with results published by other teams for these or similar sources.
\citet{Moscadelli2016} constructed SEDs for 40 high-mass YSOs, including four of our targets (G11.92$-$0.61, G14.63$-$0.58, G16.59$-$0.05, and G35.03$+$0.35), using integrated fluxes from the online image archives for the MSX \citep{Egan2003} and WISE \citep{Wright2010} surveys and the point-source catalogs of IRAS \citep{Neugebauer1984} and SCUBA \citep{DiFrancesco2008}. They  calculated bolometric luminosity for each source by directly integrating the area under the SED curve. 
The $L$ from our greybody fits for these four EGOs agree with the \citet{Moscadelli2016} luminosities within $\pm$20\%, with no trend toward over- or under-estimation. 
\citet{Urquhart2018} conducted a systematic analysis of the properties of $\sim$8,000 dense clumps in the ATLASGAL Compact Source Catalog (CSC), including deriving $L$, $M$, and temperature. 
They performed automated aperture photometry for each clump using the ATLASGAL 870~\mum\/ maps in conjunction with Hi-GAL 70 to 500~\mum\/ images, MSX emission maps at 8, 12, 14, and 21~\mum, and WISE 12 and 24~\mum\/ images, and fit the resulting SEDs to derive $T_{dust}$ and $L$. Each SED was fit with either one (greybody-only) or two (greybody+blackbody) components, depending one whether the source was best represented by a single cold component or a combination of cold and hot components. Clumps were fit with two components if they had at least two flux measurements at $\lambda < 70$~\mum.
All twelve of our sources are represented in the \citet{Urquhart2018} sample, and the median ratio of their luminosities to ours is $+$1.5. The \citet{Urquhart2018} $L$ are higher than our greybody-derived $L$ in all cases; this likely reflects the fact that our greybody fits are exclusively single-component fits, while the fitting procedure of \citet{Urquhart2018} requires that at least eleven of our sources were fit with two components in their analysis. The explicit inclusion of a hot component in the fit would be expected to increase the overall luminosity derived for a clump.

\subsection{SED Modeling}
\label{seds}
In \S~\ref{r06_models}, \ref{r17_models}, and \ref{zt18_models}, we provide brief summaries of the model assumptions and underlying physics for each of the three SED model types we used \citep{R06,R17,ZT18}. While we present the \citet{R17} results first in our figures for easy visual comparison (Figure~\ref{g1192_seds_single}, and all figures in Appendix A), we have chosen to present the models in chronological order by publication date in these summary sections, as the improvements in the \citet{R17} models were directly influenced by the \citet{R06} models.

In Figure~\ref{g1192_seds_single} and Appendix A, the \chisq\/ value shown on the plots is \chisq\/ per data point, where a ``data point'' is defined as any flux density used for the fit that is not an upper or lower limit. (For G11.92, for example, $n_{data}$ is 6.) From this point forward, any discussion of ``\chisq'' values refers to \chisq\/ per data point unless explicitly stated otherwise.

\begin{figure*}[hbt]
     \centering
     \includegraphics[width=\textwidth]{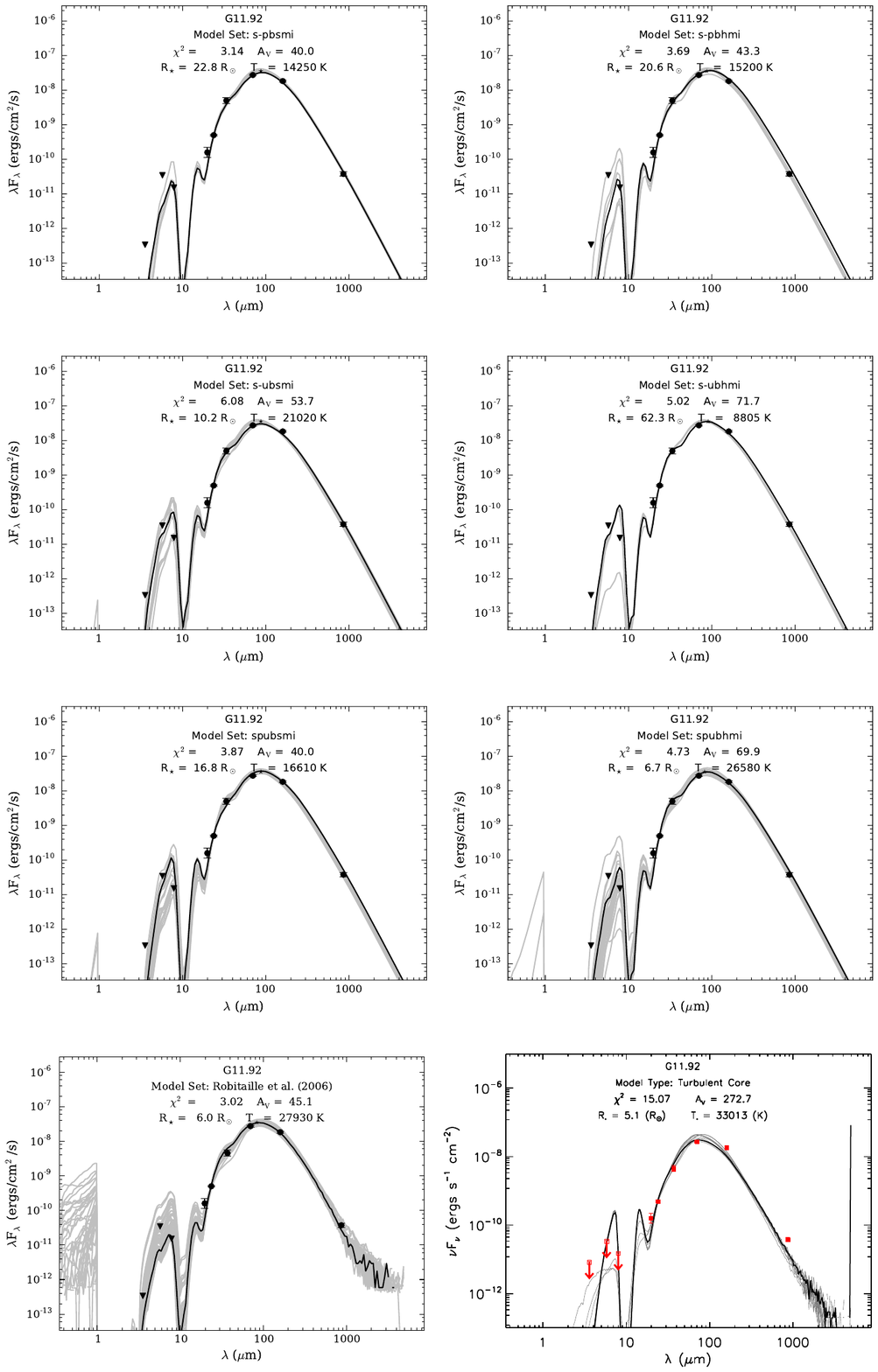}
     \caption{\citet{R17}, \citet{R06}, and \citet{ZT18} SED modeling results for G11.92$-$0.61. The top three rows are the best six model sets from the \citet{R17} model package, the bottom left panel shows the results from \citet{R06}, and the bottom right panel shows the results from \citet{ZT18}. All \chisq\/ values shown are \chisq/n$_{data}$, where n$_{data}$ is the number of data points used for the fitting that are not upper or lower limits. For this source, n$_{data}$ = 6. The best-fit model for each model set is denoted by a black line. The gray lines are SED models whose \chisq\/ per data point values were within a factor of 5 of the best-fit \chisq\/ per data point. The spike in the last wavelength bin in the \citet{ZT18} model is due to a binning error at the first and last wavelength bins: all flux above or below the longest or shortest wavelength is binned into the last or first wavelength bin. The error is present for all \citet{ZT18} models, but only produces a significant spike in a few.}
     \label{g1192_seds_single}
\end{figure*}

\subsubsection{Robitaille et al.\ (2006) Model Grid}
\label{r06_models}
The \citet{R06} models are a single grid of 200,000 model SEDs for YSOs. All \citet{R06} models include a central star, rotationally-flattened infalling (Ulrich-type) envelope, bipolar cavities, and a flared accretion disk. 
The models use the dust optical constants of \citet{Laor1993}, and neither include emission from PAHs nor account for the possibility of ice-coated grains.
The models were interpolated along evolutionary tracks in order to derive stellar radius (\rstar) and stellar temperature (\tstar) from a given combination of stellar mass (M$_{\star}$) and age (\age). 
They use two sets of evolutionary tracks: \citet{Bernasconi1996} for stars with M$_{\star}$ $>$ 9\msun, \citet{Siess2000} for stars with M$_{\star}$ $<$ 7\msun, and a combination of the two for stars with 7\msun\/ $<$ M$_{\star}$ $<$ 9\msun. 
The grid values of M$_{\star}$ and \age\/ were sampled from probability density distributions: M$_{\star}$ was sampled between M$_{min}$ = 0.1 \msun\/ and M$_{max}$ = 50 \msun\/ such that there was a constant density of models in log$_{10}$M$_{\star}$ space, and \age\/ was sampled between t$_{min}$ = 10$^3$ yr and t$_{max}$ = 10$^7$ yr such that there was a nearly-constant density of models in log$_{10}$t$_{\star}$ space, with a slight bias toward higher values of \age.

The ranges of the envelope accretion rate $\dot{M}_{env}$/$M_{\star}$, envelope outer radius, cavity opening angle $\theta_{cavity}$, and cavity density are dependent on the age of the central source.
Overall in the grid, $\dot{M}_{env}$/$M_{\star}$ varies from $\sim5\times 10^{-4}$ to $\sim10^{-9}$, and spans two orders of magnitude for any given source age. $\dot{M}_{env}$/$M_{\star}$ is sampled uniformly in logarithmic space. It is held constant for \age $< 10^4$~yr, then decreases, and finally goes to zero around $10^6$~yr. For models with $M_{\star} > 20$~\msun, $\dot{M}_{env}$/$M_{\star}$ was sampled with the same range as a $20$~\msun\/ model. That is, $\dot{M}_{env}$/$M_{\star}$ for sources with \mstar\/ $>$ 20~\msun\/ is no longer specific to each stellar mass, but a general rate used for all stars with \mstar\/ $>$ 20~\msun.

The envelope outer radius ranges from 10$^3$ to 10$^5$ AU, sampled uniformly in $log(R)$ space.
Bipolar cavities follow a conical shape described in cylindrical coordinates by $z = c\omega^d$, $\omega$ is the radial coordinate, $d = 1.5$ is a fixed value, and $c$ is a constant of proportionality defined as $c = (R_{env}^{max})/(R_{env}^{max}\tan (\theta_{cavity}))$.
$\theta_{cavity}$ is sampled from a range of values that increases with \age; values range from 0$^{\circ}$ to 60$^{\circ}$ in the grid overall, but are limited to $\sim0^{\circ}-10^{\circ}$ for the youngest sources and gradually shift to $\sim20^{\circ}-60^{\circ}$ for the oldest sources.
Envelope cavity density is sampled from a range one order of magnitude wide that decreases with evolutionary age; it ranges from 8 $\times$ 10$^{-20}$ g \cc\/ to 1 $\times$ 10$^{-22}$ g \cc, except in cases where the ambient density is greater than the cavity density. In such cases, the cavity density is reset to the density of the ambient medium, which is constant in both space and time and ranges from $\sim1.67 \times 10^{-22}$ (\mstar/\msun) g~\cc\/ to $\sim6.68 \times 10^{-22}$ (\mstar/\msun) g~\cc\/ \citep[for precise ranges and sampling conditions, see \S~2.2.2.5 of][]{R06}.

The flared accretion disk is described by five parameters: disk mass, disk outer radius, disk inner radius, disk structure, and disk accretion rate.
Disk mass is originally sampled from $\sim$0.001$-$0.1\msun for sources with \age\/ $< 1$ Myr, and then sampled over a wider range of masses for later evolutionary stages. 
Disk outer radius is usually associated with the centrifugal radius R$_C$, so R$_C$ is sampled from 1 to 10,000 AU, but is time-dependent such that earlier evolutionary stages may have smaller radii than later stages.
Disk inner radius was set to the dust sublimation radius R$_{sub}$ for one third of the models, and sampled between R$_{sub}$ and 100 AU or the disk outer radius, whichever was smaller, in the remaining two thirds of the models. Additionally, the envelope inner radius was set to the disk inner radius for all models. 
The disk structure is described by the disk flaring parameter $\beta$ and scale height factor z$_{factor}$, both of which were sampled from ranges dependent on the disk outer radius.
The disk accretion rate is calculated using the disk $\alpha_{disk}$ parameter (a unitless measure of the efficiency of angular momentum transport, which is dependent on disk radius and always less than 1) and accretion-to-$\alpha$ relations published in prior works \citep[for details, see][and references therein]{R06}. $\alpha_{disk}$ is sampled from 10$^{-3}$ to 10$^{-1}$ in log space.

\subsubsection{Robitaille (2017) Model Sets}
\label{r17_models}
The \citet{R17} models are a set of eighteen different YSO model grids, each with 10,000 to 80,000 models. 
The models were created using a similar computational method to \citet{R06}, but with different physical components and parameter ranges. 
Each grid (referred to as a ``model set'' in \citet{R17}, nomenclature that we henceforth adopt here) contains a different combination of physical components. 
Each model set includes a central stellar/protosteller source, and then may or may not include 1) an ambient medium, 2) a power-law envelope or an Ulrich-type envelope, 3) bipolar cavities, 4) a passive disk, and 5) an ``inner hole'' (gap between the stellar surface and the inner radius of the disk/envelope).
The \citet{R17} models do not assume a particular evolutionary track; it is left to the user to determine additional protostellar properties (other than those returned by the model) using the evolutionary track of their choice. 
Users are encouraged to test multiple model sets against their data and identify trends in model results in order to determine which physical components do or do not make a significant difference to the goodness-of-fit.

Given that we already have evidence that our targets have outflow activity \citep[e.g.][]{egocat,Cyganowski2009} and most are embedded in IRDCs, we chose to run all models with bipolar cavities, except for two controls (one with a disk, envelope, and no cavities and one with an envelope and neither a disk nor cavities).
However, while the outflow activity of our target sources is suggestive of the presence of disks, they have been confirmed in few of our targets (e.g. G11.92$-$0.61). Therefore, we have chosen to run all models with bipolar cavities - including those without disks - in order to avoid biasing our analysis towards only models with disks.

In all eighteen \citet{R17} model sets, T$_{star}$ varies from 2000 K to 30000 K.
In models with a disk, the disk shape varies from from hydrostatic (flared) to flat. All disks in all models are passive (accretion is not explicitly included in the model). A detailed explanation of the reasoning for using only passive disks can be found in \citet{R17}, \S~3.2.2.
The disk density distribution goes as r$^{(\beta - p)}$e$^{(z/h)^2}$, where disk flaring power (1 $<$ $\beta$ $<$ 1.3), disk surface density power ($-$2 $<$ $p$ $<$ 0), and disk scale height (1 AU $<$ $h$ $<$ 20 AU) are free parameters.
The envelope can be either Ulrich-type, in which case the centrifugal radius (R$_C$) varies from 50 to 5000 AU, or power-law, in which case the envelope power ($\gamma$) ranges from $-$2 to $-$1. 
The shape of the bipolar cavities follows a power-law, where the cavity power ($c$) varies from 1 to 2. 
The bipolar cavities are assumed to be filled with dust of a constant density, where the density ranges from 10$^{-23}$ to 10$^{-20}$ g \cc.
The ambient medium is defined as a lower limit to the density and temperature of the envelope (T$_{amb}$ = 10 K, $\rho_{amb}$ = 10$^{-23}$ g \cc\/).
The dust in the \citet{R17} models is taken from \citet{Draine2003a,Draine2003b} and \citet{Weingartner2001}, and does not include emission from PAHs.
Each SED is computed for nine viewing angles between 0$^{\circ}$ and 90$^{\circ}$, where the viewing angles are selected using stratified sampling: viewing angle is randomly chosen within a specific range, so that each SED is sampled at one random angle between 0$^{\circ}$ and 10$^{\circ}$, one random angle between 10$^{\circ}$ and 20$^{\circ}$, and so on up to 90$^{\circ}$.

The model sets we used are as follows:
\paragraph{s-pbhmi:} model contains a central star, no passive disk, a power-law envelope, a bipolar cavity, an ambient medium, and a variable inner envelope radius (rather than the inner radius being set to the dust sublimation radius). This variability has the effect of creating an ``inner hole'' between the inner radius of the envelope and the stellar surface.
\paragraph{s-ubhmi:} model contains a central star, no passive disk, an Ulrich envelope, a bipolar cavity, an ambient medium, and an inner hole.
\paragraph{s-pbsmi:} model contains a central star, no passive disk, a power-law envelope (no rotational flattening), a bipolar cavity, an ambient medium, and no inner hole (i.e. the inner radius is the dust sublimation radius).
\paragraph{s-ubsmi:} model contains a central star, no passive disk, an Ulrich (rotationally-flattened) envelope, a bipolar cavity, an ambient medium, and no inner hole.
\paragraph{spubsmi:} model contains a central star, a passive disk, an Ulrich envelope, a bipolar cavity, an ambient medium, and no inner hole.
\paragraph{spubhmi:} model contains a central star, a passive disk, an Ulrich envelope, a bipolar cavity, an ambient medium, and an inner hole.
\paragraph{spu-smi:} model contains a central star, a passive disk, an Ulrich envelope, no bipolar cavities, an ambient medium, and no inner hole.
\paragraph{s-u-smi:} model contains a central star, no passive disk, an Ulrich envelope, no bipolar cavities, an ambient medium, and no inner hole.

\vspace{3 mm}

A complete key, including diagrams, for all eight of these model sets can be found in \citet{R17}, Table~2.

\subsubsection{Zhang \& Tan (2018) Model Grid}
\label{zt18_models}
The \citet{ZT18} models are a grid of $\sim$9000 YSO model SEDs. 
These models are based on the Turbulent Core theory of high-mass star formation \citep{McKee2003}. 
The Zhang \& Tan models all assume a central source, disk, envelope, and bipolar outflow; they neither assume nor fit an ambient medium or emission from the parent clump.
The model grid is composed of five variables (three physical, two observational): core mass, mass surface density, stellar mass, A$_V$ along the line of sight, and inclination/viewing angle.
Core mass is sampled from 10 to 480 \msun, mass surface density ranges from 0.1 g \cc \/ to 3.16 g \cc, and stellar mass ranges from 0.5 to 160 \msun. 
Each model SED is sampled at 20 viewing inclinations, from cos($\theta_{view}$) = 0.975 to cos($\theta_{view}$) = 0.025.
The range of A$_V$ is set by the user; we chose to use 40 $<$ A$_V$ $<$ 1000, as we did for both types of Robitaille models.

In the \citet{ZT18} models, the initial core is assumed to have an r$^{-3/2}$ power-law density distribution, and is assumed to exhibit inside-out collapse and rotational-flattening (i.e. is assumed to be an Ulrich-type envelope).
\citet{ZT18} assume that the ratio of disk mass to protostellar mass is constant, at $M_{disk}$/\mstar $= 1/3$. 
They assume that all disks are hydrostatic (modified alpha disks).
In order to return protostellar radius, temperature, and luminosity, the models assume 
the evolutionary tracks of \citet{Hosokawa2009}.
\citet{ZT18} use the same dust models as \citet{R06}.

\subsection{Model Results: Robitaille et al. (2006), Robitaille (2017), and Zhang \& Tan (2018)}
\label{seds_per_source}
Due to the different physical assumptions and parameters fit by each model, the number of physical parameters that could be compared directly is small. Table~\ref{sed_lsun} shows the stellar radii (\rstar\/) and stellar temperatures (\tstar\/) returned by each of the three model grids, as well as the Stefan-Boltzmann luminosities ($4\pi R_{\star}^2\sigma T_{\star}^4$) calculated from those radii and temperatures. The \chisq\/ values shown are \chisq\/ per data point, as described in \S~\ref{seds}.
The individual \rstar\/ and \tstar\/ values returned by the three different sets of SED models for a particular source frequently span up to two orders of magnitude.
However, the Stefan-Boltzmann luminosities calculated from the different combinations of \rstar\/ and \tstar\/ (hereafter \lstar) tend to agree to within a factor of 3.

\citet{R17} stress that the model sets therein are best used to compare how much the presence or absence of a particular physical component (e.g. bipolar cavities) affects the accuracy of each model. In order to accomplish this, the author suggests that a Bayesian analysis (rather than \chisq\/ scores alone) are needed.
Unfortunately, we cannot compare Bayesian scores from the \citet{R17} models to Bayesian scores from the \citet{R06} and \citet{ZT18} set, as this approach of comparing probabilities assumes the models in question have similar underlying parameters and parameter ranges. The \citet{R06} and \citet{ZT18} models sample different parameter ranges and, in the case of \citet{ZT18}, different parameters altogether, so the comparison of probabilities cannot be performed. For the purposes of comparing the results from the different sets of published models, we use \chisq. A detailed discussion of the Bayesian scores for the \citet{R17} models $-$ and how this approach affects the overall trends as compared to the \chisq\/ analysis $-$ can be found in Appendix B.

\subsubsection{Expected Luminosity Sensitivity} 
\label{sensitivity}

The only source which had no 37.1~\mum\/ emission detected toward the EGO is G10.29$-$0.13 (see Table~\ref{sofia_fluxes}, Fig.~\ref{g1029_seds}).
This source is also a non-detection at 19.7 and 24~\mum, and has fairly isolated Hi-GAL and ATLASGAL emission compared to the rest of the sample. Therefore, based on the non-detection of this source and its comparative isolation and morphological simplicity, we use G10.29$-$0.13 as a test case in order to estimate the minimum luminosity sensitivity of the SOFIA observations. The 37.1~\mum\/ observations of 10.29$-$0.13 have an integration time of 502 s and $\sigma = 0.26$ \jyb, and the assumed distance is 1.9~kpc.

We construct a synthetic SED for this source by inserting the 3$\sigma$ upper limits at 37.1 and 24~\mum\/ as actual photometric measurements with uncertainties of 1$\sigma$. The 19.7~\mum\/ data remain as upper limits, while the values and treatment of flux densities from the other six wavelengths are likewise unchanged. We modeled this SED using the eight \citet{R17} model sets described in \S~\ref{r17_models}, and the best-fit model returned a Stefan-Boltzmann luminosity of $L_{min} = 1.1\times 10^{3}$\lsun.

This approximate lower limit is consistent with 92\% of the luminosities shown in Table~\ref{sed_lsun}. There are four models which produced \lstar\/ $<$ 1.1$\times$10$^{3}$, and interestingly all of these come from the \citet{ZT18} models. These four low luminosity \citet{ZT18} model results also correspond to the largest discrepancies between the three types of models assessed. See additional discussion of the trends, limitations, and overall quality of the three model packages below.

\subsubsection{Stefan-Boltzmann Luminosities}
\label{sb_lum}

\begin{deluxetable*}{l|lcccc|cccc|cccc}[!hbt]
\tablecaption{Stellar Radius, Temperature, and Luminosity Results for SED Model Grids}
\tablefontsize{\scriptsize}
\tablehead{
\colhead{Source} & \multicolumn{5}{c}{Robitaille (2017)$^{a,b}$} & \multicolumn{4}{c}{Robitaille et al. (2006)$^{a}$} & \multicolumn{4}{c}{Zhang \& Tan (2018)$^{a}$} \\
\colhead{Name} & \colhead{Model} & \colhead{\rstar\/ (\rsun)} & \colhead{\tstar\/ (K)} & \colhead{\lstar\/ (10$^3$ \lsun)} & \colhead{\chisq} & \colhead{\rstar\/ (\rsun)} & \colhead{\tstar\/ (K)} & \colhead{\lstar\/ (10$^3$ \lsun)} & \colhead{\chisq} & \colhead{\rstar\/ (\rsun)} & \colhead{\tstar\/ (K)} & \colhead{\lstar\/ (10$^3$ \lsun)} & \colhead{\chisq} 
}
\startdata  
G10.29$-$0.13   & spubhmi & 49.1    &   6640    &   4.15    & 0.0003    & 71.0  &   7405	&   13.41   & 0.002 & 7.3   &   4994	&   0.03 & 0.001\\
G10.34$-$0.14   & s-pbhmi & 27.5	&   7980	&   2.71    & 1.51      & 79.3	&   4492	&   2.27    & 1.45  & 21.7	&   6835	&   0.91 & 3.98\\
G11.92$-$0.61   & s-pbsmi & 22.8	&   14250	&   18.97   & 3.14      & 6.0	&   27930	&   19.39   & 3.02  & 5.1	&   33013	&   27.34 & 15.07\\
G12.91$-$0.03   & s-pbhmi & 6.4	    &   19940	&   5.73    & 1.51      & 118.8	&   4355	&   4.49    & 2.48  & 11.2	&   16298	&   7.83 & 20.43\\
G14.33$-$0.64   & s-pbsmi & 31.4	&   7976	&   3.53    & 0.45      & 110.7	&   4428	&   4.17    & 0.81  & 11.2	&   16298	&   7.83 & 1.89\\
G14.63$-$0.58   & s-pbhmi & 11.8	&   11330	&   2.03    & 3.68      & 68.6	&   4172	&   1.26    & 1.94  & 18.5	&   6780	&   0.64 & 20.54\\
G16.59$-$0.05   & s-pbhmi & 31.8	&   11020	&   13.20   & 1.12      & 65.0  &   7510	&   11.89   & 0.53  & 5.1	&   33013	&   27.34 & 11.84\\
G18.89$-$0.47   & s-pbhmi & 14.6	&   10610	&   2.39    & 1.00      & 96.7	&   4234	&   2.66    & 4.2   & 7.3	&   4994	&   0.03 & 37.73\\
G19.36$-$0.03   & s-pbsmi & 8.4	    &   14420	&   2.70    & 1.73      & 98.3	&   4493	&   3.48    & 1.41  & 13.6	&   14585	&   7.41 & 10.93\\
G22.04$+$0.22   & s-pbhmi & 14.3	&   12570	&   4.52    & 1.81      & 27.3	&   9345	&   5.03    & 1.92  & 11.2	&   16298	&   7.83 & 7.66\\
G28.83$-$0.25   & s-pbhmi & 74.3	&   8766	&   28.85   & 0.81      & 293.4	&   4355	&   27.40   & 1.62  & 42.9	&   12552	&   40.43 & 12.96\\
G35.03$+$0.35   & spubhmi & 69.9	&   8552	&   23.13   & 1.08      & 4.8	&   31560	&   20.23   & 1.31  & 5.1	&   33013	&   27.34 & 0.66\\
\hline\\
Median$^{c}$:         & \nodata & 25      & 10800     &   4.3     & 1.31      & 75    &   5950    &   4.8    & 1.54  & 11 &   15440 &   7.8   & 12.40
\enddata
\tablenotetext{a}{Parameters shown are the \rstar  and \tstar\/ returned by the best-fit model in each model package, where ``best-fit model'' is defined as the model with the lowest \chisq\/ per data point. \lstar\/ is the Stefan-Boltzmann luminosity (4$\pi$R$_{\star}^2\sigma$T$_{\star}^4$) calculated from each combination of \rstar\/ and \tstar.}
\tablenotetext{b}{Since the \citet{R17} model package contains multiple model grids, we include an extra column stating the name of the model set to which the best-fit model belongs.}
\tablenotetext{c}{Median values are computed for our twelve sources for a given parameter within a given model package. The median \lstar\/ of all 36 best-fit models is $5.83\times 10^{3}$ \lsun.}
\label{sed_lsun}
\end{deluxetable*}

The \lstar\/ (calculated from \rstar\/ and \tstar) are almost always larger than the $L$ returned by the greybody fits, typically by a factor of $\sim$2. This trend is consistent with the fact that our greybody fits are single-component and largely account for emission from cold dust, whereas the fits to the full SEDs can also account for emission from hotter components (e.g., hot cores) that emit predominantly in the NIR and MIR.
This is also consistent with our previous comparison to the luminosities reported in \citet{Urquhart2018}.
We find that the ratio between the \citet{Urquhart2018} luminosities and ours has decreased; the median ratio between the \citet{Urquhart2018} $L$ and those listed in Table~\ref{sed_lsun} is now $+$1.27. The median ratios between the \citet{Urquhart2018} $L$ and ours for the individual model packages are $+$1.22 for \citet{R06}, $+$1.31 for \citet{R17}, and $-$0.21 for \citet{ZT18}.

There are a few cases in which the \lstar\/ calculated from our SED model results was lower than $L$ returned by our greybody fits (hereafter $L_{grey}$, to distinguish from \lstar). We believe that these cases can be explained by flux or confusion limitations during the aperture photometry procedure, which then lead to the SED models returning low \lstar\/ results.
We identify three categories of SEDs which exhibit the $L_{grey}$ $>$ \lstar\/ discrepancy:

\paragraph{Confusion problems at 160~\mum:} 
The majority of \lstar\/$-L_{grey}$ discrepancies occur in sources which suffer from angular confusion in the 160~\mum\/ Hi-GAL data (G12.91$-$0.03, G18.89$-$0.47, G22.04$+$0.22; see Figs.~\ref{g1291_seds}, \ref{g1889_seds}, and \ref{g2204_seds}). In all cases, the measured 160~\mum\/ flux was unexpectedly high, not low. For these three sources, at least one and as many as all three SED modeling packages produced \lstar\/ $<$ $L_{grey}$. While we tried to account for the 160~\mum\/ confusion issue by significantly increasing the errors on the flux measurements for these sources, it is still possible that either a) the high 160~\mum\/ values or the large uncertainties on those values are leading to poor fits from the SED modeling packages, or b) the high 160~\mum\/ points lead to greybody fits that overestimate $L_{grey}$. Either cause (or possibly both) would result in $L_{grey}$ $>$ \lstar.

\paragraph{Poorly-constrained SEDs:}
All sources use flux measurements at eight or nine separate wavelengths in order to construct the SEDs. In most sources, three of these data points (IRAC bands) are always upper limits. However, in one source (G10.29$-$0.13), six of the nine flux measurements (67\%) are upper limits, and the \citet{ZT18} SED-derived \lstar\/ value is extremely low. We believe that this discrepancy can be explained by the very poor constraints on the MIR and NIR flux measurements, which makes it possible to fit a wide variety of models to the data (see Fig.~\ref{g1029_seds}); it is therefore unsurprising that at least one of these models produces a very low \lstar.

\paragraph{Upper limits at 19~\mum:}
There are four sources in our sample which are non-detections at 19~\mum, and so use upper limits for the 19~\mum\/ flux instead of direct measurements.
Of these four sources, two have additional issues (overall poor constrains on the SED, confusion problems at 160~\mum) that have already been discussed. However, the remaining two sources have no additional issues with flux measurements, but do still have \lstar\/ $<$ $L_{grey}$ for at least one SED-derived \lstar. It is possible that, in these cases, the use of an upper limit at 19~\mum\/ is allowing the SED modeling packages to underestimate the true 19~\mum\/ fluxes, which then leads to spuriously low \lstar\/ values. See Figs.~\ref{g1034_seds} and \ref{g1463_seds} for SEDs for these two sources.

\subsubsection{Spread in Physical Parameters Returned by the Radiative-Transfer Models}
Within our sample, the values of \rstar, \tstar, and \lstar\/ produced by a given model package typically span one order of magnitude. Exceptions are the \rstar\/ values produced by the \citet{R06} models, which span two orders of magnitude across our 12 sources, and the \lstar\/ values from the \citet{ZT18} models, which span four orders of magnitude.
Conversely, when comparing the results of all three model packages for a given source, the \rstar\/ results typically span one order of magnitude, but can span two; the \tstar\/ results typically also span one order of magnitude. The \lstar\/ results are more consistent with each other: 75\% of the \lstar\/ results from all three models agree to within a factor of 3. 
This result should not be overlooked $-$ the models rarely converge for physical parameters that assume specific geometries (i.e. \rstar), but do converge to properties that can be extracted from SED shape alone (i.e. \lstar).

The different model packages will have difficulty converging to a single combination of \rstar\/ and \tstar\/ if the geometry of a single source is not one of those assumed by the models (e.g. multiple protostellar sources, accretion disk much more or less massive than the range in the model grid, etc.).
All three of these sets of models do fit only one protostar at a time $-$ they assume only one source is contributing to the emission. However, recent research suggests that less-massive protostars may form in the accretion reservoirs of more massive companions ($<$0.2 pc separation), and in fact there is compelling evidence that this is the case for at least one of our sources \citep[G11.92$-$0.61; see][and references therein]{Cyganowski2017}.
Furthermore, the nine different wavelengths used to create these SEDs probe different spatial scales due to the angular resolution of individual telescopes (e.g. 1$\farcs$66 for {\it Spitzer} IRAC band I1 versus 19$\farcs$2 for the ATLASGAL survey). This may also contribute to the scatter in \rstar\/ and \tstar\/ for individual sources if, for instance, the NIR and MIR fluxes are correctly attributed to only one protostellar source but the FIR fluxes are instead the blended fluxes of multiple adjacent sources in a clustered environment.

Finally, it is possible that the SED fits themselves are good but the fits to individual parameters poor because the assumption by the models of hierarchical structure (i.e. central source(s), disk, envelope, cavities) is true, but the assumption that all structures are present in a single source (i.e. the same individual protostar) is false. Possible scenarios in which this could occur are the case in which multiple cores are present within a single envelope, such as for a protobinary system, or IR-bright outflows due to multiple cores, even if only one core is visible in the MIR.

\subsubsection{Trends by Model}
The trends in \rstar, \tstar, and \lstar\/ suggest that the \citet{R06} models favor cooler, larger - and hence younger - protostars to describe our data, while the \citet{ZT18} models favor smaller, hotter protostars. The \citet{R17} \rstar\/ and \tstar\/ results typically fall between the two other model packages, but the resulting Stefan-Boltzmann luminosities agree fairly well with those of \citet{R06}.
The \citet{R06} models tend to produce similar \chisq\/ values as the \chisq\/ of the best-fit \citet{R17} models, and produce \chisq\/ values that are lower than those of the \citet{ZT18} models in all but two cases.
The only notable trend in the \citet{R06} models is a slight tendency to overestimate the 37.1~\mum\/ flux (see Figs.~\ref{g1029_seds} through \ref{g3503_seds}, Appendix A). This effect is sometimes also present in the \citet{R17} models, though to a lesser degree.

We found that the \citet{R17} models without bipolar cavities routinely gave very poor results, as expected (\chisq\/ values that are factors of $\sim$10 to 40 higher than the models with bipolar cavities). 

Models s-pbsmi through spubhmi are shown Figures~\ref{g1029_seds} through \ref{g3503_seds} in Appendix A; models spu-smi and s-u-smi are not shown, as in all cases they produced significantly poorer fits than any of the other six \citet{R17} model sets.
Overall, the best-fit models (as determined by minimum \chisq\/ value) were always of the sets s-pbhmi, s-pbsmi, or spubhmi, in order of decreasing frequency.
That is, the models overall favored no disk and a power-law envelope.
For some sources, some model sets are clearly inappropriate, as they consistently underestimate long-wavelength emission or overestimate short-wavelength emission (e.g. model sets s-ubsmi and s-ubhmi for G14.33$-$0.64, Figure~\ref{g1433_seds}; model sets s-ubsmi, s-ubhmi, spubsmi, spubhmi for G14.63$-$0.58, Figure~\ref{g1463_seds}). However, in very few cases were there no \citet{R17} model sets that could reasonably fit our data.

In general, we find that the \citet{ZT18} models fit some of the mid-IR fluxes fairly well, but consistently overestimate the 37~\mum\/ and 70~\mum\/ fluxes and underestimate the emission at 160~\mum\/ and 870~\mum. 
The sources for which this is not the case are either very poorly constrained (G10.29) or contain known UC \HII\/ regions (G35.03).
The \citet{ZT18} models also have the highest \chisq\/ value in all but these two sources and, in these two sources, \chisq\/ $<$ 1, which indicates that those models may be overfit.
In general, the \citet{ZT18} \chisq\/ values are an order of magnitude higher than at least one of the other two models; in nine of twelve sources, it is higher than both.
Since the \citet{ZT18} models do not include an ambient medium/emission from the parent clump, it is likely that this omission is leading the model to underestimate our measured emission at long wavelengths, and to produce higher \chisq\/ values.
The cause of the overestimation in the MIR is currently unclear, though this trend is also present to some degree in the other two model packages as well. 
Combined with the underestimation of the FIR emission, this MIR overestimation creates a trend wherein the \citet{ZT18} models in general seem to be pushed toward SEDs with peaks at slightly shorter wavelengths than our data exhibit; this may indicate that the \citet{ZT18} models tend to produce better results for slightly older or less deeply-embedded sources.

\citet{soma} tested the \citet{ZT18} models against data from their SOFIA Massive Star Formation Survey (SOMA), and compared their results to results from the \citet{R06} models. In general, the \citet{ZT18} models produced good results for their sample, and they do not note a systematic underestimation of long-wavelength emission from these models. However, the SOMA survey identifies four source types, and \citet{soma} examines only their Type II (``Hypercompact'') sources. \citet{soma} state that these sources often have jet-like radio emission, and MIR emission that extends beyond the radio emission. While the second criterion applies to our sample, the first does not. The characteristics of our sources are a better match to their Type I (``MIR Sources in IRDCs,'' which is a quality nearly all of our sources share) or their Type IV (``Clustered Sources,'' which recent work \citep[e.g.][]{Cyganowski2017} shows is the case for at least one source in our sample, and likely more). \citet{soma} note that there is a rough evolutionary sequence from Type I to Type III sources. If this sequence is accurate, and our sources are more similar to their Type I sources, then our sample would be slightly younger than that evaluated in \citet{soma}. In this case, the MIR-emitting sources are indeed likely to be cooler and/or more deeply embedded than those in the SOMA Type II sample, and this would at least partially explain the discrepancy between the model results for our sample (i.e. consistent underestimation of FIR emission) and theirs.

\subsection{Do the \citet{R17} Model Sets Tell Us Something About Source Structure?}
\label{r17_trends}
The \citet{R17} model sets are the only models evaluated in this work which explicitly allow the user to test multiple different source geometries. We briefly evaluate the overall trends in the geometries of the best-fit model sets for our sources in order assess what, if anything, the \citet{R17} models are telling us about the structure of the protostellar sources in our sample. Model results for each individual source can be found in Appendix B, along with a discussion of how the method of evaluating which is the ``best'' fit affects these trends.

G10.29$-$0.13 is excluded from this discussion of general trends, as its SED is very poorly constrained. 
For the remaining 11 sources, the \citet{R17} model package frequently returns best-fit models which have power-law envelopes and no disks, and do not favor either the presence or absence of an inner hole. However, among model sets specifically with no disk and with a power-law envelope, there is a clear preference (64\% to 36\%, or 7 to 4) for models with an inner hole. That is, a power-law envelope favors a larger distance between the inner edge of the envelope and the stellar surface. 
Among models with no disk and an Ulrich envelope, this trend is exactly reversed, with 64\% of models (7  out of 11) preferring no inner hole and only 36\% (4 out of 11) having one; in this case, the addition of rotation seems to favor a smaller distance between the envelope inner radius and the stellar surface. 
It should also be noted that only three of our eight model sets contain a disk (spubhmi, spubsmi, spu-smi), and one of these (spu-smi) was expected to generally give poor fits to our data anyway due to its lack of bipolar cavities. It is possible that the bias against disks may be due, at least in part, to the relative dearth of individual model sets with disks compared to those without. Of the eighteen model sets available to us, we chose to run the six models with bipolar cavities and to ``control'' models. Two of the six model sets with bipolar cavities have disks, while four do not $-$ this 2-to-1 ratio is simply a feature of the model sets available to us. However, this ratio may give an unphysical ``advantage'' to the disk-lacking  models in the evaluation of model-set statistics. 

The lack of disks in the favored models is inconsistent with our more detailed knowledge of particular sources, such as G11.92$-$0.61 \citep{Ilee2016,Ilee2018} and G16.59$-$0.05 \citep{Moscadelli2016,Rosero2016}, as well as our more general knowledge of these sources based on their additional attributes (e.g. shocked H$_2$ emission (\citealt{egocat}), both Class I and Class II \methanol\/ masers (\citealt{Cyganowski2009}), etc.). For most sources, the \citet{R17} models with disks do not appear (visually) to be significantly different in the mid-infrared ($\sim$10 to 40~\mum; see Figures \ref{g1029_seds} through \ref{g3503_seds}) from models without them. This region of the SED is frequently dominated by hot dust emission from the outflow cavity and heated portions of the envelope. A disk that is small relative to the mass of the protostar, or highly extincted by embedding material, might manifest its presence less strongly in the mid-infrared portion of the SED, in which case SED modeling would not need to invoke a disk in order to reproduce the given data.

\subsection{$L/M$ and Evolutionary State}
\begin{figure*}
    \centering
    \includegraphics[angle=270,width=\textwidth]{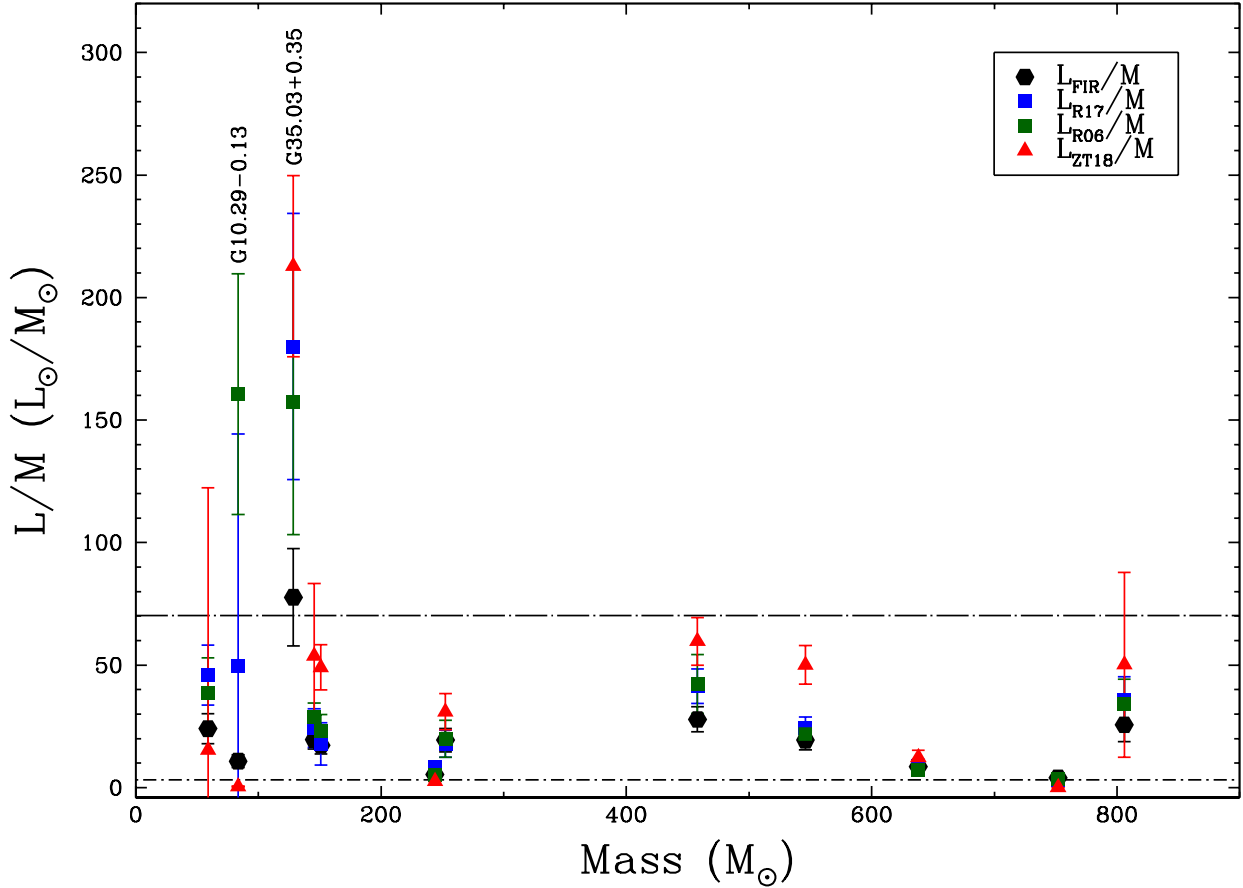}
    \caption{$L/M$ versus $M$ for all sources using all four luminosity values. Mass is the average of the \ammonia-derived and greybody-derived masses for each source. Symbols denote which luminosity value was used for L/M: black hexagons are the greybody-derived $L_{FIR}$, blue squares are the luminosity returned by the \citet{R17} models, green squares are the luminosity returned by the \citet{R06} models, and red triangles are the luminosity returned by the \citet{ZT18} models. The upper and lower dash-dotted lines are the median $L/M$ values for the \citet{Tige2017} ``IR-bright'' and ``IR-quiet'' categories, respectively. Sources G10.29$-$0.13 and G35.03$+$0.35 are labeled specifically, as they both have some $L/M$ values approximately an order of magnitude higher than the rest of our sample.}
    \label{lmm}
\end{figure*}

Figure~\ref{lmm} shows the luminosity-to-mass ratio $L/M$ versus mass $M$ for each source using each of the four derived luminosities. For both $L/M$ and $M$, $M$ is the average of the \ammonia-derived and greybody-derived masses for each source. The errors on $L/M$ are determined from the error propagation equation and the errors on $L$ and $M$, respectively. 
The uncertainty on each $L$ value, $\sigma_{L}$, is the median absolute deviation from the median (MAD) of the luminosities of all 
fits with \chisq\/ within a factor of 3 of the best-fit \chisq\/ value. 
The uncertainty on each $M$ value, $\sigma_{M}$, is calculated from the uncertainties on both the greybody-derived and \ammonia-derived temperatures, the 870~\mum\/ flux F$_{870 \mu m}$, and distance, D, using the error propagation equation. Uncertainties in temperature are reported in Table~\ref{temp_mass}, and uncertainties in F$_{870 \mu m}$ are reported in Table~\ref{fir_fluxes}. Distance uncertainties for parallax-derived distances are from \citet{Reid2014}. To estimate a distance uncertainty for the EGOs that only have kinematic distances, we assessed the percent difference between the predicted kinematic distance and the parallax distance for the five sources for which both are available, and found a median percent difference of 15\%.

The median $L/M$ for our sources is 24.7 $\pm$ 8.4 \lsun/\msun, where the uncertainty is the MAD and the $L/M$ for each source is the median of the four values shown in Figure~\ref{lmm}. Most $L/M$ values fall in the range 5 $-$ 60 \lsun/\msun, regardless of the method of deriving the luminosity. This result is in line with the results of \citet{Carpenter1990}, who studied a set of 21 molecular clouds in the Outer Milky Way whose masses, luminosities, and suspected evolutionary state are comparable to our sample. They report an $L/M$ range of 1.1 $-$ 39.2 \lsun/\msun\/ (mean 6.8 \lsun/\msun).
Typical L/M values for low- and intermediate-mass protostars, in contrast, usually span $\sim$0.1 $-$ 10 \lsun/\msun\/ \citep[see, e.g., L and M values in][]{Enoch2009}.

For most of our sources, the four $L/M$ values we derive span a range of a factor of $\sim$2.5, likely due to differences in the four different methods of deriving luminosity. 
The exceptions are G10.29$-$0.13 and G35.03$+$0.35, which span ranges of a factor of $\sim$100 and $\sim$3, respectively. The spread in $L/M$ for G10.29$-$0.13 is likely due to poorly-constrained SEDs, as has been previously discussed. G35.03$+$0.35 is discussed in the context of evolutionary stage in greater detail below.

The only notable trend in our $L/M$ values is the tendency of the \citet{ZT18} models to give $L/M$ values that are either higher than all three other methods, or lower than all three, but never in between. In particular, the \citet{ZT18} results tend to produce lower $L/M$ only when the value is very low ($<$ 20 \lsun/\msun); otherwise, the \citet{ZT18} $L/M$ is higher than both the two Robitaille- and the greybody-derived $L/M$ values. This dichotomy is entirely consistent with the trend in Stefan-Boltzmann luminosities for the \citet{ZT18} results noted in \S~\ref{sb_lum}.
We do not note any particular trend in $L/M$ with mass, though it should be noted that the mass range of our sample is small compared to that of other teams \citep{Urquhart2018,Elia2017}.

A comparison of our results with the results of other teams shows that our $L/M$ are well in line with established values for MYSOs.
Both \citet{Urquhart2018} and \citet{Elia2017} compare $L$ and $M$ values for pre- and protostellar clumps.
In both samples, massive star-forming regions are distinguished from prestellar sources by the space they occupy in $L - M$ parameter space.
\citet{Urquhart2018} note that the $L/M$ values of massive star-forming clumps (as distinct from less massive or prestellar objects) are well-described by lower and upper limits of 1 and 100, respectively.
Most $L/M$ in our sample fall in the range 5$-$60 \lsun/\msun\/ (see Figure~\ref{lmm}), which is well in line with the star-forming samples of both \citet{Urquhart2018} and \citet{Elia2017}. 
\citet{Urquhart2018} further note that compact \HII\/ regions become common in their sample at $L/M >$ 40. The median $L/M$ for our sample is $\sim$25, with a few $L>$ 40 from SED fit results. We have only one source which has $L >$ 40 for all four luminosities, G35.03$+$0.35, and this source does have a known UC \HII\/ region within the extended 4.5~\mum\/ emission of the EGO. 

Similarly, \citet{Tige2017} examine 46 high-mass pre- and protostellar cores in NGC 6334 as part of the {\it Hershel}-HOBYS program. They separate their massive dense cores (MDCs) into three categories: IR-bright MDCs, IR-quiet MDCs, and Starless MDC Candidates. 
To distinguish IR-bright and IR-quiet sources, they use flux limits of 10, 12, and 15~Jy at 21, 22, and 24~\mum, respectively. Sources with fluxes above these limits are considered IR-bright, and sources with fluxes below these limits are IR-quiet.
The $F$ values are based on the predicted mid-IR emission of a B3-type protostar and calculated for a distance of 1.75~kpc, which is the distance to NGC6334 assumed by \citet{Tige2017}. 
These values are consistent with the weak-to-bright cutoff used by \citet{Konig2017} when scaled to 4~kpc (see \S~\ref{temp_mass}).
\citet{Tige2017} determine their source masses by fitting source SEDs from 70 to 1200 \mum\/ using data from {\it Herschel}, JCMT, APEX, and SEST.
In cases of no significant mid-IR emission, \citet{Tige2017} determine bolometric luminosity by performing greybody fits to the far-IR data, similar to the process we use for our own data (see \S~4.1).
In cases of significant mid-IR emission, they determine bolometric luminosity by integrating directly under the observed flux values. In this case, they use the data sets described above as well as data from 3.6 to 24~\mum\/ from {\it Spitzer}, MSX, and WISE.
\citet{Tige2017} at all times assume optically thin emission at $\lambda$ $>$ 100~\mum\/ and use $\beta$ = 2.

When we compare our sources with the \citet{Tige2017} subsamples, an interesting feature emerges.
Our median  $L/M$ when $L$ is the greybody luminosity is 19.4 $\pm$ 7.3 \lsun/\msun, and the median $L/M$ when $L$ is the median of all four luminosities is 24.7 $\pm$ 8.4 \lsun/\msun. 
Compare this to the median $L/M$ of the IR-quiet and IR-bright cores in the \citet{Tige2017} sample (3.1 $\pm$ 2.8 \lsun/\msun\/ and 70 $\pm$ 28 \lsun/\msun, respectively, where these medians are calculated from the \lstar\/ and \mstar\/ listed in \citet{Tige2017}, Table~3).
The median $L/M$ of our sample falls neatly between the median $L/M$ of the IR-quiet and IR-bright populations in \citet{Tige2017}, whether we use $L$ from only the greybody fit or the median of all four $L$-values. Even with uncertainties, our sources are still well-separated from either category. 
While there are some differences between our methods of deriving $L/M$ and those of \citet{Tige2017}, these are unlikely to significantly change this result. Recalculating our $L$ and $M$ values using $\beta$~=~2 (the value used by \citet{Tige2017}) instead of $\beta$~=~1.7 
only increases our median $L/M$ by 7.3\%.
This finding is consistent with \citet{Tige2017}, who calculate that using $\beta$~=~1.5 instead of $\beta$~=~2 would only alter their calculated masses by 5$-$10\%.

\citet{Tige2017} suggest that IR-quiet MDCs are precursors to IR-bright MDCs; once IR-quiet cores have accreted enough mass to produce a stellar embryo with M $>$ 8 \msun, their luminosity sharply increases and they become IR-bright. (See \citet{Motte2018} for a review of the theory and current observational support for this scenario.)
This transition corresponds to the swelling phase of \citet{Hosokawa2009}, in which a massive protostar rapidly expands after reaching $M$ $\sim$ 6 \msun. The rapid expansion is driven by the sudden escape of significant entropy from the interior of the star, which can only occur after the opacity is sufficiently decreased by increasing temperature. The swelling phase is comparatively brief and lasts only until $M$ $\sim$ 10 \msun, and is immediately followed by a Kelvin-Helmholtz contraction phase in which the protostellar radius decreases again.
Given that our median $L/M$ fall between the values for the well-established ``IR-quiet'' and ``IR-bright'' categories of massive protostellar objects, it is possible that our EGO-12 sample represents a transitional stage between the IR-quiet and IR-bright phases of evolution, i.e., a phase in which an accreting protostar reaches some critical (large) mass, undergoes a concrete physical change, and consequently increases sharply in luminosity.
Given the lack of predicted observable properties for this swelling/growth phase other than an increase in luminosity, we are hesitant to suggest that this is definitively the state in which our sources exist. However, the possibility is intriguing and suggests an interesting avenue for further investigation. 

If $L/M$ is indeed a reliable indicator of evolutionary state, as asserted by other teams, then it should be unsurprising both that a) the majority of our sample, which were specifically selected due to their {\it uniformity} of evolutionary state, all exhibit very similar $L/M$ values, and b) that G35.03$+$0.35, which compared to the majority of the sample is in a very late stage of evolution, has a significantly higher $L/M$.
Such a possibility $-$ that EGOs in particular represent the stage of MYSO evolution immediately prior to the emergence of strong mid-infrared emission and subsequent \HII\/ regions $-$ warrants further investigation in future work.

\section{Conclusions \& Future Work}
\label{future}
We have conducted a multiwavelength study of twelve typical massive protoclusters in the Milky Way using SOFIA FORCAST imaging and archival infrared data.
We performed aperture photometry at each wavelength in order to construct SEDs from the near-IR (3.6~\mum) to sub-mm (870~\mum), which we then fit with one greybody and three radiative transfer models. The radiative transfer models \citep{R06,R17,ZT18} all model near-IR to sub-millimeter emission in the context of a single protostar.

The SOFIA images, in conjunction with archival data, suggest that the number of massive sources per EGO is between 0.9 and 1.9. This moderate MYSO multiplicity is in line with published values for similar samples \citep{Rosero2018} and for G11.92$-$0.61 \citep{Cyganowski2017,Ilee2018}. The multiplicity of these sources cannot be further constrained without sub-arcsecond resolution images, in either the mid-infrared or radio regimes.
\citet{Cyganowski2017} do detect a plethora of lower-mass sources in G11.92$-$0.61, none of which are indicated in the SOFIA images; this is consistent with both our angular resolution limitations and with the luminosity sensitivity limit of 1.1$\times$10$^{3}$ that we calculate in \S~\ref{sensitivity}, which indicates these SOFIA observations will not be sensitive to lower-mass, lower-luminosity YSOs.

We find that, for this sample, the temperatures derived from greybody fits to dust emission are quite similar to the temperatures derived from single-component fits to \ammonia\/ (1,1) to (3,3) emission \citep{Nobeyama}, with the dust temperatures trending slightly higher than the \ammonia\/ temperatures. While these differences fall below a level of statistical significance, this trend is in line with the published results of other teams \citep{Giannetti2017,Konig2017} which find that, at temperatures above $\sim$15 K, \ammonia\/ emission tends to probe gas that is slightly cooler than the local dust. 
We find that the median $T_{dust}$ of the EGO-12 sample is consistent with the median $T_{dust}$ of the Top100 sample \citep{Konig2017}, and most closely aligned with either the ``IR-weak'' or ``IR-bright'' subcategories. The overlap between the EGO $T_{dust}$ range and both the IR-weak and IR-bright $T_{dust}$ ranges suggests that EGO-12 sample may represent the cooler end of the IR-bright population, or possibly a separate, intermediate population between the IR-weak and IR-bright subcategories.
High-precision temperature measurements over a much larger EGO sample are needed in order to address both possibilities.

The $L$ we derive from the greybody fits agree within 20$-$50\% of other published $L$ for these sources \citep{Moscadelli2016,Urquhart2018}. We find that the greybody-derived $L$ of \citet{Urquhart2018} are greater than our greybody-derived luminosities in all cases. This is consistent with the fact that our greybody fits assume a single (cold) component, while \citet{Urquhart2018} use both a cold (greybody) and hot (blackbody) component for at least eleven of our sources. 
The Stefan-Boltzmann luminosities (\lstar), which we calculate from the \rstar\/ and \tstar\/ returned by the radiative transfer models and which do account for hot-component emission, are typically of order 2$\times$ higher that the luminosities returned by our greybody fits. They are also more in line (within $\sim$30\% rather than 50\%) with the luminosities of \citep{Urquhart2018}.

We find that the individual \rstar\/ and \tstar\/ returned by the three radiative transfer packages vary widely both within and between packages (one order of magnitude in \tstar\/ and up to two in \rstar for both cases).
This suggests that the models are having trouble converging to a single set of protostellar parameters, and that perhaps none of the model packages are fitting the sources uniquely well. 
This result is consistent with the fact that, although we assumed a single dominant source for modeling purposes, these sources are actually protoclusters rather than isolated protostars. Objects of different evolutionary states may be contributing to the total emission even if they are too embedded or too clustered to be detected individually in our SOFIA images. This possibility is further supported by the detection of multiple mid-infrared SOFIA sources within the ATLASGAL emission for most of our targets.

The specific cases of G11.92$-$0.61, G16.59$-$0.05, and G35.03$+$0.35 $-$ for which high-resolution, high-sensitivity (sub)millimeter and/or radio-wavelength data are available $-$ highlights the limitations of such (comparatively) low-resolution photometry when applied to clustered sources. In particular, it strongly suggests that SED model results should not be used to identify or describe the properties of disk candidates in high-mass protostars as is commonly done for isolated low-mass protostars \citep[e.g.][]{Spezzi2013}, or should only be used with extreme caution or in cases where the isolation of the high-mass protostar can be positively confirmed. For clustered sources, lower-resolution infrared data can be a powerful tool for describing global properties of each protocluster and testing the multiplicity of massive sources, but results from models assuming individual protostellar sources should be used with caution. 

The $L/M$ values of our sample are well in line with $L/M$ values measured by other teams \citep{Carpenter1990}, but fall between the two distinct IR-quiet and IR-bright categories suggested in \citet{Tige2017}. 
Given that the evolutionary stage in which our objects exist - MYSO outflows being powered by active protostellar accretion - is thought to be short-lived, it is possible that our sample represents the transitional stage between the IR-quiet and IR-bright phases of evolution. While intriguing, this possibility requires further investigation before any definitive statements can be made.

If $L/M$ is indeed indicative of evolutionary stage, then it likely correlates with other source properties such as outflow momentum, millimeter luminosity (e.g. as observed by ALMA, SMA, etc.), or the presence and nature of radio continuum emission. In order to assess the existence and strength of such correlations, additional centimeter-millimeter wavelength observations are needed. We have recently obtained or are in the process of obtaining ALMA 1.3 and 3.2~mm and JVLA 1.3 and 5~cm line and continuum observations for this purpose. These observations, and their correlation (or lack thereof) with the infrared and sub-mm results of this paper, will be presented in future publications.

\section{Acknowledgements}
We thank the referee for their timely, thoughtful review and helpful suggestions which have improved this paper.
This work is based in part on observations made with the NASA/DLR Stratospheric Observatory for Infrared Astronomy (SOFIA). SOFIA is jointly operated by the Universities Space Research Association, Inc. (USRA), under NASA contract NNA17BF53C, and the Deutsches SOFIA Institut (DSI) under DLR contract 50 OK 0901 to the University of Stuttgart. Financial support for this work was provided by NASA through award \#BU 29120004 issued by USRA.
This work is based in part on observations made with the Spitzer Space Telescope, which is operated by the Jet Propulsion Laboratory, California Institute of Technology under a contract with NASA.
Herschel is an ESA space observatory with science instruments provided by European-led Principal Investigator consortia and with important participation from NASA.
The ATLASGAL project is a collaboration between the Max-Planck-Gesellschaft, the European Southern Observatory (ESO) and the Universidad de Chile. It includes projects E-181.C-0885, E-078.F-9040(A), M-079.C-9501(A), M-081.C-9501(A) plus Chilean data. 
This research has made use of the NASA/ IPAC Infrared Science Archive, which is operated by the Jet Propulsion Laboratory, California Institute of Technology, under contract with the National Aeronautics and Space Administration.
The National Radio Astronomy Observatory is a facility of the National Science Foundation  operated under  agreement  by  the  Associated Universities,  Inc. 
This research made use of NASA's Astrophysics Data System Bibliographic Services, APLpy, an open-source plotting package for Python (Robitaille and Bressert, 2012), the SIMBAD database (operated at CDS, Strasbourg, France), and CASA.
CJC acknowledges support from the STFC (grant number ST/M001296/1).

\paragraph{Software}
{\tt sedfitter}, \citep{R06};
{\tt sedfitter}, \citep{R17};
{\tt sedfit}, \citep{ZT18};
APLpy, \citep{aplpy}, http://aplpy.github.com.

\newpage
\appendix
\section{A: \citet{R17}, \citet{R06}, \& \citet{ZT18} Model SED Plots for Individual Sources}
\begin{figure*}[hbt]
    \centering
    \includegraphics[width=\textwidth]{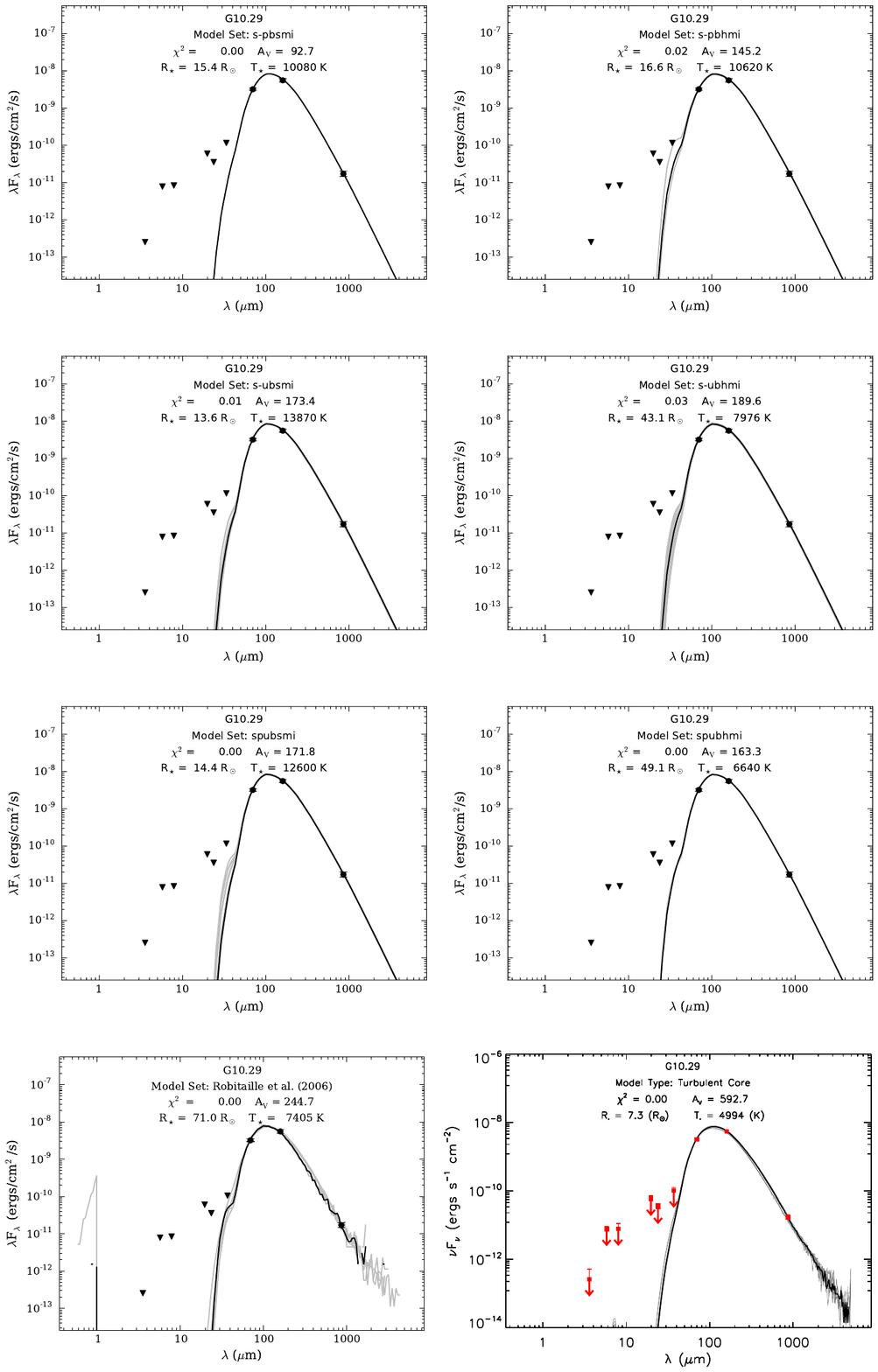}
    \caption{SED modeling results for G10.29$-$0.13, showing: (top six panels) the six best model sets from the \citet{R17} models based on \chisq\/ values, (bottom left panel) the model results from \citet{R06}, and (bottom right panel) the model results from \citet{ZT18}.}
    \label{g1029_seds}
\end{figure*}

\begin{figure*}[hbt]
    \centering
    \includegraphics[width=\textwidth]{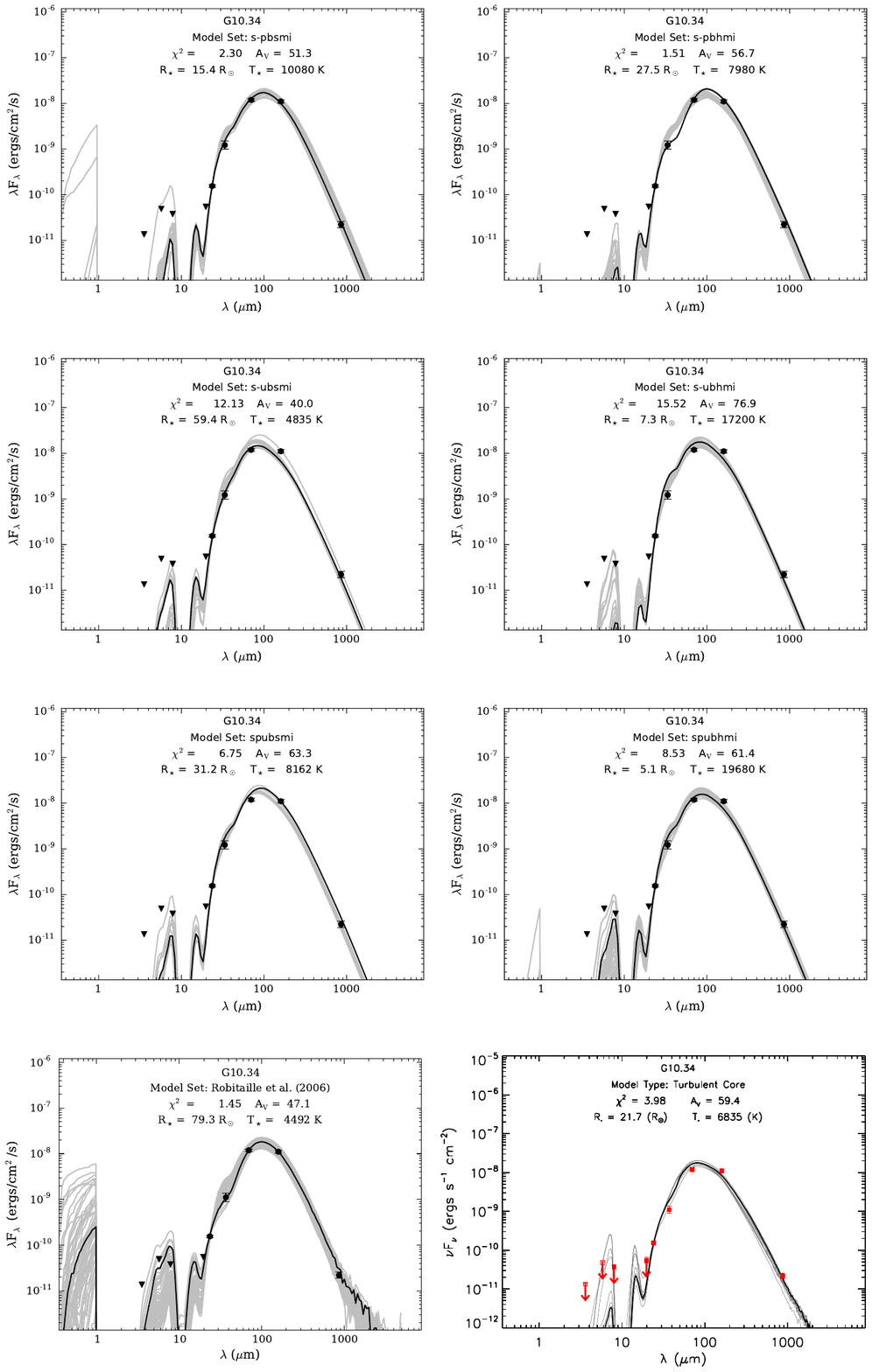}
     \caption{SED modeling results for G10.34$-$0.14, showing: (top six panels) the six best model sets from the \citet{R17} models based on \chisq\/ values, (bottom left panel) the model results from \citet{R06}, and (bottom right panel) the model results from \citet{ZT18}.}
     \label{g1034_seds}
\end{figure*}

\begin{figure*}[hbt]
    \centering
    \includegraphics[width=\textwidth]{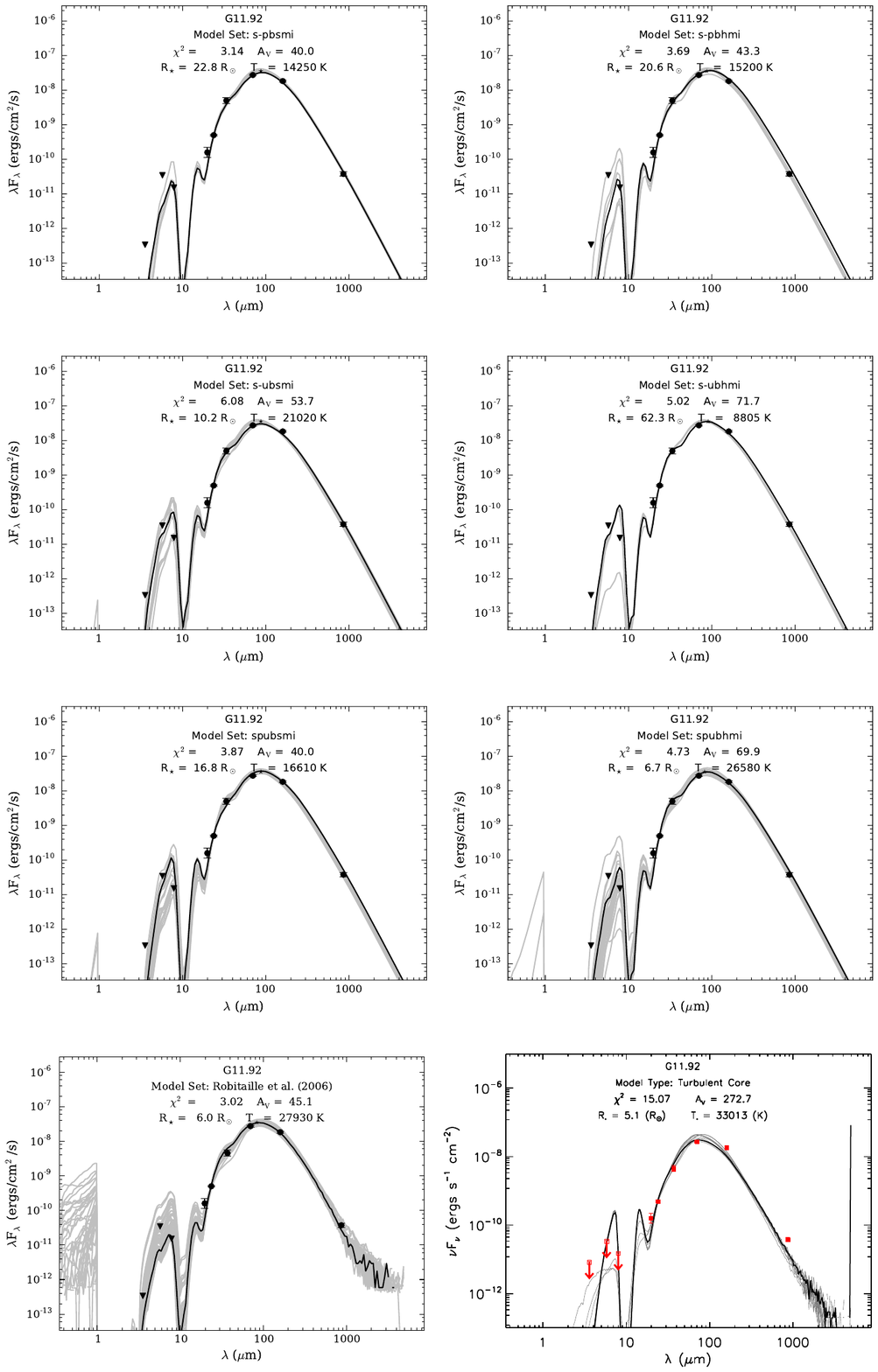}
     \caption{SED modeling results for G11.92$-$0.61, showing: (top six panels) the six best model sets from the \citet{R17} models based on \chisq\/ values, (bottom left panel) the model results from \citet{R06}, and (bottom right panel) the model results from \citet{ZT18}.}
     \label{g1192_seds}
\end{figure*}

\begin{figure*}[hbt]
    \centering
    \includegraphics[width=\textwidth]{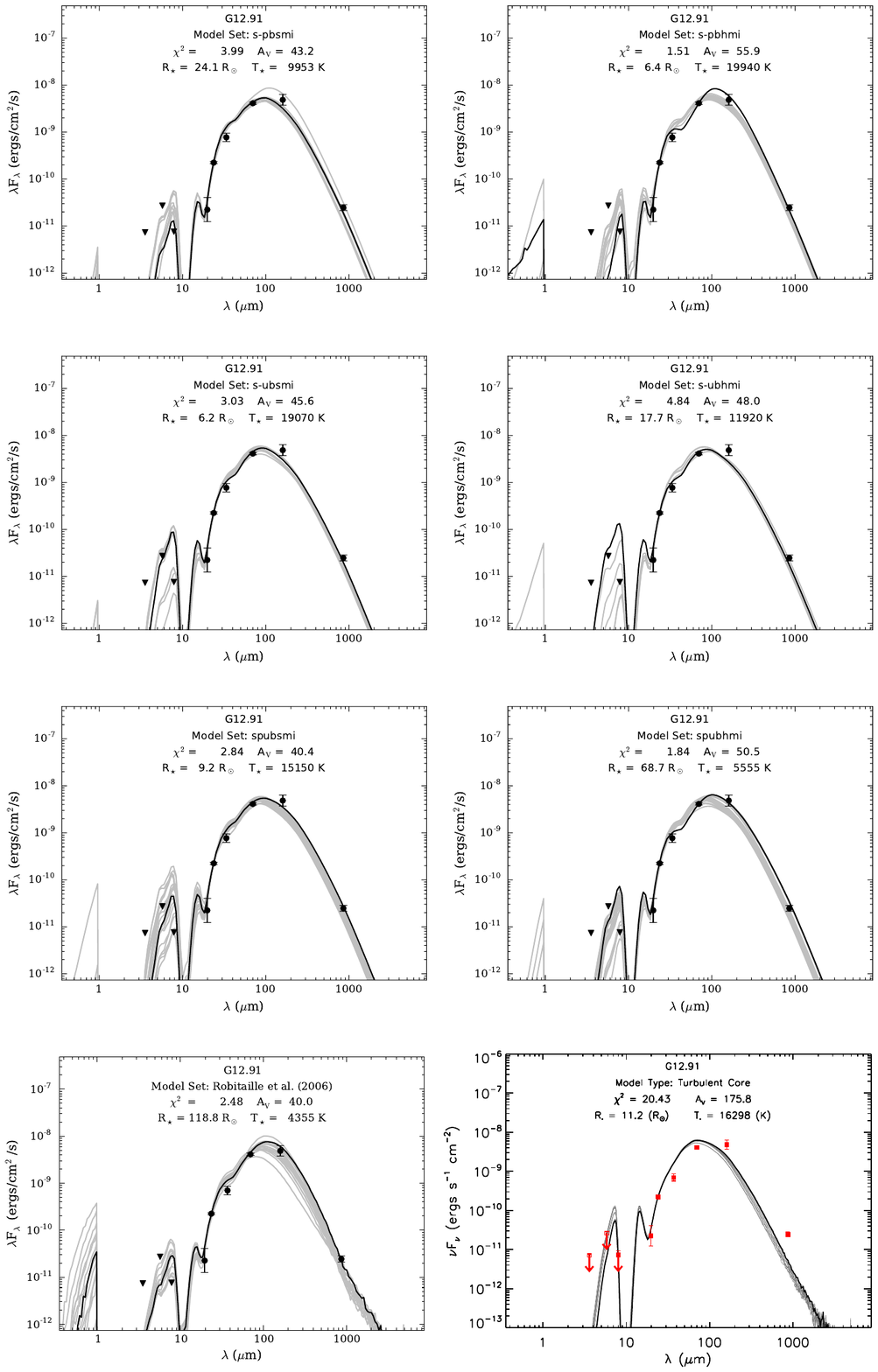}
     \caption{SED modeling results for G12.91$-$0.03, showing: (top six panels) the six best model sets from the \citet{R17} models based on \chisq\/ values, (bottom left panel) the model results from \citet{R06}, and (bottom right panel) the model results from \citet{ZT18}.}
     \label{g1291_seds}
\end{figure*}

\begin{figure*}[hbt]
     \centering
    \includegraphics[width=\textwidth]{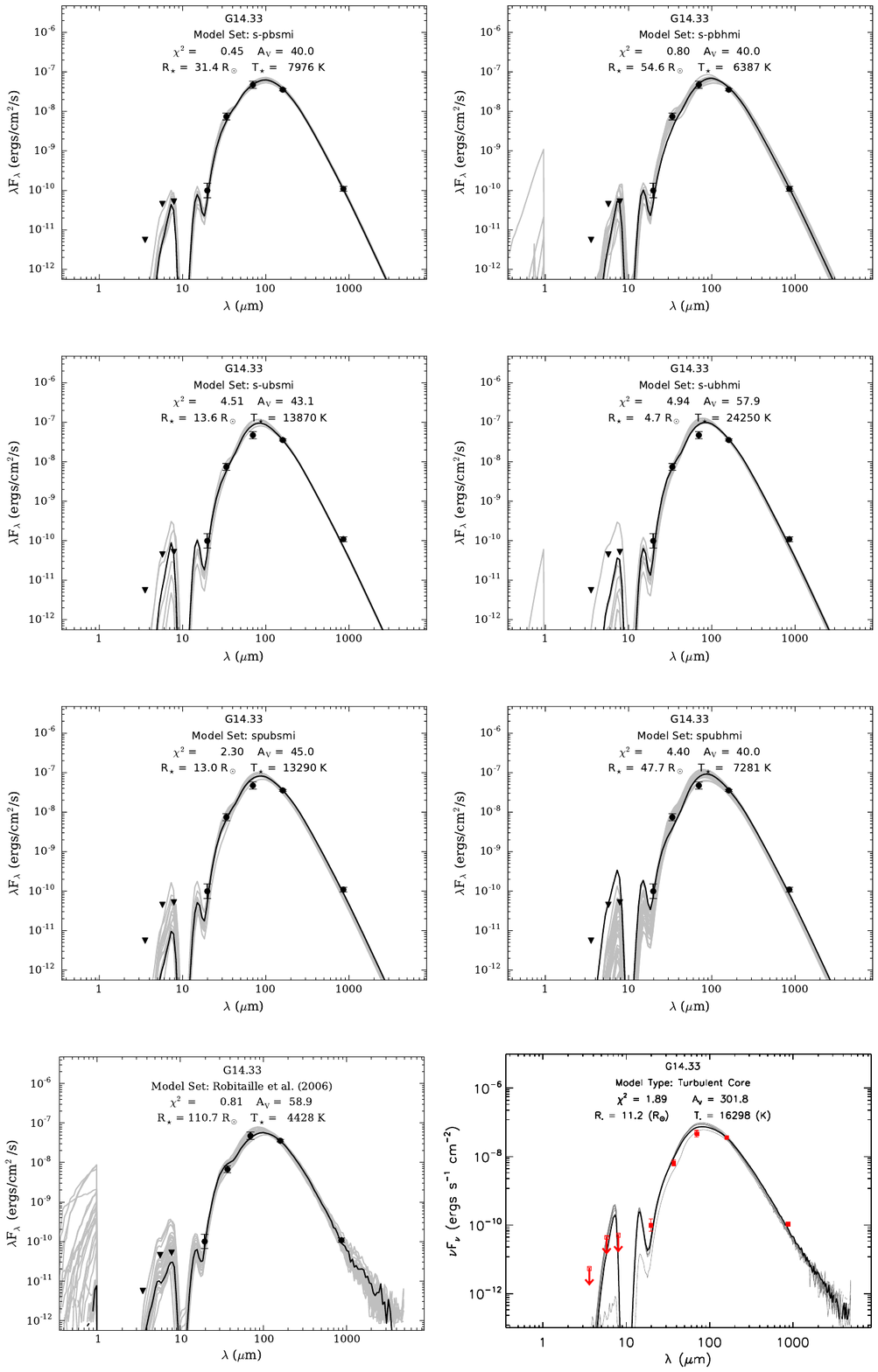}
     \caption{SED modeling results for G14.33$-$0.64, showing: (top six panels) the six best model sets from the \citet{R17} models based on \chisq\/ values, (bottom left panel) the model results from \citet{R06}, and (bottom right panel) the model results from \citet{ZT18}.}
     \label{g1433_seds}
\end{figure*}

\begin{figure*}[hbt]
     \centering
    \includegraphics[width=\textwidth]{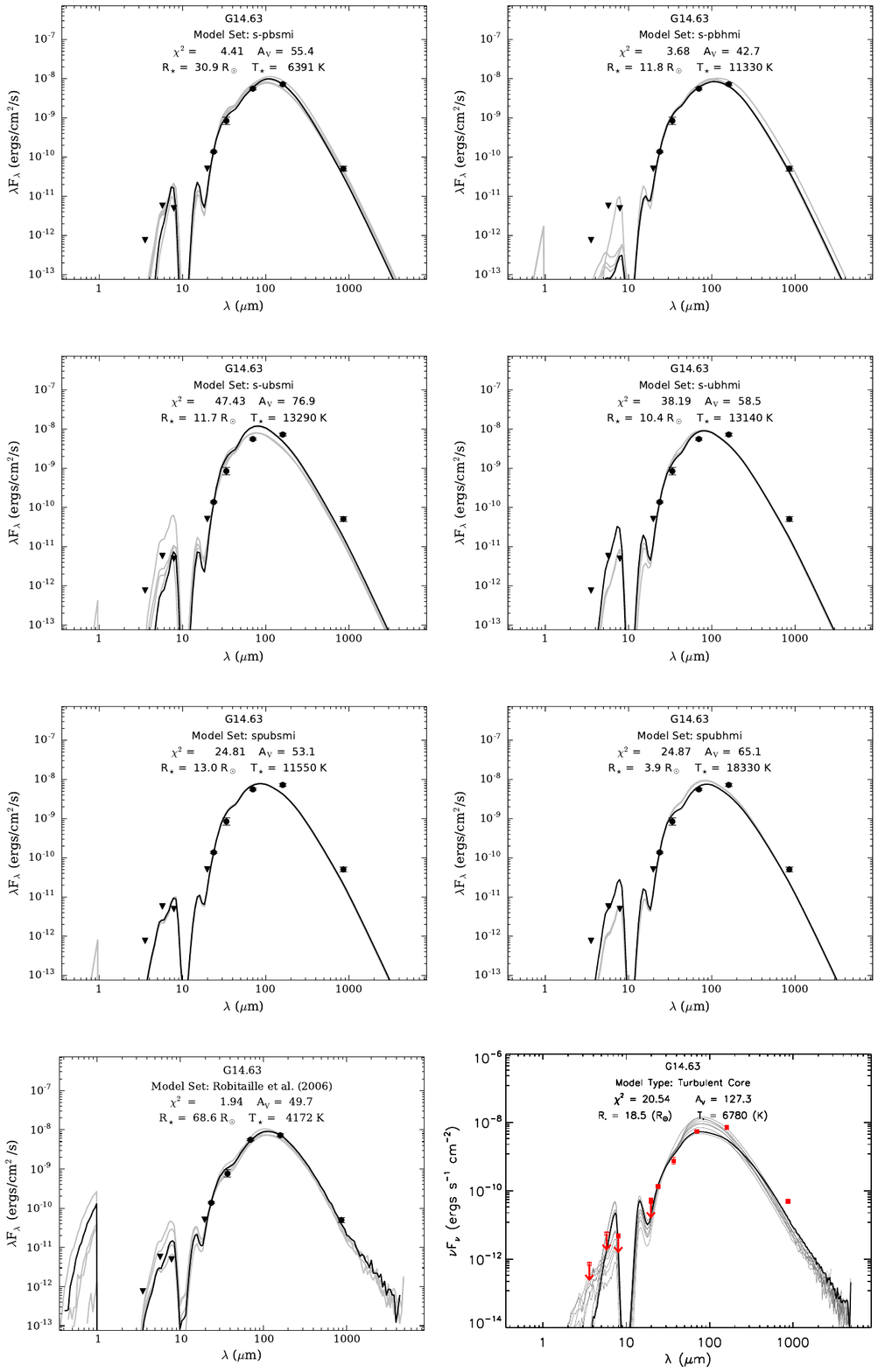}
     \caption{SED modeling results for G14.63$-$0.58, showing: (top six panels) the six best model sets from the \citet{R17} models based on \chisq\/ values, (bottom left panel) the model results from \citet{R06}, and (bottom right panel) the model results from \citet{ZT18}.}
     \label{g1463_seds}
\end{figure*}

\begin{figure*}[hbt]
     \centering
    \includegraphics[width=\textwidth]{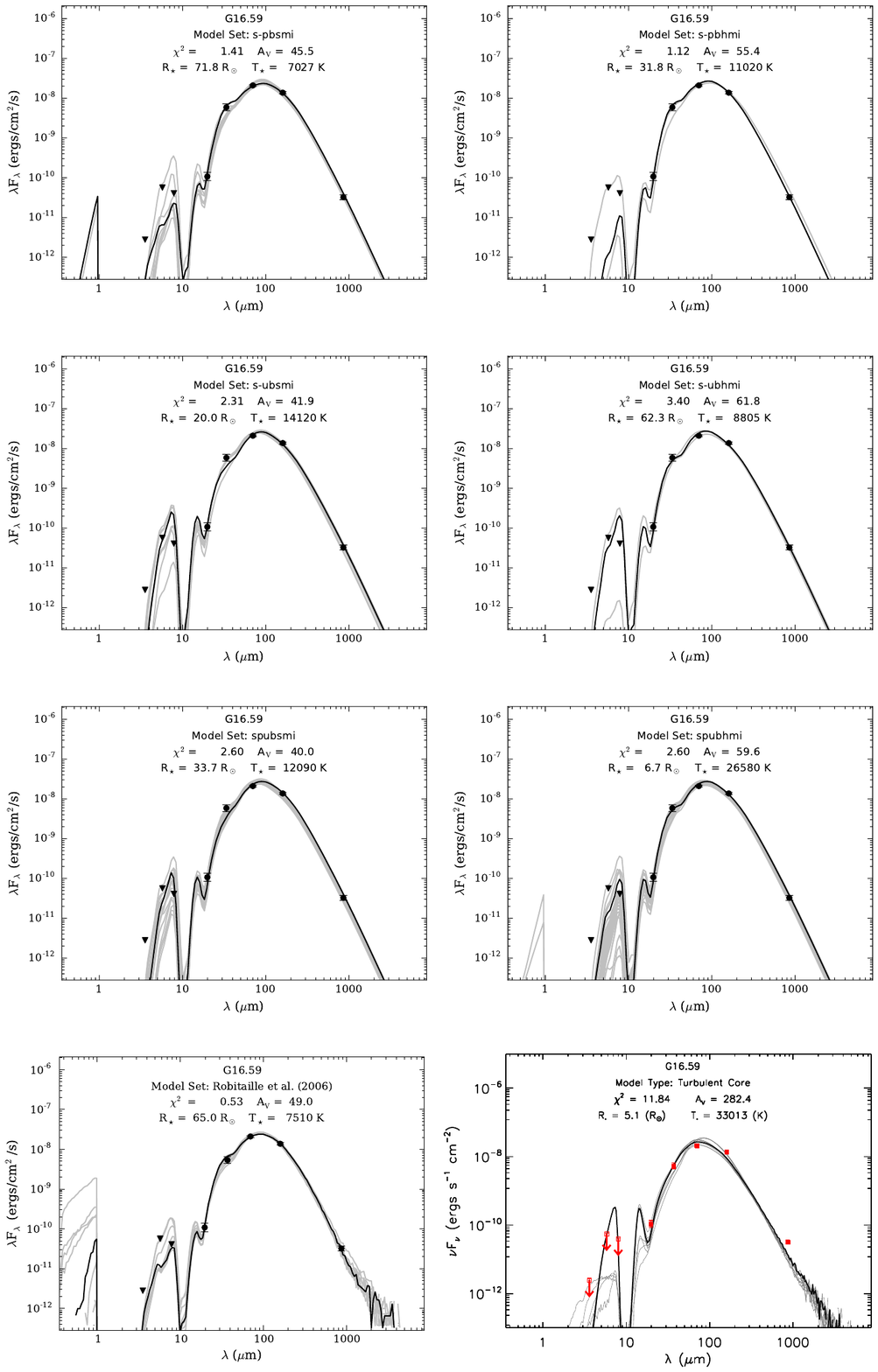}
     \caption{SED modeling results for G16.59$-$0.05, showing: (top six panels) the six best model sets from the \citet{R17} models based on \chisq\/ values, (bottom left panel) the model results from \citet{R06}, and (bottom right panel) the model results from \citet{ZT18}.}
     \label{g1659_seds}
\end{figure*}

\begin{figure*}[hbt]
     \centering
    \includegraphics[width=\textwidth]{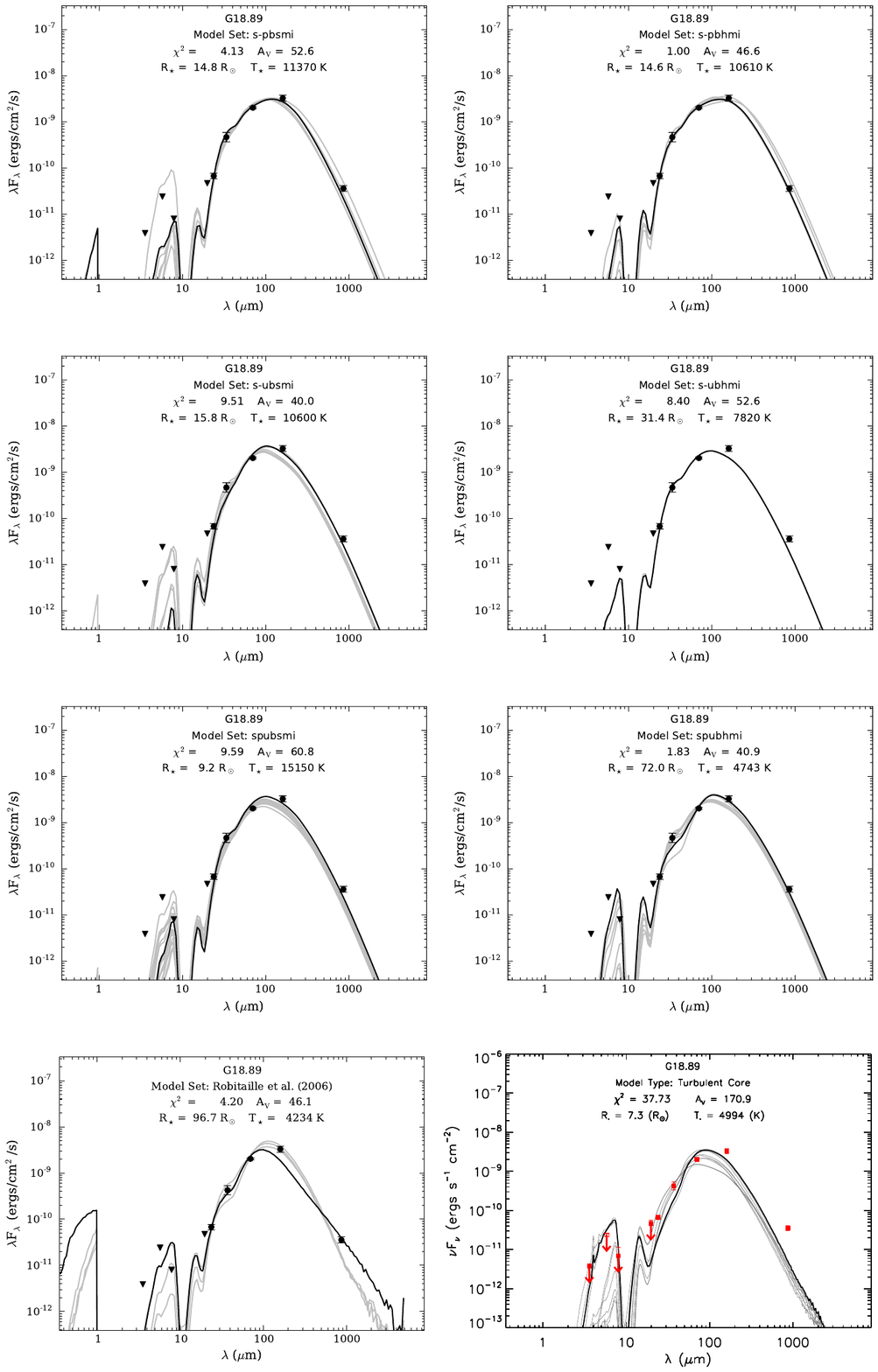}
     \caption{SED modeling results for G18.89$-$0.47, showing: (top six panels) the six best model sets from the \citet{R17} models based on \chisq\/ values, (bottom left panel) the model results from \citet{R06}, and (bottom right panel) the model results from \citet{ZT18}.}
     \label{g1889_seds}
\end{figure*}

\begin{figure*}[hbt]
     \centering
    \includegraphics[width=\textwidth]{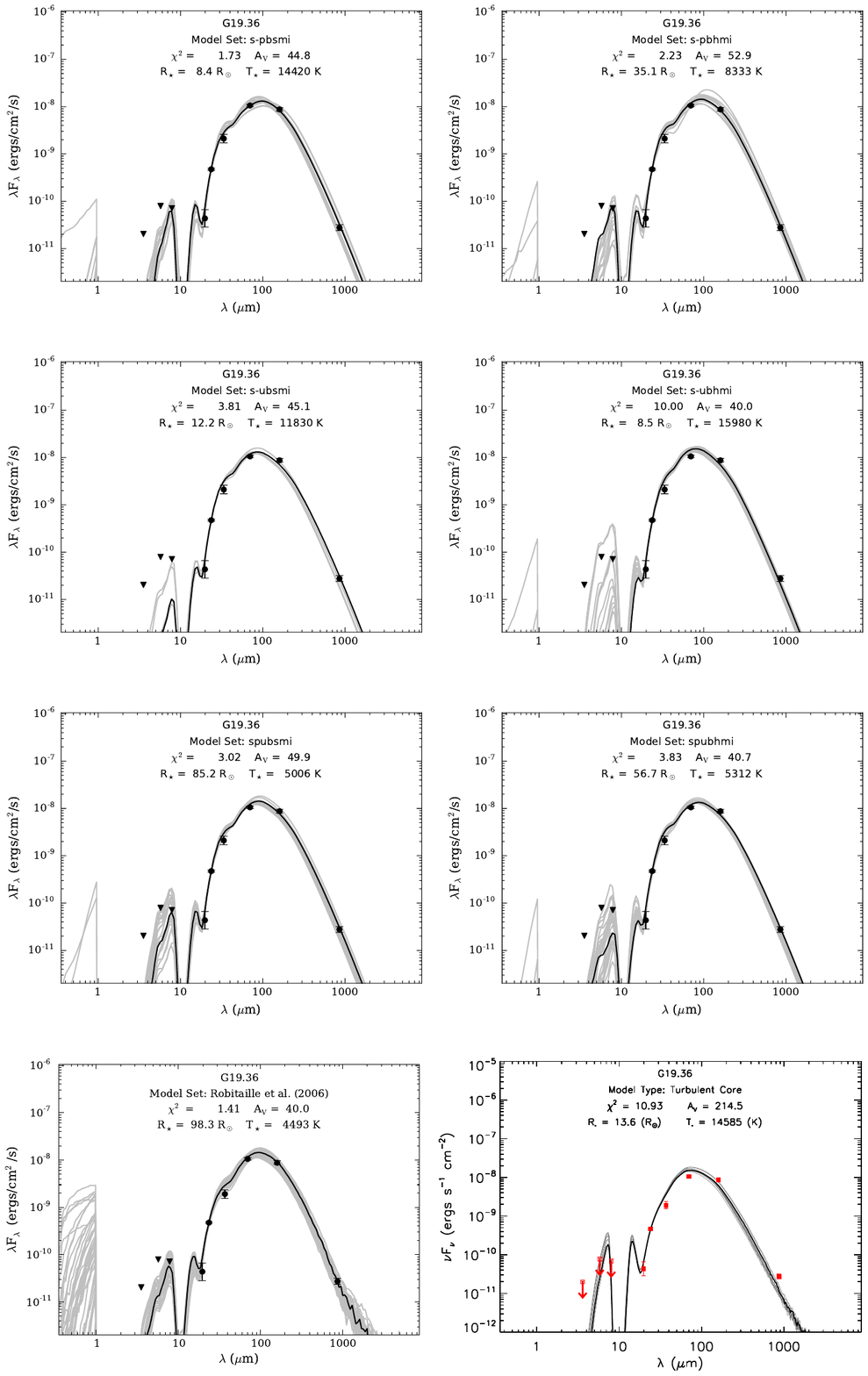}
     \caption{SED modeling results for G19.36$-$0.03 showing: (top six panels) the six best model sets from the \citet{R17} models based on \chisq\/ values, (bottom left panel) the model results from \citet{R06}, and (bottom right panel) the model results from \citet{ZT18}.}
     \label{g1936_seds}
\end{figure*}

\begin{figure*}[hbt]
     \centering
    \includegraphics[width=\textwidth]{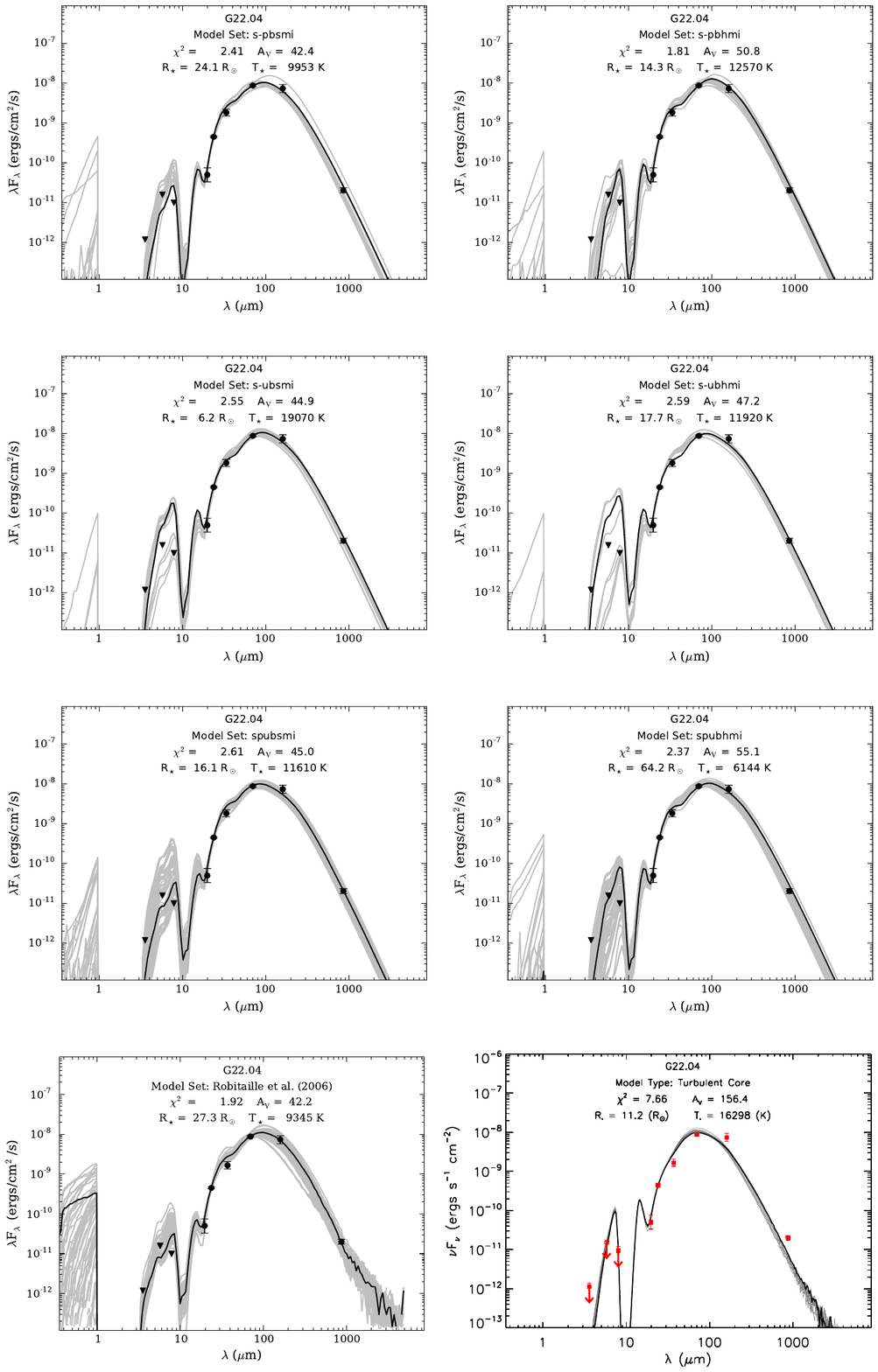}
     \caption{SED modeling results for G22.04$+$0.22, showing: (top six panels) the six best model sets from the \citet{R17} models based on \chisq\/ values, (bottom left panel) the model results from \citet{R06}, and (bottom right panel) the model results from \citet{ZT18}.}
     \label{g2204_seds}
\end{figure*}

\begin{figure*}[hbt]
    \centering
    \includegraphics[width=\textwidth]{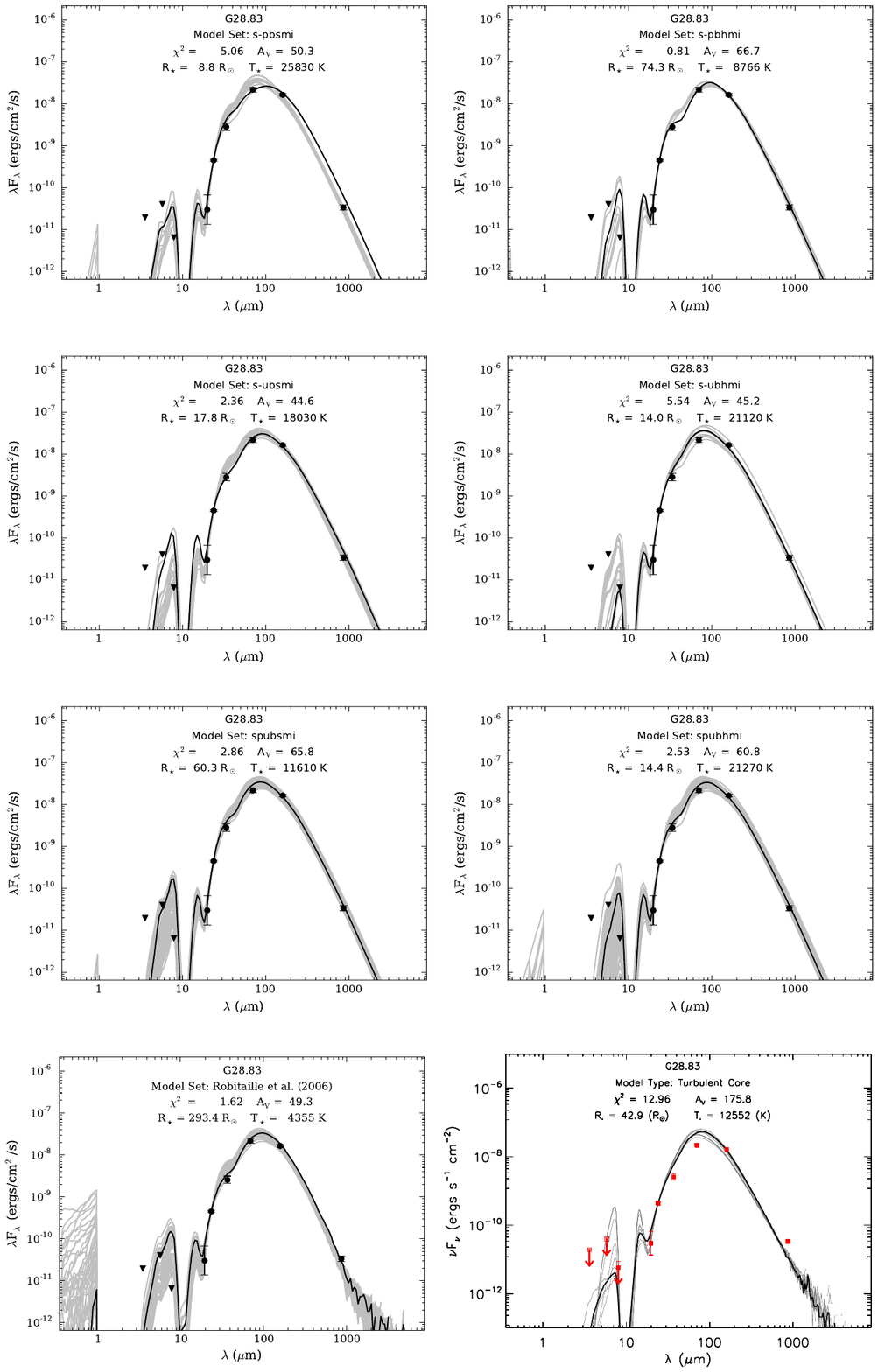}
    \caption{SED modeling results for G28.83$-$0.25, showing: (top six panels) the six best model sets from the \citet{R17} models based on \chisq\/ values, (bottom left panel) the model results from \citet{R06}, and (bottom right panel) the model results from \citet{ZT18}.}
    \label{g2883_seds}
\end{figure*}

\begin{figure*}[hbt]
    \centering
    \includegraphics[width=\textwidth]{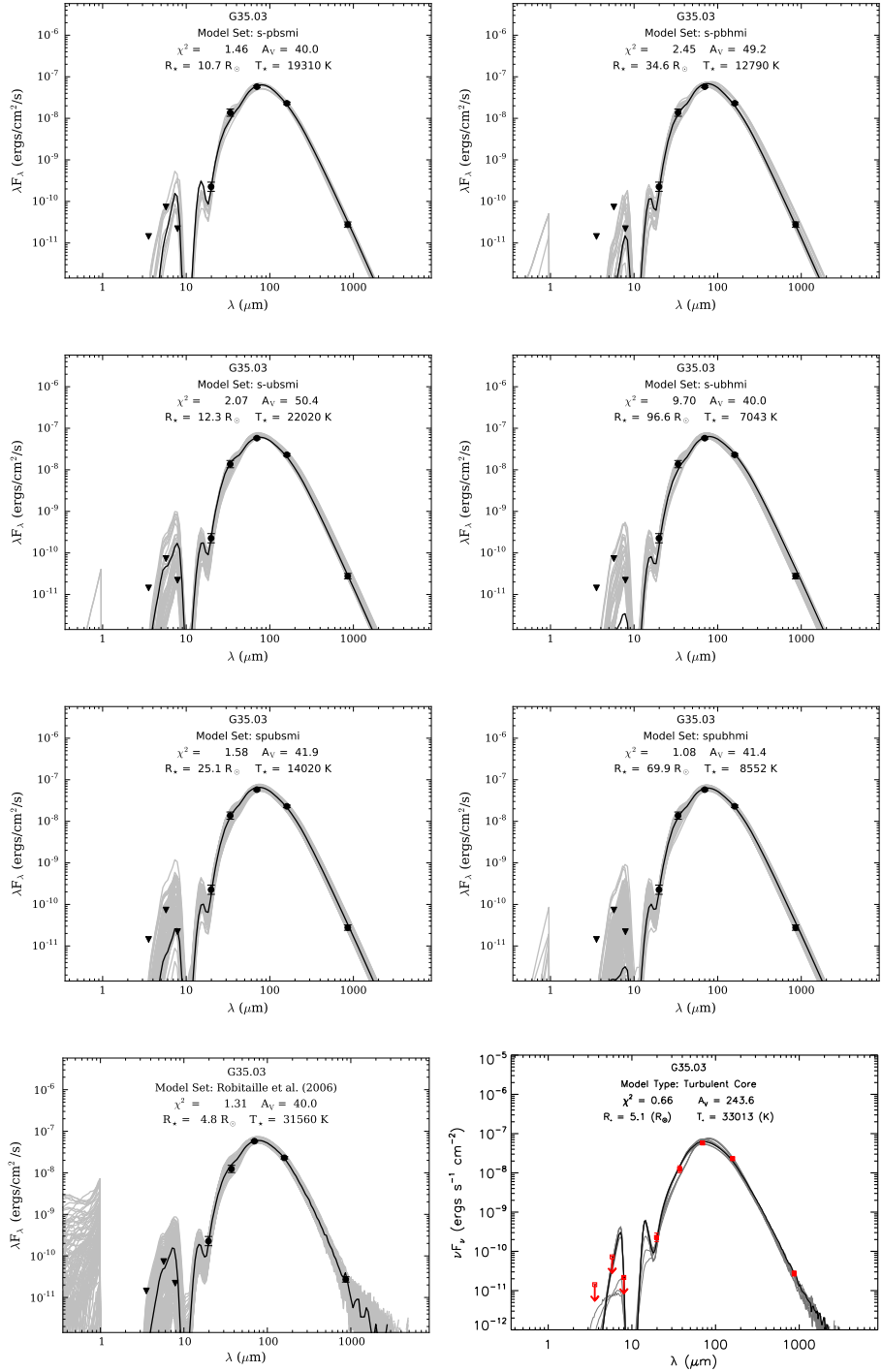}
    \caption{SED modeling results for G35.03$+$0.35, showing: (top six panels) the six best model sets from the \citet{R17} models based on \chisq\/ values, (bottom left panel) the model results from \citet{R06}, and (bottom right panel) the model results from \citet{ZT18}.}
    \label{g3503_seds}
\end{figure*}

\newpage
\section{B: Variations in the results of the \citet{R17} model packages with Bayesian versus \chisq\/ best-fit evaluations}
\citet{R17} stress that \chisq\/ values alone may not be the optimal method of evaluating which source geometry best represents one's data. 
In order to assess the model sets in comparison to each other in a statistically robust way, \citet{R17} suggest calculating P(D$|$M) $\propto N_{good}/N$, where $N_{good}$ is the total number of good models from a given model set, and $N$ is the total number of models in that set. 
This method accounts for the possibility that one model set may have produced an unusually low \chisq\/ value that is not representative of the quality of that model set overall. In this approach, the model set with the highest number of good models is the best model set for a given source. 
The definition of ``good'' in this case is determined by the user; \citet{R17} use the convention \chisq\/ $-$ \chisq$_{best}$ $<$ Xn$_{data}$, where \chisq$_{best}$ is the best \chisq\/ value across all model sets tested, n$_{data}$ is the number of flux values that are not upper or lower limits, and X is a multiplicative factor chosen by the user. For the majority of our sources, \chisq\/ $-$ \chisq$_{best}$ $<$ 3n$_{data}$ gives a reasonable split between good and bad fits; the exception is G14.63$-$0.58, for which we used $X = 5$ for reasons discussed below.
Table~\ref{bayes_chisq} shows which model set best represents each source according to both the \chisq$_{best}$ and P(D$|$M)$_{best}$ methods of determining ``best representation.''

\begin{deluxetable*}{lclc||lclc||lclc}[!hbt]   
\tablecaption{\chisq\/ and P(D$|$M) Scores for \citet{R17} Model Sets$^{a,b}$}
\tablecolumns{12}
\tablefontsize{\scriptsize}
\startdata  
\hline
\hline
G10.29$-$0.13 & & & & {\bf G10.34$-$0.14$^{c}$} & & & & G11.92$-$0.61 & & & \\
Model & \chisq\/ & Model & P(D$|$M) & Model & \chisq\/ & Model & P(D$|$M) & Model & \chisq\/ & Model & P(D$|$M) \\
\hline
spubhmi & 0.0003 & s-u-smi & 0.0519     & {\bf s-pbhmi} & 1.51  & {\bf s-pbhmi}  & 0.0029    & s-pbsmi & 3.14  & s-ubsmi & 0.0015 \\
spubsmi & 0.002  & spu-smi & 0.0422     & s-pbsmi       & 2.30  & s-pbsmi        & 0.0028    & s-pbhmi & 3.69  & s-pbsmi & 0.0009 \\
s-pbsmi & 0.004  & spubhmi & 0.0408625  & spubsmi       & 6.75  & spubhmi        & 0.0008625 & spubsmi & 3.87  & s-pbhmi & 0.0007 \\
s-ubsmi & 0.013  & s-pbhmi & 0.0384     & spubhmi       & 8.53  & s-ubsmi        & 0.0008    & spubhmi & 4.73  & spubsmi & 0.000675 \\
s-u-smi & 0.017  & s-ubhmi & 0.0359     & s-ubsmi       & 12.13 & s-ubhmi        & 0.0359    & s-ubhmi & 5.02  & s-ubhmi & 0.0006 \\
s-pbhmi & 0.022  & spubsmi & 0.02725    & s-ubhmi       & 15.52 & spubsmi        & 0.0325    & s-ubsmi & 6.08  & spubhmi & 0.0004125 \\
s-ubhmi & 0.028  & s-pbsmi & 0.0261     & spu-smi       & 21.70 & spu-smi        & 0         & s-u-smi & 34.70 & spu-smi & 0 \\
spu-smi & 0.044  & s-ubsmi & 0.0217     & s-u-smi       & 23.52 & s-u-smi        & 0         & spu-smi & 40.68 & s-u-smi & 0\\
\hline
\hline\\
{\bf G12.91$-$0.03} & & & & {\bf G14.33$-$0.64} & & & & G14.63$-$0.58 & & & \\
Model & \chisq\/ & Model & P(D$|$M) & Model & \chisq\/ & Model & P(D$|$M) & Model & \chisq\/ & Model & P(D$|$M) \\
\hline
{\bf s-pbhmi}   & 1.51  & {\bf s-pbhmi} & 0.0013   & {\bf s-pbsmi} & 0.45  & {\bf s-pbsmi} & 0.0034    & s-pbhmi & 3.68  & s-pbsmi & 0.0007 \\
spubhmi         & 1.84  & s-ubsmi       & 0.0013   & s-pbhmi       & 0.80  & s-pbhmi       & 0.003     & s-pbsmi & 4.41  & s-pbhmi & 0.0006 \\
spubsmi         & 2.84  & s-pbsmi       & 0.0011   & spubsmi       & 2.30  & s-ubhmi       & 0.0006    & spubsmi & 24.81 & spubsmi & 0.0001 \\
s-ubsmi         & 3.03  & s-ubhmi       & 0.0006   & spubhmi       & 4.40  & s-ubsmi       & 0.0006    & spubhmi & 24.87 & spubhmi & 0.00005 \\
s-pbsmi         & 3.99  & spubsmi       & 0.00055  & s-ubsmi       & 4.51  & spubhmi       & 0.000425  & s-ubhmi & 38.19 & s-ubhmi & 0 \\
s-ubhmi         & 4.84  & spubhmi       & 00003375 & s-ubhmi       & 4.94  & spubsmi       & 0.000425  & s-ubsmi & 47.43 & s-ubsmi & 0 \\
s-u-smi         & 28.58 & spu-smi       & 0        & spu-smi       & 12.78 & spu-smi       & 0.0003    & spu-smi & 67.87 & spu-smi & 0 \\
spu-smi         & 31.37 & s-u-smi       & 0        & s-u-smi       & 12.84 & s-u-smi       & 0.0002    & s-u-smi & 70.85 & s-u-smi & 0 \\
\hline
\hline\\
G16.59$-$0.05 & & & & {\bf G18.89$-$0.47} & & & & {\bf G19.36$-$0.03} & & & \\
Model & \chisq\/ & Model & P(D$|$M) & Model & \chisq\/ & Model & P(D$|$M) & Model & \chisq\/ & Model & P(D$|$M) \\
\hline
s-pbhmi & 1.12  & s-pbsmi & 0.0011   & {\bf s-pbhmi} & 1.01  & {\bf s-pbhmi} & 0.0008    & {\bf s-pbsmi} & 1.73  & {\bf s-pbsmi} & 0.0022 \\
s-pbsmi & 1.41  & s-ubsmi & 0.0008   & spubhmi       & 1.83  & s-pbsmi       & 0.0008    & s-pbhmi       & 2.23  & s-ubhmi       & 0.0019 \\
s-ubsmi & 2.32  & spubsmi & 0.00045  & s-pbsmi       & 4.13  & s-ubsmi       & 0.0007    & spubsmi       & 3.02  & s-pbhmi       & 0.0015 \\
spubsmi & 2.59  & spubhmi & 0.000425 & s-ubhmi       & 8.40  & spubsmi       & 0.000625  & s-ubsmi       & 3.81  & spubsmi       & 0.000625 \\
spubhmi & 2.60  & s-pbhmi & 0.0003   & s-ubsmi       & 9.51  & s-ubhmi       & 0.0002    & spubhmi       & 3.83  & s-ubsmi       & 0.0005 \\
s-ubhmi & 3.40  & s-ubhmi & 0.0003   & spubsmi       & 9.59  & spubhmi       & 0.0001    & s-ubhmi       & 10.00 & spubhmi       & 0.000425 \\
spu-smi & 32.60 & spu-smi & 0        & spu-smi       & 44.43 & spu-smi       & 0         & s-u-smi       & 20.86 & spu-smi       & 0 \\
s-u-smi & 35.47 & s-u-smi & 0        & s-u-smi       & 44.73 & s-u-smi       & 0         & spu-smi       & 22.85 & s-u-smi       & 0 \\
\hline
\hline\\
G22.04$+$0.22 & & & & G28.83$-$0.25 & & & & G35.03$+$0.35 & & & \\
Model & \chisq\/ & Model & P(D$|$M) & Model & \chisq\/ & Model & P(D$|$M) & Model & \chisq\/ & Model & P(D$|$M) \\
\hline
s-pbhmi & 1.81  & spubsmi & 0.001925 &   s-pbhmi & 0.81  & spubsmi & 0.00255  & spubhmi & 1.08  & s-ubsmi & 0.0056 \\
spubhmi & 2.37  & s-pbhmi & 0.0019   &   s-ubsmi & 2.36  & s-ubsmi & 0.0023   & s-pbsmi & 1.46  & s-ubhmi & 0.0034 \\
s-pbsmi & 2.41  & s-pbsmi & 0.0019   &   spubhmi & 2.53  & s-ubhmi & 0.002    & spubsmi & 1.58  & spubhmi & 0.0033 \\
s-ubsmi & 2.55  & s-ubsmi & 0.0019   &   spubsmi & 2.86  & s-pbsmi & 0.002    & s-ubhmi & 1.94  & s-pbhmi & 0.0033 \\
s-ubhmi & 2.59  & s-u-smi & 00017    &   s-pbsmi & 5.06  & spubhmi & 0.00195  & s-ubsmi & 2.07  & spubsmi & 0.002775 \\
spubsmi & 2.61  & spubhmi & 0.001025 &   s-ubhmi & 5.54  & s-pbhmi & 0.0011   & s-pbhmi & 2.45  & s-u-smi & 0.0019 \\
s-u-smi & 16.97 & s-ubhmi & 0.001    &   s-u-smi & 18.12 & s-u-smi & 0.0006   & spu-smi & 9.47  & spu-smi & 0.0017 \\
spu-smi & 19.13 & spu-smi & 0.0005   &   spu-smi & 20.02 & spu-smi & 0        & s-u-smi & 12.47 & s-pbsmi & 0.0014
\enddata
\tablenotetext{a}{The \chisq\/ values shown are \chisq\/ per data point, as defined in \S~\ref{seds}.}
\tablenotetext{b}{Model sets are shown for each source in order of best (top row) to worst (bottom row). The first two columns for each source are the best model sets and \chisq\/ values according to the \chisq\/ method, and the second two columns for each source are the best model sets and \chisq\/ values according to the Bayesian method. We show all eight model sets used in order to more clearly illustrate source trends according to both the \chisq\/ and P(D$|$M) methods $-$ that is, to illustrate trends not only in which model sets produce the best fits, but also in which model sets produce the worst fits.}
\tablenotetext{c}{In cases where the \chisq\/ and P(D$|$M) methods produce the same best-fit model set, source names and best-fit model set names are highlighted in {\bf bold}.}
\label{bayes_chisq}
\end{deluxetable*}

The \chisq\/ and P(D$|$M) approaches yield the same best-fit model set in five cases (G10.34$-$0.14, G12.91$-$0.03, G14.33$-$0.64, G18.89$-$0.47, and G19.36$-$0.03), and different model sets in seven. Sources for which both approaches yield the same best-fit model set have their source name and best model set name marked in bold in Table~\ref{bayes_chisq}. Interestingly, both the \chisq\/ and Bayesian approaches tend to yield the same or very similar overall trends, such as, e.g., a strong preference for a power-law envelope or a slight preference for an inner hole. There are four sources for which the overall trends disagree on the presence or absence of at least one physical component (e.g. disk, inner hole); for one of these sources, the \chisq\/ and Bayesian methods return results that differ on every physical component. It is worth noting that the four sources for which the trends identified by the \chisq\/ and P(D$|$M) methods show disagreement either lack a 24~\mum\/ data point (the 24~\mum\/ flux densities have the lowest uncertainties in $\lambda$F$_{\lambda}$-space, so a 24~\mum\/ non-detection has an outsize effect on the \chisq\/ values for all models for that source) or suffer from confusion problems at 160~\mum\/ as discussed in \S~\ref{fir_phot}. Interestingly, confusion problems at 70~\mum\/ do not seem to produce similar disagreements in the model results. 

Below, we discuss best-fit geometries for each source in detail, using both the \chisq\/ and Bayesian methods of determining ``best model.''

\paragraph{G10.29$-$0.13}
The \chisq\/ and Bayesian methods yield different best-fit model sets for this source (spubhmi for the former and s-u-smi for the latter), but both yield a general preference for a rotating-infalling (Ulrich-type) envelope and a slight preference for both a passive disk and no inner hole. All results for this particular source should be taken with the caveat that this source is quite poorly constrained (six of the nine flux densities used for SED modeling are upper limits).

\paragraph{G10.34$-$0.14}
Both methods yield the same trends for this source: a general preference for a power-law envelope and no passive disk for this source, with no real preference as to whether or not there is an inner hole. There is a significant increase (approximately a factor of 3) in \chisq\/ values between the two lowest-\chisq\/ model sets, which have no disk, and the third-best, which does. Likewise, there is a jump of approximately a factor of 3.5 in P(D|M) between the two best model sets (which do not have a passive disk) and the third-best, which does.

\paragraph{G11.92$-$0.61}
All preferred \citet{R17} model sets for G11.92$-$0.61 favor a power-law envelope and no disk, with no strong preference as to the presence or absence of an inner hole. This is true for both the \chisq\/ and P(D$|$M) methods of determining the best model set. The fact that the results favor not having a disk is in direct contradiction with our knowledge of this source from high-resolution centimeter- and millimeter-wavelength observations \citep[see][]{Ilee2016,Ilee2018}. This disagreement is likely a result of the comparatively poor resolution of our SOFIA and archival infrared observations ($\sim$1$\arcsec$ to 19$\farcs$2) as compared with the millimeter observations \citep[$\sim$0$\farcs$09 to $\sim$0$\farcs$75;][]{Ilee2016,Ilee2018}; the infrared observations simply do not have sufficient resolution to distinguish the necessary small-scale structure in such a clustered source.

The discrepancy between the SED modeling results and the results of \citet{Ilee2016,Ilee2018} warrant a closer look.  Based on $\sim0\farcs5$ (1550 au)-resolution Submillimeter Array (SMA) data, \citet{Ilee2016} estimated a disk gas mass of $\sim$2$-$3 \msun\/ and an enclosed mass $M_{\rm enc}\sim$30$-$60 \msun.  Using $\sim0\farcs09$ (310 au)-resolution ALMA observations, \citet{Ilee2018} find an enclosed mass $M_{\rm enc}$ of 40 $\pm$ 5 \msun\/ and a disk gas mass of $\sim$2$-$6 \msun.  The \citet{R06} best-fit model returns a central source mass of 13.4 \msun\/ and a disk gas mass of 4.6$\times$10$^{-2}$ \msun, while the \citet{ZT18} best-fit model gives a central source mass of 16 \msun\/ and a disk mass of 5.33 \msun\/ ($\frac{1}{3}$ the mass of the central source, as discussed in \S~\ref{zt18_models}). In this case, while the \citet{R17} models do not favor the known physical geometry, the \citet{R06} and \citet{ZT18} models do not reproduce the observationally-derived stellar and disk masses. That is, none of the three models accurately describes the known physical parameters of this source.

\paragraph{G12.91$-$0.03}
While the \chisq\/ and P(D$|$M) methods do produce the same best-fit model set for this source, the overall trends in the \chisq\/ and P(D$|$M) results disagree. Neither method particularly seems to favor one envelope type over another. However, the \chisq\/ results overall favor models which have an inner hole and a passive disk, whereas the P(D$|$M) evaluation shows a strong preference for having no disk, but no preference as to the presence or absence of an inner hole.

\paragraph{G14.33$-$0.64}
Both the \chisq\/ values and P(D$|$M) show a strong preference for having no disk and a power-law envelope, and no real preference as to the presence or absence of an inner hole. 

\paragraph{G14.63$-$0.58}
Both methods show, for this source, a strong preference for no passive disk, and for a power-law envelope, with no real preference for or against an inner hole. The P(D$|$M) results for this source should be considered carefully, however. G14.63$-$0.58 has few good fits in any model set, and so for this source we used the cutoff \chisq\/ $-$ \chisq$_{best}$ $<$ 5n$_{data}$ instead.

\paragraph{G16.59$-$0.05}
The \chisq\/ and P(D|M) evaluations both favor models with no inner hole and no disk, but differ as to envelope type. The \chisq\/ values suggest that a power-law envelope produces the best fit to our data, while the Bayesian approach suggests an Ulrich-type envelope instead. G16.59$-$0.05 is saturated in the MIPSGAL data, so its SED lacks a flux density at that wavelength. MIR emission in YSOs tends to be dominated by emission from the protostellar envelope and/or outflow cavities, and our fitted 24~\mum\/ flux densities were usually the best-constrained data points for a given source. It is possible that the discrepancy in preferred envelope type is due to this combination of factors.

As for the fact that both evaluations of the \citet{R17} results favor model sets with no disk, this is another source for which additional data in the literature show this implication to be incorrect.
\citet{Moscadelli2016} identify G16.59$-$0.05 as a $\sim$20 \msun\/ YSO with a disk/jet system. 
The central source appears as compact K{\it u}-  and K-band continuum emission at 0$\farcs$2 and 0$\farcs$1 resolutions, respectively, and the rotating disk is traced by multi-epoch EVN observations of 6.7~GHz \methanol\/ masers associated with the compact K{\it u}- and K-band emission \citep[for details of the EVN observations, see][and references therein]{Moscadelli2016}. 
The jet is traced by extended C-band emission ($\sim$ 6~cm, 0$\farcs$4 resolution) in both \citet{Moscadelli2016} and \citet{Rosero2016}.

Neither \citet{Moscadelli2016} nor \citet{Rosero2016} estimate disk mass or accretion rate, so we cannot asses the quality of the \citet{R06} and \citet{ZT18} results in that context. However, unlike as for G11.92$-$0.61, both the \citet{R06} and \citet{ZT18} models do give results for protostellar mass (15\msun\/ and 16\msun, respectively) that are fairly well in line with the mass reported by \citet{Moscadelli2016}. While the specific parameters of the disk cannot be explored at this time, we can state that the \citet{R06} and \citet{ZT18} \mstar\/ results are consistent, for the moment, with the results available in the literature.

\paragraph{G18.89$-$0.47}
Both methods of evaluation favor models with no disk and a power-law envelope for this source. However, the Bayesian approach suggests that models without an inner hole more accurately fit the data, whereas the \chisq\/ values favor models that do have an inner hole. G18.89$-$0.47 is one of our sources with a confusion problem at 160~\mum.

\paragraph{G19.36$-$0.03}
Both the \chisq\/ and P(D$|$M) values for this source strongly favor models with no disk, but neither shows any particular trend in envelope type or presence/absence of an inner hole.

\paragraph{G22.04$+$0.22}
Both methods of model evaluation agree for this source: models which have no disk and no inner hole are favored, but there is no strong preference as to envelope type.

\paragraph{G28.83$-$0.25}
The \chisq\/ and P(D$|$M) values for this source both favor models with an Ulrich-type envelope and no disk, with either no or a very slight preference for models with no inner hole.

\paragraph{G35.03$+$0.35}
This is the one source for which the \chisq\/ and P(D$|$M) values produce entirely different trends. While the model sets with the best \chisq\/ values notably lack trends for any particular physical components, the P(D$|$M) values show a strong preference for models with an Ulrich envelope, no disk, and an inner hole. This is the one source which has a known UC\HII\/ region in the EGO itself (as opposed to an \HII\/ region nearby but not within the ATLASGAL clump which hosts the EGO, as is the case with G14.33$-$0.64).

\end{document}